\documentclass[preprint,showpacs,preprintnumbers,amsmath,amssymb]{article}
\usepackage{txfonts}

\usepackage{graphicx}
\usepackage{dcolumn}
\usepackage{bm}

\begin{document}

\title{\textbf{\textsf{Fundamental Structure of Loop Quantum Gravity}}}

\author{\\ \\ \textsf{Muxin Han$^{1,2}$}\footnote{Email\ address:\ mhan1@lsu.edu},\ \
\textsf{Weiming Huang$^{3}$},\ and
\textsf{Yongge Ma$^{1}$}\footnote{Email\ address:\ mayg@bnu.edu.cn}\\ \\
\small \textsf{$1.$ Department of Physics, Beijing Normal University, }\\
\small \textsf{Beijing 100875, CHINA}\\ \small $2.$ \textsf{Horace
Hearne Jr.
Institute for Theoretical Physics,} \\ \small \textsf{Louisiana State University,} \\
\small \textsf{Baton Rouge, LA 70803, USA}\\
\small \textsf{$3.$ School of Mathematics, Beijing Normal University, }\\
\small \textsf{Beijing 100875, CHINA}}

\date{\textsf{\today}}

\maketitle

\newpage
\begin{abstract}
In recent twenty years, loop quantum gravity, a background
independent approach to unify general relativity and quantum
mechanics, has been widely investigated. The aim of loop quantum
gravity is to construct a mathematically rigorous, background
independent, nonperturbative quantum theory for Lorentzian
gravitational field on four-dimensional manifold. In the approach,
the principles of quantum mechanics are combined with those of
general relativity naturally. Such a combination provides us a
picture of, so-called, quantum Riemannian geometry, which is
discrete at fundamental scale. Imposing the quantum constraints in
analogy from the classical ones, the quantum dynamics of gravity is
being studied as one of the most important issues in loop quantum
gravity. On the other hand, the semi-classical analysis is being
carried out to test the classical limit of the quantum theory.

In this review, the fundamental structure of loop quantum gravity
is presented pedagogically. Our main aim is to help non-experts to
understand the motivations, basic structures, as well as general
results. It may also be beneficial to practitioners to gain
insights from different perspectives on the theory. We will focus
on the theoretical framework itself, rather than its applications,
and do our best to write it in modern and precise langauge while
keeping the presentation accessible for beginners. After reviewing
the classical connection dynamical formalism of general
relativity, as a foundation, the construction of kinematical
Ashtekar-Isham-Lewandowski representation is introduced in the
content of quantum kinematics. The algebraic structure of quantum
kinematics is also discussed. In the content of quantum dynamics,
we mainly introduce the construction of a Hamiltonian constraint
operator and the master constraint project. At last, some
applications and recent advances are outlined. It should be noted
that this strategy of quantizing gravity can also be extended to
obtain other background independent quantum gauge theories. There
is no divergence within this background independent and
diffeomorphism invariant quantization programme of matter coupled
to gravity.

\end{abstract}

{PACS number(s): 04.60.Pp, 04.60.Ds}

\newpage

\tableofcontents
\newpage

\section{Introduction}
\subsection{Motivation of Quantum Gravity}

Nowadays, in traditional view there are four elementary
interactions widely understood by community of physicists: strong
interaction, weak interaction, electromagnetic interaction and
gravitational interaction. The description for the former three
kinds of forces is quantized in the well-known standard model. The
interactions are transmitted by exchanging mediate particles.
However, the last kind of interaction, gravitational interaction,
is described by Einstein's theory of general relativity, which is
absolutely a classical theory which describes the gravitational
field as a smooth metric tensor field on a manifold, i.e., a
4-dimensional spacetime geometry. There is no $\hbar$ and hence no
discrete structure of spacetime. Thus there is a fundamental
inconsistency in our current description of the whole physical
world. Physicists widely accept the assumption that our world is,
so called, quantized at fundamental level. So all interactions
should be brought into the framework of quantum mechanics
fundamentally. As a result, the gravitational field should also
have "quantum structure".

Throughout the last century, our understanding of the nature has
considerably improved from macroscale to microscale, including the
phenomena in molecule scale, atom scale, sub-atom scale, and
elementary particle scale. The standard model of particle physics
coincides almost with all present experimental tests in laboratory
(see e.g. \cite{weinberg}). However, because unimaginably large
amount of energy would be needed, no one has understood how the
physical process happens near the Planck scales
$\ell_p\equiv(G\hbar/{c^3})^{1/2}\sim10^{-33}cm$ and
$t_p\equiv(G\hbar/{c^5})^{1/2}\sim10^{-43}s $, which are viewed as
the most fundamental scales. The Planck scale arises naturally in
attempts to formulate a quantum theory of gravity, since $\ell_p$
and $t_p$ are unique combinations of speed of light $c$, Planck
constant $\hbar$, and gravitational constant $G$, which have the
dimensions of length and time respectively. The dimensional
arguments suggest that at Planck scale the smooth structure of
spacetime should break down, where the well-known quantum field
theory is invalid since it depends on a fixed smooth background
spacetime. Hence we believe that physicists should go beyond the
successful standard model to explore the new physics near Planck
scale, which is, perhaps, a quantum field theory without a
background spacetime, and this quantum field theory should include
the quantum theory of gravity. Moreover, current theoretical
physics is thirsting for a quantum theory of gravity to solve at
least the following fundamental difficulties.
\begin{itemize}
\item {\it Classical Gravity - Quantum Matter Inconsistency}

The most crucial equation to perform the relation between the
matter field and gravitational field is the famous Einstein field
equation:
\begin{equation}
R_{\alpha\beta}[g]-\frac{1}{2}R[g]g_{\alpha\beta}=\kappa
T_{\alpha\beta}[g],\label{ein}
\end{equation}
where the left hand side of the equation concerns spacetime
geometry which has classical smooth structure, while the right
hand side concerns also matter field which is fundamentally
quantum mechanical in standard model. In quantum field theory the
energy-momentum tensor of matter field should be an
operator-valued tensor $\hat{T}_{\alpha\beta}$. One possible way
to keep classical geometry consistent with quantum matter is to
replace $T_{\alpha\beta}[g]$ by the expectation value
$<\hat{T}_{\alpha\beta}[g]>$ with respect to some quantum state of
the matter on a fixed spacetime. A primary attempt is to consider
the vacuum expectation. However, in the solution of this equation
the background $g_{\alpha\beta}$ has to be changed due to the
non-vanishing of $<\hat{T}_{\alpha\beta}[g]>$. So one has to feed
back the new metric into the definition of the vacuum expectation
value etc. The result of the iterations does not converge in
general \cite{FM}. This inconsistency motivates us to quantize the
background geometry to arrive at an operator formula also on the
left hand side of Eq.(\ref{ein}).

\item {\it Singularity in General Relativity}

Einstein's theory of General Relativity is considered as one of
the most elegant theories in 20th century. Many experimental tests
confirm the theory in classical domain \cite{will}. However,
Penrose and Hawking proved that singularities are inevitable in
general spacetimes with several tempered conditions on energy and
causality by the well-known singularity theorem (for a summary,
see \cite{hawking}\cite{wald}). Thus general relativity can not be
valid unrestrictedly. One naturally expects that, in extra strong
gravitational field domains near the singularities, the
gravitational theory would probably be replaced by an unknown
quantum theory of gravity.

\item {\it Infinity in Quantum Field Theory}

It is well known that there are infinity problems in quantum field
theory in Minkowski spacetime. In curved spacetime, the problem of
UV divergence is even more serious because of the interacting
fields. Although much progress on the renormalization for
interacting fields have been made \cite{wald2}\cite{wald1}, a
fundamentally satisfactory theory is still far from reaching. So it
is expected that some quantum gravity theory, playing a fundamental
role at Planck scale, would provide a natural cut-off to cure the UV
singularity in quantum field theory. The situation of quantum field
theory on a fixed spacetime looks just like that of quantum
mechanics for particles in electromagnetic field before the
establishing of quantum electrodynamics, where the particle
mechanics (actress) is quantized but the background electromagnetic
field (stage) is classical. The history suggests that such a
semi-classical situation is only an expedient which should be
replaced by a much more fundamental and satisfactory theory.

\end{itemize}

\subsection{Purpose of Loop Quantum Gravity}

The research on quantum gravity theory is rather active. Many
quantization programmes for gravity are being carried out (for a
summary see e.g. \cite{thiemann2}). In these different kinds of
approaches, the idea of loop quantum gravity is motivated by
researchers in the community of general relativity. It follows
closely the thoughts of general relativity, and hence it is a
quantum theory born with background independency. Roughly
speaking, loop quantum gravity is an attempt to construct a
mathematically rigorous, non-perturbative, background independent
quantum theory of four-dimensional, Lorentzian general relativity
plus all known matter in the continuum. The project of loop
quantum gravity inherits the basic idea of Einstein that gravity
is fundamentally spacetime geometry. Here one believes in that the
theory of quantum gravity is a quantum theory of spacetime
geometry with diffeomorphism invariance (this legacy is discussed
comprehensively in Rovelli's book \cite{rovelli}). To carry out
the quantization procedure, one first casts general relativity
into the Hamiltonian formalism of a diffeomorphism invariant
Yang-Mills gauge field theory with a compact internal gauge group.
Thus the construction of loop quantum gravity is valid to all
background independent gauge field theories. So the theory can
also be called as a background independent quantum gauge field
theory.

All classical fields theories, other than gravitational field, are
defined on a fixed spacetime, which provides a foundation to the
perturbative Fock space quantization. However general relativity
is only defined on a manifold and hence is the unique background
independent classical field theory, since gravity itself is the
background. So the situation for gravity is much different from
other fields by construction \cite{rovelli}, namely gravity is not
only the background stage, but also the dynamical actress. Such a
double character for gravity leads to many difficulties in the
understanding of general relativity and its quantization, since we
cannot analog the strategy in ordinary quantum theory of matter
fields. However, an amazing result in loop quantum gravity is that
the background independent programme can even enlighten us to
avoid the difficulties in ordinary quantum field theory. In
perturbative quantum field theory in curved spacetime, the
definition of some basic physical quantities, such as the
expectation value of energy-momentum, is ambiguous and it is
difficult to calculate the back-reaction of quantum fields to the
background spacetime \cite{wald1}. One could speculate on that the
difficulty is related to the fact that the present formulation of
quantum field theories is background dependent. For instance, the
vacuum state of a quantum field is closed related to spacetime
structure, which plays an essential role in the description of
quantum field theory in curved spacetime and their renormalization
procedures. However, if the quantization programme is by
construction background independent and non-perturbative, it is
possible to solve the problems fundamentally. In loop quantum
gravity there is no assumption of a priori background metric at
all and the gravitational field and matter fields are coupled and
fluctuating naturally with respect to each other on a common
manifold.

In the following sections, we will review pedagogically the basic
construction of a completely new, background independent quantum
field theory, which is completely different from the known quantum
fields theory. For completeness and accuracy, we will not avoid
mathematical terminologies. While, for simplicity, we will skip the
complicated proofs of many important statements. One may find the
missing details in the references cited. Thus our review will not be
comprehensive. We refer to Ref.\cite{thiemann2} for a more detailed
exploration, Refs. \cite{AL} and \cite{lecture} for more advanced
topics. It turns out that in the framework of loop quantum gravity
all theoretical inconsistencies introduced in the previous section
are likely to be cured. More precisely, one will see that there is
no UV divergence in quantum fields of matter if they are coupled
with gravity in the background independent approach. Also recent
works show that the singularities in general relativity can be
smeared out in the symmetry-reduced models
\cite{bojowald}\cite{singular}\cite{boj5}. The crucial point is that
gravity and matter are coupled and consistently quantized
non-perturbatively so that the problems of classical gravity and
quantum matter inconsistency disappear.

\section{Classical Framework of Connection Dynamics}
\subsection{Lagrangian Formalism}
In order to canonically quantize the classical system of gravity,
Hamiltonian analysis has to be performed to obtain a canonical
formalism of the classical theory suitable to be represented on
certain Hilbert space. The first canonical formalism of general
relativity is the ADM formalism (Geometric dynamics) from the
Einstein-Hilbert action\cite{wald}\cite{Liang}, which by now has
not been cast into a quantum theory rigorously. Another well-known
action of general relativity is the Palatini formalism, where the
tetrad and the connection are regarded as independent dynamical
variables. However, unluckily the dynamics of Palatini action is
the same with the Einstein-Hilbert action for the gravitational
field without fermion coupling \cite{Ash}\cite{han}. In 1986,
Ashtekar gave a formalism of true connection dynamics with a
relatively simple Hamiltonian constraint, and thus opens the door
to apply quantization techniques from gauge fields theory
\cite{Ash1}\cite{Ash2}\cite{RS}. However the weakness of that
formalism is that the canonical variables are complex variables,
which needs a complicated real section condition. Moreover, the
quantization based on the complex connection could not be carried
out rigorously, since the internal gauge group is noncompact. In
1995, Barbero modified the Ashtekar new variables to give a system
of real canonical variables for dynamical theory of connections
\cite{barbero}. Then Holst constructed a generalized Palatini
action to support Barbero's real connection dynamics \cite{holst}.
Although there is a free parameter (Barbero-Immirzi parameter
$\beta$) in generalized Palatini action and the Hamiltonian
constraint is more complicated than the Ashtekar one, now the
generalized Palatini Hamiltonian with the real connections is
widely accepted by loop theorists for the quantization
programme\footnote{One may take the other viewpoint that the
transition from complex connection to real variables is only a
mathematical convenience at the present stage, since we do not
have a rigorous framework to deal with the infinite dimensional
space of connections with non-compact internal group. Some
researchers are working on this generalization
\cite{FLncs}\cite{okolow1}\cite{okolow2}.}. All the following
analysis is based on the generalized Palatini formalism.

Consider an 4-manifold, $M$, on which the basic dynamical
variables in the generalized Palatini framework are tetrad
$e_I^{\alpha}$ and $so(1,3)$-valued connection $\omega_{\alpha}^{\
IJ}$(not necessarily torsion-free), where the capital Latin
indices $I,\ J,...$ denote the internal $SO(1,3)$ group and the
Greek indices $\alpha,\ \beta,...$ denote spacetime indices. A
tensor with both spacetime indices and internal indices is named
as a generalized tensor. The internal space is equipped with a
Minkowskian metric $\eta_{IJ}$ (of signature $-, +, +, +$) fixed
once for all, such that the spacetime metric reads:
\begin{displaymath}
g_{\alpha\beta}=\eta_{IJ}e^{I}_{\alpha}e^{J}_{\beta}.
\end{displaymath}
The generalized Palatini action in which we are interested is
given by \cite{AL}:
\begin{eqnarray}
S_{p}[e_{K}^{\beta},\omega_{\alpha}^{\ IJ}]
=\frac{1}{2\kappa}\int_{M}d^4x(e)
e_{I}^{\alpha}e_{J}^{\beta}(\Omega_{\alpha\beta}^{\ \
IJ}+\frac{1}{2\beta}\epsilon^{IJ}_{\ \ KL}\Omega_{\alpha\beta}^{\ \
KL}) \label{action},
\end{eqnarray}
where $e$ is the square root of the determinant of the metric
$g_{\alpha\beta}$, $\epsilon^{IJ}_{\ \ KL}$ is the internal
Levi-Civita symbol, $\beta$ is the Barbero-Immirzi parameter, which
we fix to be real, and the $so(1,3)$-valued curvature 2-form
$\Omega_{\alpha\beta}^{\ \ IJ}$ of the connection
$\omega_{\alpha}^{\ IJ}$ reads:
\begin{eqnarray}
\Omega_{\alpha\beta}^{\ \
IJ}:=2\mathcal{D}_{[\alpha}\omega_{\beta]}^{\ IJ} =
\partial_\alpha\omega^{\ IJ}_{\beta}-\partial_\beta\omega^{\ IJ}_{\alpha}+\omega_{\alpha}^{\
IK}\wedge\omega_{\beta K}^{\ \ \ J},\nonumber
\end{eqnarray}
here $\mathcal{D}_\alpha$ denote the $so(1,3)$ generalized covariant
derivative with respect to $\omega_{\alpha}^{\ IJ}$ acting on both
spacetime and internal indices. Note that the generalized Palatini
action returns to the Palatini action when $\frac{1}{\beta}=0$ and,
if a complex Barbero-Immirzi parameter is assumed, gives the
(anti)self-dual Ashtekar formalism when one sets
$\frac{1}{\beta}=\pm i$. Moreover, besides spacetime diffeomorphism
transformations, the action is also invariant under internal
$SO(1,3)$ rotations:
\begin{eqnarray}
(e,\omega)\mapsto({e'},{\omega'})=(b^{-1}{e},b^{-1}{\omega}b+b^{-1}{d}b),
\label{gauge}\nonumber
\end{eqnarray}
for any $SO(1,3)$ valued function $b$ on $M$. The gravitational
field equations are obtained by varying this action with respect
to $e_{I}^{\alpha}$ and $\omega_{\alpha}^{\ IJ}$. We first study
the variation with respect to the connection $\omega_{\alpha}^{\
IJ}$. One has
\begin{eqnarray}
\delta\Omega_{\alpha\beta}^{\ \ IJ}=(d\ \delta\omega^{\
IJ})_{\alpha\beta}+\delta\omega_{\alpha}^{\ IK}\wedge\omega_{\beta
K}^{\ \ \ J}+\omega_{\alpha}^{\ IK}\wedge\delta\omega_{\beta K}^{\
\ \ J}=2\mathcal{D}_{[\alpha}\delta\omega_{\beta]}^{\ IJ}\nonumber
\end{eqnarray}
by the definition of covariant generalized derivative
$\mathcal{D}_\alpha$. Note that $\delta\omega_\alpha^{\ IJ}$ is a
Lorentz covariant generalized tensor field since it is the
difference between two Lorentz connections
\cite{peldan}\cite{nakahara}. One thus obtains
\begin{eqnarray}
\delta S_{p}&=&\frac{1}{2\kappa}\int_{M}d^4x(e)
e_{I}^{\alpha}e_{J}^{\beta}(\delta\Omega_{\alpha\beta}^{\ \
IJ}+\frac{1}{2\beta}\epsilon^{IJ}_{\ \
KL}\delta\Omega_{\alpha\beta}^{\ \ KL})\nonumber\\
&=&-\frac{1}{\kappa}\int_M(\delta\omega_\beta^{\ \
IJ}+\frac{1}{2\beta}\epsilon^{IJ}_{\ \ KL}\delta\omega_{\beta}^{\ \
KL})\mathcal{D}_\alpha[(e)e_{I}^{\alpha}e_{J}^{\beta}],\nonumber
\end{eqnarray}
where we have used the fact that $\mathcal{D}_\alpha
\widetilde{\lambda}^\alpha=\partial_\alpha\widetilde{\lambda}^\alpha$
for all vector density $\widetilde{\lambda}^\alpha$ of weight $+1$
and neglected the surface term. Then it gives the equation of
motion:
\begin{eqnarray}
\mathcal{D}_\alpha[(e)e_{I}^{\alpha}e_{J}^{\beta}]
=-\frac{1}{4}\mathcal{D}_\alpha[\widetilde{\eta}^{\alpha\beta\gamma\delta}\epsilon_{IJKL}e^K_\gamma
e^L_\delta]=0,\nonumber
\end{eqnarray}
where $\widetilde{\eta}^{\alpha\beta\gamma\delta}$ is the
spacetime Levi-Civita symbol. This equation leads to the
torsion-free Cartan's first equation:
\begin{eqnarray}
\mathcal{D}_{[\alpha} e^I_{\beta]}=0,\nonumber
\end{eqnarray}
which means that the connection $\omega_\alpha^{\ IJ}$ is the
unique torsion-free Levi-Civita spin connection compatible with
the tetrad $e^\alpha_I$. As a result, the second term in the
action (\ref{action}) can be calculated as:
\begin{eqnarray}
(e)e_{I}^{\alpha}e_{J}^{\beta}\epsilon^{IJKL}\Omega_{\alpha\beta
KL}=\eta^{\alpha\beta\gamma\delta}R_{\alpha\beta\gamma\delta},\nonumber
\end{eqnarray}
which is exactly vanishing, because of the symmetric properties of
Riemann tensor. So the generalized Palatini action returns to the
Palatini action, which will certainly give the Einstein field
equation.
\subsection{Hamiltonian Formalism}
To carry out the Hamiltonian analysis of action (\ref{action}),
suppose the spacetime $M$ is topologically
$\Sigma\times\mathbf{R}$ for some 3-dimensional compact manifold
$\Sigma$ without boundary. We introduce a foliation parameterized
by a smooth function $t$ and a time-evolution vector field
$t^\alpha$ such that $t^\alpha (dt)_\alpha=1$ in $M$, where
$t^\alpha$ can be decomposed with respect to the unit normal
vector $n^\alpha$ of $\Sigma$ as:
\begin{equation}
t^\alpha=Nn^\alpha+N^\alpha, \nonumber
\end{equation}
here $N$ is called the \textit{lapse function} and $N^\alpha$ the
\textit{shift vector}\cite{wald}\cite{Liang}. The internal normal
vector is defined as $n_I\equiv n_\alpha e^\alpha_I$. It is
convenient to carry out a partial gauge fixing, i.e., fix a internal
constant vector field $n^I$ with $\eta_{IJ}n^In^J=-1$. Note that the
gauge fixing puts no restriction on the real
dynamics\footnote{However, there are some arguments that such a
gauge fixing is a non-natural way to break the internal Lorentz
symmetry (see e.g. \cite{samuel}).}. Then the internal vector space
$V$ is 3+1 decomposed with a 3-dimensional subspace $W$ orthogonal
to $n^I$, which will be the internal space on $\Sigma$. With respect
to the internal normal $n^I$ and spacetime normal $n^\alpha$, the
internal and spacetime projection maps are denoted by $q_i^I$ and
$q_a^\alpha$ respectively, where we use $i,j,k,...$ to denote the
3-dimensional internal space indices and $a,b,c,...$ to denote the
indices of space $\Sigma$. Then an internal reduced metric
$\delta_{ij}$ and a reduced spatial metric on $\Sigma$, $q_{ab}$,
are obtained by these two projection maps. The two metrics are
related by:
\begin{equation}
q_{ab}=\delta_{ij}e^i_ae^j_b, \nonumber
\end{equation}
where the orthonormal co-triad on $\Sigma$ is defined by
$e^i_a:=e^I_\alpha q^i_Iq^\alpha_a$. Now the internal gauge group
$SO(1,3)$ is reduced to its subgroup $SO(3)$ which leaves $n^I$
invariant. Finally, two Levi-Civita symbols are obtained
respectively as
\begin{eqnarray}
\epsilon_{ijk}&:=&q_i^Iq_j^Jq_k^Kn^L\epsilon_{LIJK},\nonumber\\
\underline{\eta}_{abc}&:=&q_a^\alpha q_b^\beta q_c^\gamma
t^\mu\underline{\eta}_{\mu\alpha\beta\gamma}, \nonumber
\end{eqnarray}
where the internal Levi-Civita symbol $\epsilon_{ijk}$ is an
isomorphism of Lie algebra $so(3)$. Using the connection 1-form
$\omega_\alpha^{\ IJ}$, one can defined two $so(3)$-valued 1-form
on $\Sigma$:
\begin{eqnarray}
\Gamma_a^i&:=&\frac{1}{2}q^\alpha_aq^i_I\epsilon^{IJ}_{\ \ KL}n_J\omega_\alpha^{\ KL},\nonumber\\
K_a^i&:=&q^i_Iq^\alpha_a\omega_\alpha^{\ IJ}n_J, \nonumber
\end{eqnarray}
where $\Gamma$ is a spin connection on $\Sigma$ and $K$ will be
related to the extrinsic curvature of $\Sigma$ on shell. After the
3+1 decomposition and the Legendre transformation, action
(\ref{action}) can be expressed as \cite{holst}:
\begin{eqnarray}
S_{p}=\int_\mathbf{R}dt\int_{\Sigma}d^3x[\widetilde{P}^a_i\mathcal{L}_tA^i_a-\mathcal{H}_{tot}
(A^i_a,\widetilde{P}^b_j,\Lambda^i,N,N^c)] \label{action2},
\end{eqnarray}
from which the symplectic structure on the classical phase space
is obtained as
\begin{eqnarray}
\{A^i_a(x),\widetilde{P}^b_j(y)\}:=\delta^i_j\delta^a_b\delta^3(x-y),
\label{symplectic}
\end{eqnarray}
where the configuration and conjugate momentum are defined
respectively by:
\begin{eqnarray}
A_a^i&:=&\Gamma^i_a+\beta K^i_a,\nonumber\\
\widetilde{P}^a_i&:=&\frac{1}{2\kappa\beta}\widetilde{\eta}^{abc}\epsilon_{ijk}e^j_be^k_c
\ =\ \frac{1}{\kappa\beta}\sqrt{|\det q|}e^a_i,\nonumber
\end{eqnarray}
here $\det q$ is the determinant of the 3-metric $q_{ab}$ on
$\Sigma$ and hence $\det q=(\kappa\beta)^3 \det P$. In the
definition of the configuration variable $A^i_a$, we should
emphasize that $\Gamma_a^i$ is restricted to be the unique torsion
free $so(3)$-valued spin connection compatible with the triad
$e^a_i$. This conclusion is obtained by solving a second class
constraint in the Hamiltonian analysis \cite{holst}. In the
Hamiltonian formalism, one starts with the fields
$(A_a^i,\widetilde{P}^a_i)$. Then neither the basic dynamical
variables nor their Poisson brackets depend on the Barbero-Immirzi
parameter $\beta$. Hence, for the case of pure gravitational field,
the dynamical theories with different $\beta$ are simplectic
equivalent. However, as we will see, the spectrum of geometric
operators are modified by different value of $\beta$, and the
non-perturbative calculation of black hole entropy is compatible
with Bekenstein-Hawking's formula only for a specific value of
$\beta$ \cite{lewandowski1}. In addition, it is argued that the
Barbero-Immerzi parameter $\beta$ may lead to observable effects in
principle when the gravitational field is coupled with fermions
\cite{rovelli7}. In the decomposed action (\ref{action2}), the
Hamiltonian density $\mathcal{H}_{tot}$ is a linear combination of
constraints:
\begin{eqnarray}
\mathcal{H}_{tot}=\Lambda^iG_i+N^aC_a+NC,\nonumber
\end{eqnarray}
where $\Lambda^i\equiv-\frac{1}{2}\epsilon^i_{\ jk}\omega^{\
jk}_t$, $N^a$ and $N$ are Lagrange multipliers. The three
constraints in the Hamiltonian are expressed as \cite{AL}:
\begin{eqnarray}
G_i&=&D_a\widetilde{P}^a_i\ :=\
\partial_a\widetilde{P}^a_i+\epsilon_{ij}^{\
\ k}A_a^j\widetilde{P}^a_k,\nonumber\\
C_a&=&\widetilde{P}^b_iF_{ab}^i-\frac{1+\beta^2}{\beta}K^i_aG_i,\nonumber\\
C&=&\frac{\kappa\beta^2}{2\sqrt{|\det
q|}}\widetilde{P}^a_i\widetilde{P}^b_j[\epsilon^{ij}_{\ \
k}F^k_{ab}-2(1+\beta^2)K^i_{[a}K^j_{b]}]\nonumber\\
&+&\kappa(1+\beta^2)\partial_a\big(\frac{\widetilde{P}^a_i}{\sqrt{|\det
q|}}\big)G^i,\label{constraint}
\end{eqnarray}
where the configuration variable $A_a^i$ performs as a
$so(3)$-valued connection on $\Sigma$ and $F_{ab}^i$ is the
$so(3)$-valued curvature 2-form of $A_a^i$ with the well-known
expression:
\begin{eqnarray}
F_{ab}^i:=2D_{[a}A^i_{b]}=\partial_aA^i_b-\partial_bA^i_a+\epsilon^i_{\
jk}A^j_aA^k_b.\nonumber
\end{eqnarray}
In any dynamical system with constraints, the constraint analysis
is essentially important because they reflect the gauge invariance
of the system. From the above three constraints of general
relativity, one can know the gauge invariance of the theory. The
Gaussian constraint $G_i=0$ has crucial importance in formulating
the general relativity into a dynamical theory of connections. The
corresponding smeared constraint function,
$\mathcal{G}(\Lambda):=\int_\Sigma d^3x\Lambda^i(x)G_i(x)$,
generates a transformation on the phase space as:
\begin{eqnarray}
\{A^i_a(x),\ \mathcal{G}(\Lambda)\}&=&-D_a\Lambda^i(x)\nonumber\\
\{\widetilde{P}^a_i(x), \
\mathcal{G}(\Lambda)\}&=&\epsilon_{ij}^{\ \
k}\Lambda^j(x)\widetilde{P}^a_k(x),\nonumber
\end{eqnarray}
which are just the infinitesimal versions of the following gauge
transformation for the $so(3)$-valued connection 1-form
$\textbf{A}$ and internal rotation for the $so(3)$-valued
densitized vector field $\widetilde{\textbf{P}}$ respectively:
\begin{eqnarray}
(\textbf{A}_a,\
\widetilde{\textbf{P}}^b)\mapsto(g^{-1}\textbf{A}_ag+g^{-1}(d
g)_a,\ g^{-1}\widetilde{\textbf{P}}^bg).\nonumber
\end{eqnarray}
To display the meaning of the vector constraint $C_a=0$, one may
consider the smeared constraint function:
\begin{eqnarray}
\mathcal{V}(\vec{N}):=\int_\Sigma
d^3x(N^a\widetilde{P}^b_iF^i_{ab}-(N^aA^i_a)G_i).\nonumber
\end{eqnarray}
It generates the infinitesimal spatial diffeomorphism by the
vector field $N^a$ on $\Sigma$ as:
\begin{eqnarray}
\{A^i_a(x),\ \mathcal{V}(\vec{N})\}=\mathcal{L}_{\vec{N}}A^i_a(x),\nonumber\\
\{\widetilde{P}^a_i(x),\
\mathcal{V}(\vec{N})\}=\mathcal{L}_{\vec{N}}\widetilde{P}^a_i(x).\nonumber
\end{eqnarray}
The smeared scalar constraint is weakly equivalent to the
following function, which is re-expressed for quantization purpose
as
\begin{eqnarray}
\mathcal{S}(N)&:=&\int_\Sigma d^3x
N(x)\widetilde{C}(x)\nonumber\\
&=&\frac{\kappa\beta^2}{2}\int_\Sigma
d^3xN\frac{\widetilde{P}^a_i\widetilde{P}^b_j}{\sqrt{|\det
q|}}[\epsilon^{ij}_{\ \
k}F^k_{ab}-2(1+\beta^2)K^i_{[a}K^j_{b]}].\label{scalar}
\end{eqnarray}
It generates the infinitesimal time evolution off $\Sigma$. The
constraints algebra, i.e., the Poisson brackets between these
constraints, play a crucial role in the quantization programme. It
can be shown that the constraints algebra of (\ref{constraint})
has the following form:
\begin{eqnarray}
\{\mathcal{G}(\Lambda),\ \mathcal{G}(\Lambda')\}&=&\mathcal{G}([\Lambda,\ \Lambda']),\nonumber\\
\{\mathcal{G}(\Lambda),\ \mathcal{V}(\vec{N})\}&=&-\mathcal{G}(\mathcal{L}_{\vec{N}}\Lambda),\nonumber\\
\{\mathcal{G}(\Lambda),\ \mathcal{H}(N)\}&=&0,\nonumber\\
\{\mathcal{V}(\vec{N}),\ \mathcal{V}(\vec{N}')\}&=&\mathcal{V}([\vec{N},\ \vec{N}']),\nonumber\\
\{\mathcal{V}(\vec{N}),\ \mathcal{S}(M)\}&=&-\mathcal{S}(\mathcal{L}_{\vec{N}}M),\nonumber\\
\{\mathcal{S}(N),\ \mathcal{S}(M)\}&=&-\mathcal{V}((N\partial_bM-M\partial_bN)q^{ab})\nonumber\\
&&-\mathcal{G}((N\partial_bM-M\partial_bN)q^{ab}A_a))\nonumber\\
&&-(1+\beta^2)\mathcal{G}(\frac{[\widetilde{P}^a\partial_aN,\widetilde{P}^b\partial_bM]}{|\det
q|}),\label{constraint algebra}
\end{eqnarray}
where $|\det
q|q^{ab}=\kappa^2\beta^2\widetilde{P}^a_i\widetilde{P}^b_j\delta^{ij}$.
Hence the constraints algebra is closed under the Poisson brackets,
i.e., the constraints are all of first class. Note that the
evolution of constraints is consistent since the Hamiltonian
$H=\int_\Sigma d^3x \mathcal{H}_{tot}$ is the linear combination of
the constraints functions. The evolution equations of the basic
canonical pair read
\begin{eqnarray}
\mathcal{L}_tA^i_a=\{A^i_a,\ H\},\ \ \ \ \ \
\mathcal{L}_t\widetilde{P}^a_i=\{\widetilde{P}^a_i,\ H\}.\nonumber
\end{eqnarray}
Together with the three constraint equations, they are completely
equivalent to the Einstein field equations. Thus general relativity
is cast as a dynamical theory of connections with a compact
structure group. Before finishing the discussion of this section,
several remarks should be emphasized.
\begin{itemize}
\item {\textit{Canonical Transformation Viewpoint}}

The above construction can be reformulated in the language of
canonical transformation, since the phase space of connection
dynamics is the same as that of triad geometrodynamics. In the triad
formalism the basic conjugate pair consists of densitized triad
$\widetilde{E}^a_i=\beta\widetilde{P}^a_i$ and "extrinsic curvature"
$K^i_a$. The Hamiltonian and constraints read
\begin{eqnarray}
\mathcal{H}_{tot}&=&\Lambda^iG'_{i}+N^aC_a+NC\nonumber\\
G'_{i}&=&\epsilon_{ij}^{\ \ k}K_a^j\widetilde{E}^a_{k},\label{g}\\
C_a&=&\widetilde{E}^b_j\nabla_{[a}K_{b]}^j,\\
C&=&\frac{1}{\sqrt{|\det q|}}[\frac{1}{2}|\det
q|R+\widetilde{E}_i^{[a}\widetilde{E}_j^{b]}K^i_aK^j_b],
\end{eqnarray}
where $\nabla_a$ is the $SO(3)$ generalized derivative operator
compatible with triad $e^a_i$ and $R$ is the scalar curvature with
respect to it. Since $\widetilde{E}^a_i$ is a vector density of
weight one, we have
\begin{eqnarray}
\nabla_a\widetilde{E}^a_i=\partial_a\widetilde{E}^a_i+\epsilon_{ij}^{\
\ k}\Gamma_a^j\widetilde{E}_{k}^{a}=0.\nonumber
\end{eqnarray}
One can construct the desired Gaussian constraint by
\begin{eqnarray}
G_i&:=&\frac{1}{\beta}\nabla_a\widetilde{E}^a_i+
G'_i,\nonumber\\
&=&\partial_a\widetilde{P}^a_i+\epsilon_{ij}^{\ \
k}(\Gamma_a^j+\beta K_a^j)\widetilde{P}_{k}^{a},\nonumber
\end{eqnarray}
which is weakly zero by construction. This motivates us to define
the connection $A^a_i=\Gamma_a^i+\beta K_a^i$. Moreover, the
transformation from the pair $(\widetilde{E}^a_i,K_b^j)$ to
$(\widetilde{P}^a_i,A^j_b)$ can be proved to be a canonical
transformation \cite{barbero}\cite{thiemann2}, i.e., the Poisson
algebra of the basic dynamical variables is preserved under the
transformation:
\begin{eqnarray}
\widetilde{E}^a_i&\mapsto&\widetilde{P}^a_i=\widetilde{E}^a_i/\beta\nonumber\\
K_b^j&\mapsto& A^j_b=\Gamma_b^j+\beta K_b^j,\nonumber
\end{eqnarray}
as
\begin{eqnarray}
\{\widetilde{P}^a_i(x),\ A^j_b(y)\}\ =&\{\widetilde{E}^a_i(x),K^j_b(y)\}\ =
&\delta^a_b\delta^j_i\delta(x-y),\nonumber\\
\{A^i_a(x),\ A^j_b(y)\}\ =&\{K^i_a(x),K^j_b(y)\}\ =&0,\nonumber\\
\{\widetilde{P}^a_i(x),\ \widetilde{P}^b_j(y)\}\
=&\{\widetilde{E}^a_i(x),\widetilde{E}^b_j(y)\}\ =&0.\nonumber
\end{eqnarray}

\item {\textit{The Preparation for Quantization}}

The advantage of a dynamical theory of connections is that it is
convenient to be quantized background independently. In the
following procedure of quantization, the quantum algebra of the
elementary observables will be generated by \textit{Holonomy},
i.e., connection smeared on a curve, and \textit{Electric Flux},
i.e., densitized triad smeared on a 2-surface. So no information
of background would affect the definition of the quantum algebra.
In the remainder of the paper, in order to incorporate also
spinors, we will enlarge the internal gauge group to be $SU(2)$.
This does not damage the prior constructions because the Lie
algebra of $SU(2)$ is the same as that of $SO(3)$. Due to the
well-known nice properties of compact Lie group $SU(2)$, such as
the Haar measure and Peter-Weyl theorem, one can obtain the
background independent representation of the quantum algebra and
the spin-network decomposition of the kinematic Hilbert space.

\item {\textit{Analysis on Constraint Algebra}}

The classical constraint algebra (\ref{constraint algebra}) is an
infinite dimensional Poisson algebra. However, it is not a Lie
algebra unfortunately, because the Poisson bracket between two
scalar constraints has structure function depending on dynamical
variables. This character causes much trouble in solving the
constraints quantum mechanically. On the other hand, one can see
from Eq.(\ref{constraint algebra}) that the algebra generated by
Gaussian constraints forms not only a subalgebra but also a 2-side
ideal in the full constraint algebra. Thus one can first solve the
Gaussian constraints independently. It is convenient to find the
quotient algebra with respect to the Gaussian constraint
subalgebra as
\begin{eqnarray}
\{\mathcal{V}(\vec{N}),\ \mathcal{V}(\vec{N}')\}&=&\mathcal{V}([\vec{N},\vec{N}']),\nonumber\\
\{\mathcal{V}(\vec{N}),\ \mathcal{S}(M)\}&=&-\mathcal{S}(\mathcal{L}_{\vec{N}}M),\nonumber\\
\{\mathcal{S}(N),\
\mathcal{S}(M)\}&=&-\mathcal{V}((N\partial_bM-M\partial_bN)q^{ab}),\nonumber
\end{eqnarray}
which plays a crucial role in solving the constraints quantum
mechanically. But the subalgebra generated by the diffeomorphism
constraints can not form an ideal. Hence the procedures of solving
the diffeomorphism constraints and solving Hamiltonian constraints
are entangled with each other. This leads to certain ambiguity in
the construction of a Hamiltonian constraint operator
\cite{thiemann1}\cite{thiemann3}. Fortunately, \textbf{Master
Constraint Project} addresses the above two problems as a whole by
introducing a new classical constraint algebra \cite{thiemann3}. The
new algebra is a Lie algebra where the diffeomorphism constraints
form a 2-side ideal. We will come back to this point in the
discussion on quantum dynamics of loop quantum gravity.

\end{itemize}

\section{Quantum Kinematics}

In this section, we will begin to quantize the above classical
dynamics of connections as a background independent quantum field
theory. The main purpose is to construct a suitable kinematical
Hilbert space $\mathcal{H}_{kin}$ for the representation of
quantum observables. We would like to first construct the Hilbert
space in a more concrete and straightforward way (constructive
quantum field theory aspect \cite{glimm}) in present section. Then
we will reformulate the construction in the language of
$GNS$-construction (algebraic quantum field theory aspect
\cite{haag}) in the next section. It should be emphasized that
both constructions are completely equivalent and can be
generalized to all background independent non-purturbative
Yang-Mills gauge field theories with compact gauge groups.

\subsection{Quantum Configuration Space}

In quantum mechanics, the kinematical Hilbert space is
$L^2(\mathbf{R}^3,d^3x)$, where the simple $\mathbf{R}^3$ is the
classical configuration space of free particle which has finite
degrees of freedom, and $d^3x$ is the Lebesgue measure on
$\mathbf{R}^3$. In quantum field theory, it is expected that the
kinematical Hilbert space is also the $L^2$ space on the
configuration space of the field, which is infinite dimensional,
with respect to some Borel measure naturally defined. However, it
is often hard to define concretely a Borel measure on the
classical configuration space, since the integral theory on
infinite dimensional space is involved \cite{dewitt}. Thus the
intuitive expectation should be modified, and the concept of
quantum configuration space should be introduced as a suitable
enlargement of the classical configuration space so that an
infinite dimensional measure, often called cylindrical measure,
can be well defined on it. The example of a scalar field can be
found in the references \cite{AL}\cite{Ash3}. For quantum gravity,
it should be emphasized that the construction for quantum
configuration space must be background independent. Fortunately,
general relativity has been reformulated as a dynamical theory of
$SU(2)$ connections, which would be great helpful for our further
development.

The classical configuration space for gravitational field, which
is denoted by $\mathcal{A}$, is a collection of the $su(2)$-valued
connection 1-form field smoothly distributed on $\Sigma$. The idea
of the construction for quantum configuration
is due to the concept of \textit{Holonomy}.\\ \\
\textbf{Definition 3.1.1}:  \textit{Given a smooth $SU(2)$
connection field $A_a^i$ and an analytic curve $c$ with the
parameter $t\in[0,1]$ supported on a compact subset $($compact
support $)$ of $\Sigma$, the corresponding holonomy is defined by
the solution of the parallel transport equation \cite{nakahara}
\begin{eqnarray}
\frac{d}{dt}A(c,t)=-[A_a^i\dot{c}^a\tau_i]A(c,t),\label{transport}
\end{eqnarray}
with the initial value $A(c,0)=1$, where $\dot{c}^a$ is the tangent
vector of the curve and $\tau_i\in su(2)$ constitute an orthonormal
basis with respect to the Killing-Cartan metric
$\eta(\xi,\zeta):=-2\mathrm{Tr}(\xi\zeta)$, which satisfy
$[\tau_i,\tau_j]=\epsilon^k_{\ ij}\tau_k$ and are fixed once for
all. Thus the holonomy is an element in $SU(2)$, which can be
expressed as
\begin{eqnarray}
A(c)=\mathcal{P}\exp\big(-\int_0^1[A_a^i\dot{c}^a\tau_i]\ dt
\big),\label{holonomy}
\end{eqnarray}
where $A(c)\in SU(2)$ and $\mathcal{P}$ is a path-ordering
operator along the curve $c$ (see the footnote at p382 in
\cite{nakahara}).}\\ \\
The definition can be well extended to the case of piecewise
analytic curves via the relation:
\begin{eqnarray}
A(c_1\circ c_2)=A(c_1)A(c_2),\label{1}
\end{eqnarray}
where $\circ$ stands for the composition of two curves. It is easy
to see that a holonomy is invariant under the re-parametrization
and is covariant under changing the orientation, i.e.,
\begin{eqnarray}
A(c^{-1})=A(c)^{-1}.\label{2}
\end{eqnarray}
So one can formulate the properties of holonomy in terms of the
concept of the equivalent classes of curves.\\ \\
\textbf{Definition 3.1.2}:  \textit{Two analytic curves $c$ and
$c'$ are said to be equivalent if and only if they have the same
source $s(c)$ $($beginning point $)$ and the same target $t(c)$
$($end point $)$, and the holonomies of the two curves are equal
to each other, i.e., $A(c)=A(c')$ $\forall A\in\mathcal{A}$. A
equivalent class of analytic curves is defined to be an edge, and
a piecewise analytic path is an composition of edges.}\\ \\
To summarize, the holonomy is actually defined on the set
$\mathcal{P}$ of piecewise analytic paths with compact supports.
The two properties (\ref{1}) and (\ref{2}) mean that each
connection in $\mathcal{A}$ is a homomorphism from $\mathcal{P}$,
which is so-called a groupoid by definition \cite{velh}, to our
compact gauge group $SU(2)$. Note that the internal gauge
transformation and spatial diffeomorphism act covariantly on a
holonomy as
\begin{eqnarray}
A(e)\mapsto g(t(e))^{-1}A(e)g(s(e))\ \ \ \mathrm{and} \ \ \
A(e)\mapsto A(\varphi\circ e),\label{trans}
\end{eqnarray}
for any $SU(2)$-valued function $g(x)$ on $\Sigma$ and spatial
diffeomorphism $\varphi$. All above discussion is for classical
smooth connections in $\mathcal{A}$. The quantum configuration
space for loop quantum gravity can be constructed by extending the
concept of holonomy, since its definition does not depend on an
extra background. One thus obtains the quantum configuration space
$\overline{\mathcal{A}}$ of loop quantum gravity as the following.\\ \\
\textbf{Definition 3.1.3}:  \textit{The quantum configuration
space $\overline{\mathcal{A}}$ is a collection of all quantum
connections $A$, which are algebraic homomorphism maps without any
continuity assumption from the collection of piecewise analytic
paths with compact supports, $\mathcal{P}$, on $\Sigma$ to the
gauge group $SU(2)$, i.e.,
$\overline{\mathcal{A}}:=\mathrm{Hom}(\mathcal{P},SU(2))$\footnote{It
is easy to see that the definition of $\overline{\mathcal{A}}$
does not depend on the choice of local section in $SU(2)$-bundle,
since the internal gauge transformations leave
$\overline{\mathcal{A}}$ invariant.}. Thus for any $A\in
\overline{\mathcal{A}}$ and edge $e$ in $\mathcal{P}$,
\begin{eqnarray}
A(e_1\circ e_2)=A(e_1)A(e_2)\ \ \ \mathrm{and}\ \ \
A(e^{-1})=A(e)^{-1}.\nonumber
\end{eqnarray}
The transformations of quantum connections under internal gauge
transformations and diffeomorphisms are defined by Eq.(\ref{trans}).}\\ \\
The above discussion on the smooth connections shows that the
classical configuration space $\mathcal{A}$ can be understood as a
subset in the quantum configuration space
$\overline{\mathcal{A}}$. Moreover, the Giles theorem \cite{giles}
shows precisely that a smooth connection can be recovered from its
holonomies by varying the length and location of the paths. On the
other hand, It was shown in Refs. \cite{AL1}\cite{AL3}\cite{velh}
that the quantum configuration space $\overline{\mathcal{A}}$ can
be constructed via a projective limit technique and admits a
natural defined topology. To make the discussion precise, we begin
with a
few definitions.\\ \\
\textbf{Definition 3.1.4}: \textit{\begin{enumerate}
\item A finite set $\{e_1,...,e_N\}$ of edges is
said to be independent if the edges $e_i$ can only intersect each
other at their sources $s(e_i)$ or targets $t(e_i)$.
\item A finite graph is a collection of a finite set $\{e_1,...,e_N\}$ of independent edges and their
vertices, i.e. their sources $s(e_i)$ and targets $t(e_i)$. We
denote by $E(\gamma)$ and $V(\gamma)$ respectively as the sets of
independent edges and vertices of a given finite graph $\gamma$.
$N_\gamma$ denotes the number of elements in $E(\gamma)$.
\item A subgroupoid $\alpha(\gamma)\subset\mathcal{P}$ can be generated
from $\gamma$ by identifying $V(\gamma)$ as the set of objects and
all $e\in E(\gamma)$ together with their inverses and finite
compositions as the set of homomorphisms. This kind of subgoupoid in
$\mathcal{P}$ is called tame subgroupoid. $\alpha(\gamma)$ is
independent of the orientation of $\gamma$, so the graph $\gamma$
can be recovered from tame subgroupoid $\alpha$ up to the
orientations on the edges. We will also denote by $N_\alpha$ the
number of elements in $E(\gamma)$ where $\gamma$ is recovered by the
tame subgroupoid $\alpha$.
\item $\mathcal{L}$ denotes the set of all tame subgroupoids in
$\mathcal{P}$.
\end{enumerate}}

One can equip a partial order relation $\prec$ on $\mathcal{L}$
\footnote{A partial order on $\mathcal{L}$ is a relation, which is
reflective ($\alpha\prec\alpha$), symmetric ($\alpha\prec\alpha',\
\alpha'\prec\alpha\Rightarrow\alpha'=\alpha$) and transitive
($\alpha\prec\alpha',\
\alpha'\prec\alpha''\Rightarrow\alpha'\prec\alpha''$). Note that not
all pairs in $\mathcal{L}$ need to have a relation.}, defined by
$\alpha\prec\alpha'$ if and only if $\alpha$ is a subgroupoid in
$\alpha'$. Obviously, for any two tame subgroupoids
$\alpha\equiv\alpha(\gamma)$ and $\alpha'\equiv\alpha(\gamma')$ in
$\mathcal{L}$, there exists
$\alpha''\equiv\alpha(\gamma'')\in\mathcal{L}$ such that
$\alpha,\alpha'\prec\alpha''$, where
$\gamma''\equiv\gamma\cup\gamma'$. Define
$\overline{\mathcal{A}}_\alpha\equiv Hom(\alpha, SU(2))$ as the set
of all homomorphisms from the subgroupoid $\alpha(\gamma)$ to the
group $SU(2)$. Note that an element
$A_\alpha\in\overline{\mathcal{A}}_{\alpha}$
($\alpha=\alpha(\gamma)$) is completely determined by the $SU(2)$
group elements $A(e)$ where $e\in E(\gamma)$, so that one has a
bijection $\lambda: \overline{\mathcal{A}}_\alpha \rightarrow
SU(2)^{N_\alpha}$, which induces a topology on
$\overline{\mathcal{A}}_\alpha$ such that $\lambda$ is a topological
homomorphism. Then for any pair $\alpha\prec\alpha'$, one can define
a surjective projection map $P_{\alpha'\alpha}$ from
$\overline{\mathcal{A}}_{\alpha'}$ to
$\overline{\mathcal{A}}_\alpha$ by restricting the domain of the map
$A_{\alpha'}$ from $\alpha'$ to the subgroupoid $\alpha$, and these
projections satisfy the consistency condition
$P_{\alpha'\alpha}\circ P_{\alpha''\alpha'}=P_{\alpha''\alpha}$.
Thus a projective family
$\{\overline{\mathcal{A}}_{\alpha},P_{\alpha'\alpha}\}_{\alpha\prec\alpha'}$
is obtained by above constructions. Then the projective limit
$\lim_{\alpha}(\overline{\mathcal{A}}_{\alpha})$ is naturally
obtained.\\ \\
\textbf{Definition 3.1.5}:  \textit{The projective limit
$\lim_{\alpha}(\overline{\mathcal{A}}_{\alpha})$ of the projective
family
$\{\overline{\mathcal{A}}_{\alpha},P_{\alpha'\alpha}\}_{\alpha\prec\alpha'}$
is a subset of the direct product space
$\overline{\mathcal{A}}_{\infty}:=\prod_{\alpha\in\mathcal{L}}\overline{\mathcal{A}}_{\alpha}$
defined by
\begin{eqnarray}
\lim_{\alpha}(\overline{\mathcal{A}}_{\alpha}):=
\{\{A_{\alpha}\}_{\alpha\in\mathcal{L}}|P_{\alpha'\alpha}A_{\alpha'}=A_{\alpha},\
\forall\ \alpha\prec\alpha'\}.\nonumber
\end{eqnarray}}\\
Note that the projection $P_{\alpha'\alpha}$ is surjective and
continuous with respect to the topology of
$\overline{\mathcal{A}}_{\alpha}$. One can equip the direct product
space
$\overline{\mathcal{A}}_{\infty}:=\prod_{\alpha\in\mathcal{L}}\overline{\mathcal{A}}_{\alpha}$
with the so-called Tychonov topology. Since any
$\overline{\mathcal{A}}_{\alpha}$ is a compact Hausdorff space, by
Tychonov theorem $\overline{\mathcal{A}}_{\infty}$ is also a compact
Hausdorff space. One then can prove that the projective limit,
$\lim_{\alpha}(\overline{\mathcal{A}}_{\alpha})$, is a closed subset
in $\overline{\mathcal{A}}_{\infty}$ and hence a compact Hausdorff
space with respect to the topology induced from
$\overline{\mathcal{A}}_{\infty}$. At last, one can find the
relation between the projective limit and the prior constructed
quantum configuration space $\overline{\mathcal{A}}$. As one might
expect, there is a bijection $\Phi$ between $\overline{\mathcal{A}}$
and $\lim_{\alpha}(\overline{\mathcal{A}}_{\alpha})$
\cite{thiemann2}:
\begin{eqnarray}
\Phi:\ \
\overline{\mathcal{A}}&\rightarrow&\lim_{\alpha}(\overline{\mathcal{A}}_{\alpha});\nonumber\\
A&\mapsto&\{A|_{\alpha}\}_{\alpha\in\mathcal{L}},\nonumber\nonumber
\end{eqnarray}
where $A|_{\alpha}$ means the restriction of the domain of the map
$A\in\overline{\mathcal{A}}=\mathrm{Hom}(\mathcal{P},SU(2))$. As a
result, the quantum configuration space is identified with the
projective limit space and hence can be equipped with the topology.
In conclusion, the quantum configuration space
$\overline{\mathcal{A}}$ is constructed to be a compact Hausdorff
topological space.

\subsection{Cylindrical Functions on Quantum Configuration Space}

Given the projective family
$\{\overline{\mathcal{A}}_{\alpha},P_{\alpha'\alpha}\}_{\alpha\prec\alpha'}$,
the cylindrical function on its projective limit
$\overline{\mathcal{A}}$ is well defined as follows.\\ \\
\textbf{Definition 3.2.1}:  \textit{Let
$C(\overline{\mathcal{A}}_{\alpha})$ be the set of all continuous
complex functions on $\overline{\mathcal{A}}_{\alpha}$, two
functions $f_{\alpha}\in C(\overline{\mathcal{A}}_{\alpha})$ and
$f_{\alpha'}\in C(\overline{\mathcal{A}}_{\alpha'})$ are said to
be equivalent or cylindrically consistent, denoted by
$f_{\alpha}\sim f_{\alpha'}$, if and only if
$P^*_{\alpha''\alpha}f_\alpha=P^*_{\alpha''\alpha'}f_{\alpha'}$,
$\forall\alpha''\succ\alpha,\alpha'$, where $P^*_{\alpha''\alpha}$
denotes the pullback map induced from $P_{\alpha''\alpha}$. Then
the space $Cyl(\overline{\mathcal{A}})$ of cylindrical functions
on the projective limit $\overline{\mathcal{A}}$ is defined to be
the space of equivalent classes $[f]$, i.e.,
\begin{eqnarray}
Cyl(\overline{\mathcal{A}}):=\big[\cup_\alpha
C(\overline{\mathcal{A}}_{\alpha})\big]/\sim.\nonumber
\end{eqnarray}}\\
One then can easily prove the following proposition by
definition.\\ \\
\textbf{Proposition 3.2.1}: \\
\textit{All continuous functions $f_{\alpha}$ on
$\overline{\mathcal{A}}_{\alpha}$ are automatically cylindrical
since each of them can generate a equivalent class $[f_{\alpha}]$
via the pullback map $P^*_{\alpha'\alpha}$ for all
$\alpha'\succ\alpha$, and the dependence of
$P^*_{\alpha'\alpha}f_{\alpha}$ on the groups associated to the
edges in $\alpha'$ but not in $\alpha$ is trivial, i.e., by the
definition of the pull back map,
\begin{eqnarray}
(P^*_{\alpha'\alpha}f_{\alpha})(A(e_1),...,A(e_{N_\alpha}),...,A(e_{N_{\alpha'}}))=
f_{\alpha}(A(e_1),...,A(e_{N_\alpha})).\label{func}
\end{eqnarray}
On the other hand, by definition, given a cylindrical function $f\in
Cyl(\overline{\mathcal{A}})$ there exists a suitable groupoid
$\alpha$ such that $f=[f_{\alpha}]$, so one can identify $f$ with
$f_\alpha$. Moreover, given two cylindrical functions $f,\ f'\in
Cyl(\overline{\mathcal{A}})$, by definition of cylindrical functions
and the property of projection map, there exists a common groupoid
$\alpha$ and $f_{\alpha},\ f'_{\alpha}\in
C(\overline{\mathcal{A}}_{\alpha})$ such that $f=[f_{\alpha}]$ and
$f'=[f'_{\alpha}]$.}\\ \\
Let $f$, $f'\in Cyl(\overline{\mathcal{A}})$, there exists groupoid
$\alpha$ such that $f=[f_{\alpha}]$, and $f'=[f'_{\alpha}]$, then
the following operations are well defined
\begin{eqnarray}
f+f':=[f_\alpha+f'_\alpha],\ ff':=[f_\alpha f'_\alpha],\
zf:=[zf_\alpha],\ \bar{f}:=[\bar{f}_\alpha],\nonumber
\end{eqnarray}
where $z\in\mathbf{C}$ and $\bar{f}$ denotes complex conjugate. So
we construct $Cyl(\overline{\mathcal{A}})$ as an Abelian
$*$-algebra. In addition, there is a unital element in the algebra
because $Cyl(\overline{\mathcal{A}})$ contains constant functions.
Moreover, we can well define the sup-norm for $f=[f_{\alpha}]$ by
\begin{eqnarray}
\|f\|:=\sup_{A_\alpha\in\overline{\mathcal{A}}_\alpha}|f_\alpha(A_\alpha)|,\label{norm}
\end{eqnarray}
which satisfies the $C^*$ property $\|f\bar{f}\|=\|f\|^2$. Then
$\overline{Cyl(\overline{\mathcal{A}})}$ is a unital Abelian
$C^*$-algebra, after the completion with respect to the norm. From
the theory of $C^*$-algebra, it is known that a unital Abelian
$C^*$-algebra is identical to the space of continuous functions on
its spectrum space via an isometric isomorphism, the so-called
Gel'fand transformation (see e.g. \cite{thiemann2}). So one has
the following theorem \cite{AL1}\cite{AL3}, which finishes this section.\\
\\
\textbf{Theorem 3.2.1}:  \\
\textit{$(1)$ The space $Cyl(\overline{\mathcal{A}})$ has the structure of a unital Abelian $C^*$-algebra
after completion with respect to the sup-norm.\\
$(2)$ Quantum configuration space $\overline{\mathcal{A}}$ is the
spectrum space of completed
$\overline{Cyl(\overline{\mathcal{A}})}$ such that
$\overline{Cyl(\overline{\mathcal{A}})}$ is identical to the space
$C(\overline{\mathcal{A}})$ of continuous functions on
$\overline{\mathcal{A}}$. }\\ \\

\subsection{Kinematical Hilbert Space}

The main purpose of this subsection is to construct a kinematical
Hilbert space $\mathcal{H}_{kin}$ for loop quantum gravity, which is
a $L^2$ space on the quantum configuration space
$\overline{\mathcal{A}}$ with respect to some measure $d\mu$. There
is a well-defined probability measure on $\overline{\mathcal{A}}$
originated from the Haar measure on the compact group $SU(2)$, which
is named as the Ashtekar-Lewandowski Measure for loop quantum
gravity. Consider the simplest case where the groupoid is generated
by one edge $e$ only . Then the corresponding quantum configuration
space $\overline{\mathcal{A}}_e$, being trivial elsewhere, is
identical to the group $SU(2)$. The continuous functions on
$\overline{\mathcal{A}}_e$ is certainly contained in
$Cyl(\overline{\mathcal{A}})$. Due to the compactness of $SU(2)$,
there exists a unique probability measure, namely the \textit{Haar
measure} on it, which is invariant under right and left
translations and inverse of the group elements.\\ \\
\textbf{Theorem 3.3.1} \cite{dieck}:  \\
\textit{Given a compact group $G$ and an automorphism $\varphi:\
G\rightarrow G$ on it, there exists a unique measure $d\mu_H$ on
$G$, named as Haar measure, such that:
\begin{eqnarray}
&&\int_G d\mu_H=1 ,\label{normal}\\
\int_G f(g)d\mu_H&=&\int_G f(hg)d\mu_H\ =\ \int_G
f(gh)d\mu_H\nonumber\\
&=&\int_G f(g^{-1})d\mu_H\ =\ \int_G f\circ\varphi(g)d\mu_H,
\end{eqnarray}
for all continuous functions $f$ on $G$ and for all $h\in G$.}\\
\\
Thus one equips $\overline{\mathcal{A}}_e$ with the measure
$\mu_{e}\equiv\mu_H$. Similarly, a probability measure can be
defined on any graph with finite number of edges by the direct
product of Haar measure, since
$\overline{\mathcal{A}}_{\alpha(\gamma)}=SU(2)^{N_\gamma}$. Then for
any groupoid $\alpha$, a Hilbert space is defined on
$\overline{\mathcal{A}}_\alpha$ as
$\mathcal{H}_{\alpha}=L^2(\overline{\mathcal{A}}_\alpha,
d\mu_{\alpha})=\otimes_{e\in\alpha}L^2(\overline{\mathcal{A}}_e,
d\mu_e)$. Moreover, the family of measures
$\{\mu_\alpha\}_{\alpha\in\mathcal{L}}$ defined on the projective
family
$\{\overline{\mathcal{A}}_{\alpha},P_{\alpha'\alpha}\}_{\alpha\prec\alpha'}$
are cylindrically consistent, since
\begin{eqnarray}
&&\int_{\overline{\mathcal{A}}_{\alpha'}}(P^*_{\alpha'\alpha}f_{\alpha})
d\mu_{\alpha'}\nonumber\\
&=&\int_{\overline{\mathcal{A}}_{\alpha'}}(P^*_{\alpha'\alpha}f_{\alpha})
(A(e_1),...,A(e_{N_\alpha}),...,A(e_{N_{\alpha'}})) d\mu_{e_1}...
d\mu_{e_{N_\alpha}}...
d\mu_{e_{N_{\alpha'}}}\nonumber\\
&=&\int_{\overline{\mathcal{A}}_{\alpha}}f_{\alpha}(A(e_1),...,A(e_{N_\alpha}))d\mu_{e_1}...
d\mu_{e_{N_\alpha}}\nonumber\\
&=&\int_{\overline{\mathcal{A}}_{\alpha}}f_{\alpha} d\mu_{\alpha}\
,\nonumber
\end{eqnarray}
due to Eqs. (\ref{func}) and (\ref{normal}). Given such a
cylindrically consistent family of measures
$\{\mu_\alpha\}_{\alpha\in\mathcal{L}}$, a probability measure
$d\mu$ is uniquely well defined on the quantum configuration space
$\overline{\mathcal{A}}$ \cite{AL1}, which is described precisely
in the
theorem below.\\ \\
\textbf{Theorem 3.3.2} \cite{thiemann2}:  \\
\textit{Given the projective family
$\{\overline{\mathcal{A}}_{\alpha},P_{\alpha'\alpha}\}_{\alpha\prec\alpha'}$,
whose projective limit is $\overline{\mathcal{A}}$, and the
cylindrically consistent family of measures
$\{\mu_\alpha\}_{\alpha\in\mathcal{L}}$ constructed from the Haar
measure on the compact group, there exists a unique regular Borel
probability measure $d\mu$ on the projective limit
$\overline{\mathcal{A}}$ such that
\begin{eqnarray}
\int_{\overline{\mathcal{A}}}fd\mu=\int_{\overline{\mathcal{A}}_\alpha}f_\alpha
d\mu_\alpha, \ \forall f=[f_\alpha]\in
Cyl(\overline{\mathcal{A}}),\nonumber
\end{eqnarray}
which is
guaranteed by proposition 3.2.1.}\\ \\
Then $\overline{\mathcal{A}}$ is equipped with the
Ashtekar-Lewandowski measure $d\mu$ and becomes a topological
measure space \cite{AI}\cite{AL1}. This measure will help us define
a state on the quantum holonomy-flux algebra for gauge field theory,
which is called Ashtekar-Isham-Lewandowski state in the language of
$GNS$-construction. Moreover the two important gauge invariant
properties of Ashtekar-Lewandowski measure make it well
suitable for the diffeomorphism invariant gauge field theory.\\ \\
\textbf{Theorem 3.3.3}:  \\
\textit{The Ashtekar-Lewandowski measure is invariant under internal
gauge transformations $g(x)$ and spatial diffeomorphisms $\varphi$,
i.e.,
\begin{eqnarray}
\int_{\overline{\mathcal{A}}}g\circ
fd\mu=\int_{\overline{\mathcal{A}}}f d\mu\ \ \ \mathrm{and}\ \ \
\int_{\overline{\mathcal{A}}}\varphi\circ
fd\mu=\int_{\overline{\mathcal{A}}}f d\mu,\nonumber
\end{eqnarray}
$\forall f\in Cyl(\overline{\mathcal{A}})$.}\\ \\
\textbf{Proof}:  \\
(1) (Internal gauge invariance)
\begin{eqnarray}
\int_{\overline{\mathcal{A}}}g\circ
fd\mu=\int_{\overline{\mathcal{A}}_\alpha}g\circ f_\alpha
d\mu_\alpha=\int_{\overline{\mathcal{A}}_\alpha}f_\alpha
d\mu_\alpha=\int_{\overline{\mathcal{A}}}fd\mu\nonumber
\end{eqnarray}
$\forall f=[f_\alpha]\in Cyl(\overline{\mathcal{A}})$, where we
used
\begin{eqnarray}
&&g\circ
f_\alpha\big(A(e_1),...,A(e_{N_\alpha})\big)\nonumber\\&=&
f_\alpha\big(g(t(e_1))^{-1}A(e_1)g(s(e_1)),...,g(t(e_{N_\alpha}))^{-1}A(e_{N_\alpha})g(s(e_{N_\alpha}))\big),\nonumber
\end{eqnarray}
since Haar measure is invariant under right and left translations.\\
(2) (Diffeomorphism invariance)
\begin{eqnarray}
\int_{\overline{\mathcal{A}}}\varphi\circ
fd\mu=\int_{\overline{\mathcal{A}}_{\varphi\circ\alpha}}f_{\varphi\circ\alpha}
d\mu_{\varphi\circ\alpha}=\int_{\overline{\mathcal{A}}_\alpha}f_\alpha
d\mu_\alpha=\int_{\overline{\mathcal{A}}}fd\mu,\nonumber
\end{eqnarray}
where $f_{\varphi\circ\alpha}\equiv f_\alpha\big(A(\varphi\circ
e_1),...,A(\varphi\circ e_{N_{\alpha}})\big)$ and we relabel
$A(\varphi\circ e_i)\mapsto A(e_i)$ in the second step.\\ \\
With the above constructed measure on $\overline{\mathcal{A}}$,
the kinematical Hilbert space $\mathcal{H}_{kin}$ is obtained
straight-forwardly as
\begin{eqnarray}
\mathcal{H}_{kin}:=L^2(\overline{\mathcal{A}},d\mu).\label{kin}
\end{eqnarray}
Thus, given any $f=[f_\alpha], f'=[f'_{\alpha'}]\in
Cyl(\overline{\mathcal{A}})$, the $L^2$ inner product of them is
expressed as
\begin{eqnarray}
<f|f'>_{kin}:=\int_{\overline{\mathcal{A}}_{\alpha''}}(P^*_{\alpha''\alpha}\overline{f}_\alpha)
(P^*_{\alpha''\alpha'}f_{\alpha'}) d\mu_{\alpha''}, \label{inner}
\end{eqnarray}
for any groupoid $\alpha''$ containing both $\alpha$ and $\alpha'$.
It should be noted that the cylindrical functions in
$\mathcal{H}_{kin}$ is dense with respect to the $L^2$ inner
product, as they are dense in $C(\overline{\mathcal{A}})$ with
respect to the sup-norm. As a result, the kinematical Hilbert space
can be viewed as the completion of $Cyl(\overline{\mathcal{A}})$
with respect to the inner product (\ref{inner}), i.e.,
\begin{eqnarray}
\mathcal{H}_{kin}=\langle \ Cyl(\overline{\mathcal{A}})\
\rangle=\langle\ \cup_{\alpha\in \mathcal{L}}\mathcal{H}_\alpha\
\rangle,\label{<>}
\end{eqnarray}
here the $\langle\ \cdot\ \rangle$ means the completion with
respect to the inner product (\ref{inner}). Later we will show
that $\mathcal{H}_{kin}$ is a non-separable Hilbert space. It is
important to note that all the above constructions are background
independent.

\subsection{Spin-network Decomposition of Kinematical Hilbert Space}

Up to now, the kinematical Hilbert space $\mathcal{H}_{kin}$ for
loop quantum gravity has been well defined. In this subsection, it
will be shown that $\mathcal{H}_{kin}$ can be decomposed into the
orthogonal direct sum of 1-dimensional subspaces. One can thus
find a system of basis, named as spin-network basis, in the
Hilbert space, which consists of uncountably infinite elements. So
the kinematic Hilbert space is non-separable. In the following, we
will do the decomposition in three steps.

\begin{itemize}

\item {\textit{Spin-network Decomposition on Single Edge}}

Given a groupoid of one edge $e$, which naturally associates with a
group $SU(2)=\overline{\mathcal{A}}_e$, the elements of
$\overline{\mathcal{A}}_e$ are the quantum connections only taking
nontrivial values on $e$. Then we consider the decomposition of the
Hilbert space $\mathcal{H}_e=L^2(\overline{\mathcal{A}}_e,
d\mu_e)=L^2(SU(2), d\mu_H)$, which is nothing but the space of
square integrable functions on the compact group $SU(2)$ with the
natural $L^2$ inner product. It is natural to define several
operators on $\mathcal{H}_e$. First, the so-called configuration
operator $\hat{f}\big(A(e)\big)$ whose operation on any $\psi$ in a
dense domain of $L^2(SU(2), d\mu_H)$ is nothing but multiplication
by the function $f\big(A(e)\big)$, i.e.,
\begin{eqnarray}
\hat{f}\big(A(e)\big)\psi\big(A(e)\big):=f\big(A(e)\big)\psi\big(A(e)\big),\nonumber
\end{eqnarray}
where $A(e)\in SU(2)$. Second, given any vector $\xi\in su(2)$, it
generates left invariant vector field $L^{(\xi)}$ and right
invariant vector field $R^{(\xi)}$ on $SU(2)$ by
\begin{eqnarray}
L^{(\xi)}\psi\big(A(e)\big):=\frac{d}{dt}|_{t=0}\psi\big(A(e)\exp(t\xi)\big),\nonumber\\
R^{(\xi)}\psi\big(A(e)\big):=\frac{d}{dt}|_{t=0}\psi\big(\exp(-t\xi)A(e)\big),\nonumber
\end{eqnarray}
for any function $\psi\in C^1(SU(2))$. Then one can define the
so-called momentum operators on $\mathcal{H}_e$ by
\begin{eqnarray}
\hat{J}_i^{(L)}=iL^{(\tau_i)}\ \ \ \mathrm{and}\ \ \
\hat{J}_i^{(R)}=iR^{(\tau_i)},\nonumber
\end{eqnarray}
where the generators $\tau_i\in su(2)$ constitute an orthonormal
basis with respect to the Killing-Cartan metric. The momentum
operators have the well-known commutation relation of the angular
momentum operators in quantum mechanics:
\begin{eqnarray}
[\hat{J}^{(L)}_i,\hat{J}^{(L)}_j]=i\epsilon^k_{\
ij}\hat{J}^{(L)}_k,\
[\hat{J}^{(R)}_i,\hat{J}^{(R)}_j]=i\epsilon^k_{\
ij}\hat{J}^{(R)}_k,\ [\hat{J}^{(L)}_i,\hat{J}^{(R)}_j]=0.\nonumber
\end{eqnarray}
Thirdly, the Casimir operator on $\mathcal{H}_e$ can be expressed
as
\begin{eqnarray}
\hat{J}^2:=\delta^{ij}\hat{J}^{(L)}_i\hat{J}^{(L)}_j=\delta^{ij}\hat{J}^{(R)}_i\hat{J}^{(R)}_j.\label{casimir}
\end{eqnarray}

The decomposition of $\mathcal{H}_e=L^2(SU(2),d\mu_H)$ is provided
by
the following Peter-Weyl Theorem.\\ \\
\textbf{Theorem 3.4.1} \cite{dieck}:  \\
\textit{Given a compact group $G$, the function space
$L^2(G,d\mu_H)$ can be decomposed as an orthogonal direct sum of
finite dimensional Hilbert space, and the matrix elements of the
equivalent classes of finite dimensional irreducible representations
of $G$ form an orthogonal basis in
$L^2(G,d\mu_H)$.}\\

Note that a finite dimensional irreducible representation of $G$
can be regarded as a matrix-valued function on $G$, so the matrix
elements are functions on $G$. Using this theorem, one can find
the decomposition of the Hilbert space:
\begin{eqnarray}
L^2(SU(2),d\mu_H)=\oplus_j[\mathcal{H}_j\otimes
\mathcal{H}^*_j],\nonumber
\end{eqnarray}
where $j$, labelling irreducible representations of $SU(2)$, are
the half integers, $\mathcal{H}_j$ denotes the carrier space of
the $j$-representation of dimension $2j+1$, and $ \mathcal{H}^*_j$
is its dual space. The basis $\{\mathbf{e}^j_m\otimes
\mathbf{e}_n^{j*}\}$ in $\mathcal{H}_j\otimes \mathcal{H}^{*}_j$
maps a group element $g\in SU(2)$ to a matrix $\{\pi^j_{mn}(g)\}$,
where $m,n=-j,...,j$. Thus the space $\mathcal{H}_j\otimes
\mathcal{H}^*_j$ is spanned by the matrix element functions
$\pi^j_{mn}$ of equivalent $j$-representations. Moreover, the
spin-network basis can be defined.\\

\textbf{Proposition 3.4.1} \cite{carmeli}\\
\textit{The system of spin-network functions on $\mathcal{H}_e$,
consisting of matrix elements $\{\pi^j_{mn}\}$ in finite
dimensional irreducible representations labelled by half-integers
$\{j\}$, satisfies
\begin{eqnarray}
\hat{J}^2\pi^j_{mn}=j(j+1)\pi^j_{mn},\
\hat{J}^{(L)}_3\pi^j_{mn}=m\pi^j_{mn},\
\hat{J}^{(R)}_3\pi^j_{mn}=n\pi^j_{mn}\nonumber,
\end{eqnarray}
where $j$ is called angular momentum quantum number and
$m,n=-j,...,j$ magnetic quantum number. The normalized functions
$\{\sqrt{2j+1}\pi^j_{mn}\}$ form a system of complete orthonormal
basis in $\mathcal{H}_e$ since
\begin{eqnarray}
\int_{\overline{\mathcal{A}}_e}\overline{\pi^{j'}_{m'n'}}\pi^j_{mn}d\mu_e
=\frac{1}{2j+1}\delta^{j'j}\delta_{m'm}\delta_{n'n},\nonumber
\end{eqnarray}
which is called the spin-network basis on $\mathcal{H}_e$. So the
Hilbert space on a single edge
has been decomposed into one dimensional subspaces. }\\

Note that the system of operators
$\{\hat{J}^2,\hat{J}^{(R)}_3,\hat{J}^{(L)}_3\}$ forms a complete
set of commutable operators in $\mathcal{H}_e$. There is a cyclic
"vacuum state" in the Hilbert space, which is the
$(j=0)$-representation $\Omega_e=\pi^{j=0}=1$, representing that
there is no geometry on the edge.

\item {\textit{Spin-network Decomposition on Finite Graph}}

Given a groupoid $\alpha$ generated by a graph $\gamma$ with $N$
oriented edges $e_i$ and $M$ vertices, one can define the
configuration operators on the corresponding Hilbert space
$\mathcal{H}_\alpha$ by
\begin{eqnarray}
\hat{f}\big(A(e_i)\big)\psi_\alpha\big(A(e_1),...,A(e_{N})\big):=f\big(A(e_i)\big)\psi_\alpha
\big(A(e_1),...,A(e_{N})\big).\nonumber
\end{eqnarray}
The momentum operators $\hat{J_i}^{(e,v)}$ associated with an edge
$e$ connecting a vertex $v$ are defined as
\begin{eqnarray}
\hat{J_i}^{(e,v)}:=(1\otimes...\otimes
\hat{J}_i\otimes...\otimes1),\nonumber
\end{eqnarray}
where we set $\hat{J}_i=\hat{J}_i^{(L)}$ if $v=s(e)$ and
$\hat{J}_i=\hat{J}_i^{(R)}$ if $v=t(e)$. Note that the choice is
based on the definition of gauge transformations (\ref{trans}).
Note also that $\hat{J_i}^{(e,v)}$ only acts nontrivially on the
Hilbert space associated with the edge $e$. Then one can define a
vertex operator associated with vertex $v$ in analogy with the
total angular momentum operator via
\begin{eqnarray}
[\hat{J}^v]^2:=\delta^{ij}\hat{J}_i^v\hat{J}_j^v,\nonumber
\end{eqnarray}
where
\begin{eqnarray}
\hat{J}_i^v:=\sum_{e'\ at\ v}\hat{J}^{(e',v)}_i.\nonumber
\end{eqnarray}
Obviously, $\mathcal{H}_\alpha$ can be firstly decomposed by the
representations on each edge $e$ of $\alpha$ as:
\begin{eqnarray}
\mathcal{H}_\alpha&=&\otimes_{e}\mathcal{H}_e=\otimes_{e}[\oplus_j(\mathcal{H}^e_j\otimes
\mathcal{H}^{e*}_j)]=\oplus_{\mathbf{j}}[\otimes_e(\mathcal{H}^e_j\otimes
\mathcal{H}^{e*}_j)]\nonumber\\
&=&\oplus_{\mathbf{j}}[\otimes_v(\mathcal{H}^{v=s(e)}_{\mathbf{j}(s)}\otimes
\mathcal{H}^{v=t(e)}_{\mathbf{j}(t)})],\nonumber
\end{eqnarray}
where $\mathbf{j}:=(j_1,...,j_N)$ assigns to each edge an
irreducible representation of $SU(2)$, in the fourth step the
Hilbert spaces associated with the edges are allocated to the
vertexes where these edges meet so that for each vertex $v$,
\begin{eqnarray}
\mathcal{H}^{v=s(e)}_{\mathbf{j}(s)}\equiv\otimes_{s(e)=v}\mathcal{H}^e_j
\ \ \ \mathrm{and}\ \ \
\mathcal{H}^{v=t(e)}_{\mathbf{j}(t)}\equiv\otimes_{t(e)=v}\mathcal{H}^{e*}_j.\nonumber
\end{eqnarray}
The group of internal gauge transformations $g(v)\in SU(2)$ at each
vertex is reducibly represented on the Hilbert space
$\mathcal{H}^{v=s(e)}_{\mathbf{j}(s)}\otimes
\mathcal{H}^{v=t(e)}_{\mathbf{j}(t)}$ in a natural way. So this
Hilbert space can be decomposed as a direct sum of irreducible
representation spaces via Clebsch-Gordon decomposition:
\begin{eqnarray}
\mathcal{H}^{v=s(e)}_{\mathbf{j}(s)}\otimes
\mathcal{H}^{v=t(e)}_{\mathbf{j}(t)}=\oplus_l\mathcal{H}^v_{\mathbf{j}(v),l}\
.\nonumber
\end{eqnarray}
As a result, $\mathcal{H}_\alpha$ can be further decomposed as:
\begin{eqnarray}
\mathcal{H}_\alpha=\oplus_{\mathbf{j}}[\otimes_v(\oplus_l\mathcal{H}^v_{\mathbf{j}(v),l})]
=\oplus_{\mathbf{j}}[\oplus_{\mathbf{l}}(\otimes_v\mathcal{H}^v_{\mathbf{j}(v),l})]\equiv
\oplus_{\mathbf{j}}[\oplus_{\mathbf{l}}\mathcal{H}_{\alpha,\mathbf{j},\mathbf{l}}].\label{decomposition}
\end{eqnarray}
It can also be viewed as the eigenvector space decomposition of the
commuting operators $[\hat{J}^v]^2$ (with eigenvalues $l(l+1)$) and
$[\hat{J}^e]^2\equiv \delta^{ij}\hat{J}^e_i\hat{J}^e_j$. Note that
$\mathbf{l}:=(l_1,...,l_M)$ assigns to each vertex of $\alpha$ an
irreducible representation of gauge transformation. One may also
enlarge the set of commuting operators to further refine the
decomposition of the Hilbert space. Note that the subspace of
$\mathcal{H}_\alpha$ with $\mathbf{l}=0$ is gauge invariant, since
the representation of gauge transformations is trivial.

\item {\textit{Spin-network Decomposition of $\mathcal{H}_{kin}$}}

Since $\mathcal{H}_{kin}$ has the structure
$\mathcal{H}_{kin}=\langle\ \cup_{\alpha\in
\mathcal{L}}\mathcal{H}_\alpha\ \rangle$, one may consider to
construct it as a direct sum of $\mathcal{H}_\alpha$. The
construction is precisely described as a theorem below.\\ \\
\textbf{Theorem 3.4.2}:  \\
\textit{Consider assignments $\mathbf{j}=(j_1,...,j_N)$ to the edges
of any groupoid $\alpha\in\mathcal{L}$ and assignments
$\mathbf{l}=(l_1,...,l_M)$ to the vertices. The edge representation
$j$ is non-trivial on each edge, and the vertex representation $l$
is non-trivial at each spurious\footnote{A vertex $v$ is spurious if
it is bivalent and $e\circ e'$ is itself analytic edge with $e,\ e'$
meeting at $v$.} vertex, unless it is the base point of a close
analytic loop. Let $\mathcal{H}'_{\alpha}$ be the Hilbert space
composed by the subspaces
$\mathcal{H}_{\alpha,\mathbf{j},\mathbf{l}}$ (assigned the above
conditions) according to Eq.(\ref{decomposition}). Then
$\mathcal{H}_{kin}$ can be decomposed as the direct sum of the
Hilbert spaces $\mathcal{H}'_{\alpha}$, i.e.,
\begin{eqnarray}
\mathcal{H}_{kin}=\oplus_{\alpha\in
\mathcal{L}}\mathcal{H}'_{\alpha}\oplus\mathbf{C}.\nonumber
\end{eqnarray}  }\\
\textbf{Proof}:  \\
Since the representation on each edge is non-trivial, by definition
of the inner product, it is easy to see that $\mathcal{H}'_\alpha$
and $\mathcal{H}'_{\alpha'}$ are mutual orthogonal if one of the
groupoids $\alpha$ and $\alpha'$ has at leat an edge $e$ more than
the other due to
\begin{eqnarray}
\int_{\overline{\mathcal{A}}_e}\pi^j_{mn}d\mu_e=\int_{\overline{\mathcal{A}}_e}1\cdot\pi^j_{mn}d\mu_e=0\nonumber
\end{eqnarray}
for any $j\neq0$. Now consider the case of the spurious vertex. An
edge $e$ with $j$-representation in a graph is assigned the
Hilbert space $\mathcal{H}^e_j\otimes\mathcal{H}^{e*}_j$.
Inserting a vertex $v$ into the edge, one obtains two edges $e_1$
and $e_2$ split by $v$ both with $j$-representations, which belong
to a different graph. By the decomposition of the corresponding
Hilbert space,
\begin{eqnarray}
\mathcal{H}^{e_1}_j\otimes\mathcal{H}^{e_1*}_j\otimes\mathcal{H}^{e_2}_j\otimes\mathcal{H}^{e_2*}_j
=\mathcal{H}^{e_1}_j\otimes(\oplus_{l=0...2j}\mathcal{H}^v_l)\otimes\mathcal{H}^{e_2*}_j,\nonumber
\end{eqnarray}
the subspace for all $l\neq0$ are orthogonal to the space
$\mathcal{H}^e_j\otimes\mathcal{H}^{e*}_j$, while the subspace for
$l=0$ coincides with $\mathcal{H}^e_j\otimes\mathcal{H}^{e*}_j$
since $\mathcal{H}^v_{l=0}=\mathbf{C}$ and $A(e)=A(e_1)A(e_2)$.
This completes the proof.\\ \\
Since there are uncountable many graphs on $\Sigma$, the kinematical
Hilbert $\mathcal{H}_{kin}$ is non-separable. We denote the
spin-network basis in $\mathcal{H}_{kin}$ by the vacuum state
$\Pi_0\equiv\Omega=1$ and $\Pi_s,\
s=(\gamma(s),\mathbf{j}_s,\mathbf{m}_s,\mathbf{n}_s)$, i.e.,
\begin{eqnarray}
\Pi_s:=\prod_{e\in E(\gamma(s))}\sqrt{2j_e+1}\pi^{j_e}_{m_e n_e}\ \
\ \ \ (j_e\neq0),\nonumber
\end{eqnarray}
which form a orthonormal basis with the relation
$<\Pi_s|\Pi_{s'}>_{kin}=\delta_{ss'}$. We further denote the
subset $Cyl_\gamma(\overline{\mathcal{A}})\subset
Cyl(\overline{\mathcal{A}})$ as the linear span of the
spin-network functions $\Pi_s$ for $\gamma(s)=\gamma$.
\end{itemize}
The spin-network basis can be used to construct the so-called spin-network representation of loop quantum gravity.\\ \\
\textbf{Definition 3.4.1}:  \textit{The spin-network representation
is a vector space $\widetilde{\mathcal{H}}$ of complex valued
functions
\begin{eqnarray}
\widetilde{\Psi}:\ S\ \rightarrow\ \mathbf{C}; \ \ s\ \mapsto\
\widetilde{\Psi}(s),\nonumber
\end{eqnarray}
where $S$ is the set of the labels $s$ for the spin-network states.
$\widetilde{\mathcal{H}}$ is equipped with the scalar product
\begin{eqnarray}
<\widetilde{\Psi},\ \widetilde{\Psi}'>:=\sum_{s\in
S}\overline{\widetilde{\Psi}(s)}\widetilde{\Psi}(s)'\nonumber
\end{eqnarray}
between square summable functions.}\\ \\
The relation between the Hilbert spaces $\widetilde{\mathcal{H}}$
and
$\mathcal{H}_{kin}$ is clarified by the following proposition \cite{thiemann2}.\\
\\
\textbf{Proposition 3.4.2:} \\
\textit{The spin-network transformation
\begin{eqnarray}
T:\ \mathcal{H}_{kin}\ \rightarrow\ \widetilde{\mathcal{H}}; \ \
\Psi\ \mapsto\ \widetilde{\Psi}(s):=<\Pi_s,\ \Psi>_{kin}\nonumber
\end{eqnarray}
is a unitary transformation with inverse
\begin{eqnarray}
T^{-1}\Psi=\sum_{s\in S}\widetilde{\Psi}(s)\Pi_s.\nonumber
\end{eqnarray} }\\
Thus the connection representation and the spin-network
representation are "Fourier transforms" of each other, where the
role of the kernel of the transform is played by the spin-network
basis. Note that, in the gauge invariant Hilbert space of loop
quantum gravity (see section 5.1), the Fourier transform with
respect to the gauge invariant spin network basis is the so-called
loop transform, which leads to the unitary equivalent loop
representation of the theory \cite{rov2}\cite{gp1}\cite{rovelli}.

\subsection{Holonomy-Flux Algebra and Quantum Operators}

The central aim of quantum kinematics for loop quantum gravity is to
look for a proper representation of quantum algebra of elementary
observables. In the classical theory, the basic dynamic variables
are $su(2)$-valued connection field $A^i_a$ and densitized triad
field $\widetilde{P}^a_i$ on $\Sigma$. However, these two basic
variables are not in the algebra of elementary classical observables
which will be represented in the quantum theory, whence they do not
have direct quantum analogs in loop quantum gravity. The elementary
classical observables in our representation theory are the complex
valued functions (cylindrical functions) $f_e$ of holonomies $A(e)$
along paths $e$ in $\Sigma$, and fluxes $P_i(S)$ of triad field
across 2-surfaces $S$, which is defined as
\begin{eqnarray}
P_i(S):=\int_S\underline{\eta}_{abc}\widetilde{P}^c_i,\nonumber
\end{eqnarray}
where $\underline{\eta}_{abc}$ is the Levi-Civita tensor density on
$\Sigma$. In the simplest case where a single edge $e$ intersects a
2-surface $S$ at a point $p$, one can calculate the Poisson bracket
between two functions $f\big(A(e)\big)$ and $P_i(S)$ on classical
phase space $\mathcal{M}$ as \cite{thiemann2},
\begin{eqnarray}
&&\{P_i(S),\ f(A(e))\}\ =\ \big[\frac{\partial}{\partial
A(e)_{mn}}f(A(e))\big]\cdot\{P_i(S),\
A(e)_{mn}\}\ \nonumber\\
&=&\big[\frac{\partial}{\partial
A(e)_{mn}}f(A(e))\big]\cdot\big[\frac{\kappa(S,\ e)}{2}\big]\cdot\left\{%
\begin{array}{ll}
    \sum_k A(e)_{mk}(\tau_i)_{kn} & \hbox{if $p=s(e)$} \\
    -\sum_k (\tau_i)_{mk} A(e)_{kn} & \hbox{if $p=t(e)$}, \\
\end{array}%
\right.\nonumber
\end{eqnarray}
where $\{A(e)_{mn}\}_{m,n=1,2}$ is the matrix elements of $A(e)\in
SU(2)$ and
\begin{eqnarray}
\kappa(S,\ e)=\left\{%
\begin{array}{ll}
    0, & \hbox{if $e\cap S=\emptyset$, or $e$ lies in $S$;} \\
    1, & \hbox{if $e$ lies above $S$ and $e\cap S=p$;} \\
    -1, & \hbox{if $e$ lies below $S$ and $e\cap S=p$.} \\
\end{array}%
\right.\nonumber
\end{eqnarray}
Since the surface $S$ is oriented with normal $n_a$, "above" means
$n_a(\partial/\partial t)^a|_p>0$, and "below" means
$n_a(\partial/\partial t)^a|_p<0$, where $(\partial/\partial
t)^a|_p$ is the tangent vector of $e$ at $p$. Then as one might
expect, each flux $P_i(S)$ is associated with a flux vector field
$Y_i(S)$ on the quantum configuration space
$\overline{\mathcal{A}}$, algebraically introduced by the
cylindrically consistent action on cylindrical functions
$\psi_\gamma\in Cyl_\gamma(\overline{\mathcal{A}})$ as:
\begin{eqnarray}
Y_i(S)\circ\psi_\gamma(\{A(e)\}_{e\in E(\gamma)})=\{P_i(S),
\psi_\gamma\}(\{A(e)\}_{e\in E(\gamma)}),\nonumber
\end{eqnarray}
where $E(\gamma)$ is the collection of all edges of the graph
$\gamma$. The corresponding momentum operator associated with $S$ is
defined by
\begin{eqnarray}
\hat{P}_i(S):=i\hbar Y_i(S)=i\hbar\{P_i(S),\ \ \cdot\ \
\},\nonumber
\end{eqnarray}
which is essentially self-adjoint on $\mathcal{H}_{kin}$
\cite{thiemann2}. Its action on differentiable cylindrical functions
can be expressed explicitly as
\begin{eqnarray}
\hat{P}_i(S)\psi_\gamma(\{A(e)\}_{e\in
E(\gamma)})&=&\frac{\hbar}{2}\sum_{v\in V(\gamma)\cap
S}\big[\sum_{e\ at\ v}\kappa(S,\
e)\hat{J}_i^{(e,v)}\big]\psi_\gamma(\{A(e)\}_{e\in
E(\gamma)})\nonumber\\
&=&\frac{\hbar}{2}\sum_{v\in V(\gamma)\cap
S}\big[\hat{J}^{(S,v)}_{i(u)}-\hat{J}^{(S,v)}_{i(d)}\big]\psi_\gamma(\{A(e)\}_{e\in
E(\gamma)}),\label{momentum}
\end{eqnarray}
where $V(\gamma)$ is the collection of all vertices of $\gamma$, and
\begin{eqnarray}
\hat{J}^{(S,v)}_{i(u)}&\equiv&\hat{J}^{(e_1,v)}_{i}+...+\hat{J}^{(e_u,v)}_{i},\nonumber\\
\hat{J}^{(S,v)}_{i(d)}&\equiv&\hat{J}^{(e_{u+1},v)}_{i}+...+\hat{J}^{(e_{u+d},v)}_{i},\label{updown}
\end{eqnarray}
for the edges $e_1,...,e_u$ lying above $S$ and
$e_{u+1},...,e_{u+d}$ lying below $S$. Note that $\hat{J_i}^{(e,v)}$
is the momentum operator, defined in the last section, associated
with an edge $e$ connecting a vertex $v$. On the other hand, it is
obvious to construct configuration operators by cylindrical
functions $f_{\gamma\ '}\in Cyl_{\gamma\ '}(\overline{\mathcal{A}})$
as:
\begin{eqnarray}
\hat{f}_{\gamma\ '}\psi_\gamma(\{A(e)\}_{e\in
E(\gamma)}):=f_{\gamma\ '}(\{A(e)\}_{e\in E(\gamma\
')})\psi_\gamma(\{A(e)\}_{e\in E(\gamma)}).\nonumber
\end{eqnarray}
Note that $\hat{f}_{\gamma\ '}$ may change the graph, i.e.,
$\hat{f}_{\gamma\ '}$:
$Cyl_\gamma(\overline{\mathcal{A}})\rightarrow
Cyl_{\gamma\cup\gamma\ '}(\overline{\mathcal{A}})$. So far, the
elementary operators of quantum kinematics have been well defined on
$\mathcal{H}_{kin}$. One can calculate the elementary canonical
commutation relations between these operators as:
\begin{eqnarray}
&&[\hat{f}_e(A(e)),\ \hat{f}'_{e'}(A(e'))]\ =\ 0, \nonumber\\
&&[\hat{P}_i(S),\
\hat{f}_e(A(e))]\nonumber\\&=&i\hbar\big[\frac{\partial}{\partial
A(e)_{mn}}f_e(A(e))\big]\cdot\big[\frac{\kappa(S,\ e)}{2}\big]\cdot\left\{%
\begin{array}{ll}
    \sum_k A(e)_{mk}(\tau_i)_{kn} & \hbox{if $p=s(e)$} \\
    -\sum_k (\tau_i)_{mk} A(e)_{kn} & \hbox{if $p=t(e)$}, \\
\end{array}%
\right.\nonumber\\
&&[\hat{P}_i(S),\ \hat{P}_j(S')]f_e(A(e))\nonumber\\
&=&i\hbar\big[\frac{\kappa(S',\ e)}{2}\big]\epsilon^k_{\
ij}\hat{P}_k(S)]f_e(A(e)),\nonumber
\end{eqnarray}
where we assume the simplest case of one edge graphs. From the
commutation relations, one can see that the commutators between
momentum operators do not necessarily vanish if $S\cap
S'\neq\emptyset$. This unusual property reflects the
non-commutativity of quantum Riemannian structures \cite{noncommut}.
We conclude that the quantum algebra of elementary observables
(holonomy-flux algebra) has been well represented on
$\mathcal{H}_{kin}$ background-independently. So the construction of
quantum kinematics is finished. Two important remarks on the quantum
kinematics are listed below.
\begin{itemize}

\item \textit{Kinematical Vacuum and Polymer Representation}

The constant function $\Omega=1$ has the physical meaning of a
kinematical vacuum state in Hilbert space $\mathcal{H}_{kin}$ due
to its following characters. First, $\Omega=1$ is the unique state
in $\mathcal{H}_{kin}$ with maximal gauge symmetry under
Yang-Mills gauge transformations and spatial diffeomorphisms.
Secondly, $\Omega=1$ means that there is no geometry at all on the
3-manifold $\Sigma$, since the elementary operators
$\hat{A}(e)_{mn}$ and $\hat{P}_i(S)$, corresponding to connections
and triad fields in classical sense, have vanishing expectation
values on constant function. Hence it implies that the vacuum of
quantum geometry is no geometry but a bare manifold. While the
constant funciton serves as a ground state in the kinematical
Hilbert space, the low excited states (cylindrical functions) are
only excited on graphes with finite edges. There is only
1-dimensional geometry living on these graphs, so the quantum
geometry is polymer-like object. When one increases the amount of
edges and graphs such that the graphs are densely distributed in
$\Sigma$, the quantum state is highly excited and the quantum
geometry can weave the classical smooth one
\cite{ARS}\cite{AG}\cite{ma1}. Because of this picture, the
quantum kinematical representation which we obtain is also called
polymer representation for background-independent quantum
geometry.

\item \textit{Quantum Geometric Operator and Quantum Riemannnian
Geometry}

The well-established quantum kinematics of loop quantum gravity is
now in the status just like the Riemannian geometry before the
appearance of general relativity and Einstein's equation, which
gives general relativity mathematical foundation and offers living
place to the Einstein equation. Instead of classical geometric
quantities, such as scalar, vector, tensor etc., the quantities in
quantum geometry are operators on the kinematical Hilbert space
$\mathcal{H}_{kin}$, and their spectrum serve as the possible values
of the quantities in measurements. So far, the kinematical quantum
geometric operators constructed properly in loop quantum gravity
include area operators \cite{rovelli8}\cite{area}, volume operators
\cite{AL3}\cite{rovelli8}\cite{volume}, length operator
\cite{length}, $\hat{Q}$ operator \cite{ma2} etc.. Recently there
are discussions on the consistency check to different regularization
approaches for volume operators with the triad operator
\cite{thiemann17}\cite{thiemann18}. We thus will only introduce the
volume operator defined by Ashtekar and Lewandowski \cite{volume},
which is shown to be correct in the consistency check.

First, we define the area operator with respect to a 2-surface $S$
by the elementary operators. Given a closed 2-surface or a surface
$S$ with boundary, we can divide it into a large number $N$ of
small area cells $S_I$. Taking account of the classical expression
of an area, we set the area of the 2-surface to be the limit of
the Riemannian sum
\begin{eqnarray}
A_S:=\lim_{N\rightarrow\infty}[A_S]_N=\lim_{N\rightarrow\infty}\kappa\beta\sum_{I=1}^N\sqrt{P_i(S_I)P_j(S_I)
\delta^{ij}}.\nonumber
\end{eqnarray}
Then one can unambiguously obtain a quantum operator of area from
the momentum operators $\hat{P}_i(S)$. Given a cylindrical function
$\psi_\gamma\in Cyl_\gamma(\overline{\mathcal{A}})$ which has second
order derivative, the action of the area operator on $\psi_\gamma$
is defined in the limit by requiring that each area cell contains at
most only one intersecting point $v$ of the graph $\gamma$ and $S$
as
\begin{eqnarray}
\hat{A}_S\psi_\gamma:=\lim_{N\rightarrow\infty}[\hat{A}_S]_N\psi_\gamma
=\lim_{N\rightarrow\infty}\kappa\beta\sum_{I=1}^N
\sqrt{\hat{P}_i(S_I)\hat{P}_j(S_I)\delta^{ij}}\
\psi_\gamma.\nonumber
\end{eqnarray}
The regulator $N$ is easy to be removed, since the result of the
operation of the operator $\hat{P}_i(S_I)$ does not change when
$S_I$ shrinks to a point. Since the refinement of the partition does
not affect the result of action of $[\hat{A}_S]_N$ on $\psi_\gamma$,
the limit area operator $\hat{A}_S$, which is shown to be
self-adjoint \cite{area}, is well defined on $\mathcal{H}_{kin}$ and
takes the explicit expression as:
\begin{eqnarray}
\hat{A}_S\psi_\gamma=4\pi\beta \ell_p^2\sum_{v\in V(\gamma\cap
S)}\sqrt{(\hat{J}^{(S,v)}_{i(u)}-\hat{J}^{(S,v)}_{i(d)})
(\hat{J}^{(S,v)}_{j(u)}-\hat{J}^{(S,v)}_{j(d)})\delta^{ij}}\
\psi_\gamma,\nonumber
\end{eqnarray}
where $\hat{J}^{(S,v)}_{i(u)}$ and $\hat{J}^{(S,v)}_{i(d)}$ have
been defined in Eq.(\ref{updown}). It turns out that for a given $S$
one can find some finite linear combinations of spin network basis
in $\mathcal{H}_{kin}$ which diagonalize $\hat{A}_S$ with
eigenvalues given by finite sums,
\begin{eqnarray}
a_S=4\pi\beta \ell_p^2 \sum_I \sqrt{2j^{(u)}(j^{(u)}+1) +
2j^{(d)}(j^{(d)}+1) - j^{(u+d)}(j^{(u+d)}+1)},\label{spectrum}
\end{eqnarray}
where $j^{(u)}, j^{(d)}$ and $j^{(u+d)}$ are arbitrary
half-integers subject to the standard condition
\begin{eqnarray}
j^{(u+d)}\ \in \{|j^{(u)}- j^{(d)}|, |j^{(u)}- j^{(d)}|+1, ... ,
j^{(u)}+j^{(d)}\}. \label{jconstraint}
\end{eqnarray}
Hence the spectrum of the area operator is fundamentally pure
discrete, while its continuum approximation becomes excellent
exponentially rapidly for large eigenvalues. However, in
fundamental level, the area is discrete and so is the quantum
geometry. One has seen that the eigenvalue of $\hat{A}_S$ does not
vanish even in the case where only one edge intersects the surface
at a single point, whence the quantum geometry is distributional.

The form of Ashtekar and Lewandowski's volume operator was
introduced for the first time in Ref.\cite{AL3}. Then its detail
properties were discussed in Ref.\cite{volume}. Given a region $R$
with a fixed coordinate system $\{x^a\}_{a=1,2,3}$ in it, one can
introduce a partition of $R$ in the following way. Divide $R$ into
small volume cells $C$ such that, each cell $C$ is a cube with
coordinate volume less than $\epsilon$ and two different cells only
share the points on their boundaries. In each cell $C$, we introduce
three 2-surfaces $s=(S^1,S^2,S^3)$ such that $x^a$ is constant on
the surface $S^a$. We denote this partition $(C,s)$ as
$\mathcal{P}_\epsilon$. Then the volume of the region $R$ can be
expressed classically as
\begin{eqnarray}
V_R^{s}&=&\lim_{\epsilon\rightarrow0}\sum_C\sqrt{|q_{C,s}|},\nonumber
\end{eqnarray}
where
\begin{eqnarray}
q_{C,s}&=&\frac{(\kappa\beta)^3}{3!}\epsilon^{ijk}\underline{\eta}_{abc}P_i(S^a)P_j(S^b)P_k(S^c).\nonumber
\end{eqnarray}
This motivates us to define the volume operator by naively
changing $P_i(S^a)$ to $\hat{P}_i(S^a)$:
\begin{eqnarray}
\hat{V}_R^{s}&=&\lim_{\epsilon\rightarrow0}\sum_C\sqrt{|\hat{q}_{C,s}|},\nonumber\\
 \hat{q}_{C,s}&=&\frac{(\kappa\beta)^3}{3!}\epsilon^{ijk}\underline{\eta}_{abc}
\hat{P}_i(S^a)\hat{P}_j(S^b)\hat{P}_k(S^c).\nonumber
\end{eqnarray}
Note that, given any cylindrical function $\psi_\gamma\in
Cyl_\gamma(\overline{\mathcal{A}})$, we require the vertexes of the
graph $\gamma$ to be at the intersecting points of the triples of
2-surfaces $s=(S^1,S^2,S^3)$ in corresponding cells. Thus the limit
operator will trivially exist due to the same reason in the case of
the area operator. However, the volume operator defined here depends
on the choice of orientations for the triples of surfaces
$s=(S^1,S^2,S^3)$, or essentially, the choice of coordinate systems.
So it is not uniquely defined. Since, for all choice of
$s=(S^1,S^2,S^3)$, the resulting operators have correct
semi-classical limit, one settles up the problem by averaging
different operators labelled by different $s$ \cite{volume}. The
process of averaging removes the freedom in defining the volume
operator up to an overall constant $\kappa_0$. The resulting
self-adjoint operator acts on any cylindrical function
$\psi_\gamma\in Cyl_\gamma(\overline{\mathcal{A}})$ as
\begin{eqnarray}
\hat{V}_R\ \psi_\gamma&=&\kappa_0\sum_{v\in
V(\alpha)}\sqrt{|\hat{q}_{v,\gamma}|}\ \psi_\gamma, \nonumber
\end{eqnarray}
where
\begin{eqnarray}
\hat{q}_{v,\gamma}&=&(8\pi\beta\ell_p^2)^3
\frac{1}{48}\sum_{e,e',e''\ at\ v}\epsilon^{ijk}\epsilon(e,e',e'')
\hat{J}^{(e,v)}_i\hat{J}^{(e',v)}_j\hat{J}^{(e'',v)}_k, \nonumber
\end{eqnarray}
here
$\epsilon(e,e',e'')\equiv\mathrm{sgn}(\epsilon_{abc}\dot{e}^a\dot{e}'^b\dot{e}''^c)|_v$
with $\dot{e}^a$ as the tangent vector of edge $e$ and
$\epsilon_{abc}$ as the orientation of $\Sigma$. The only
unsatisfactory point in the present volume operator is the choice
ambiguity of $\kappa_0$. However, fortunately, the most recent
discussion shows that the overall undetermined constant $\kappa_0$
can be fixed to be $\sqrt{6}$ by the consistency check between the
volume operator and the triad operator
\cite{thiemann17}\cite{thiemann18}.

\end{itemize}

\section{Algebraic Aspects of Quantum Gauge Field Theory}

In this section, we would like to reformulate the theory of loop
quantum kinematics in an algebraic approach. The kinematical
Hilbert space can be obtained via $GNS$-construction for the
quantum holonomy-flux algebra. In the following, we will cast the
general programme for canonical quantization into the algebraic
framework. The formulation of loop quantum kinematics can then be
regarded as a specific application of the general algebraic
quantization programme.

\subsection{General Programme for Algebraic Quantization}

In the strategy of loop quantum gravity, a canonical programme is
performed to quantize general relativity, which has been cast into
a diffeomorphism invariant gauge field theory, or more generally,
a dynamical system with constraints. The following is a summary
for a general procedure to quantize a dynamical system with first
class constraints\footnote{Thanks to the enlightening lectures
given by Prof. T. Thiemann at Beijing Normal University.}.
\begin{itemize}
\item {\it Algebra of Classical Elementary Observables}

One starts with the classical phase space $(\mathcal{M}, \{,\})$
and $R$ ($R$ can be countable infinity\footnote{This includes the
case of field theory with infinite many degree of freedom, since
one can introduce the expression
$C_{n,\mu}=\int_{\Sigma}d^3x\phi_n(x)C_\mu(x)$, where
$\{\phi_n(x)\}^\infty_{n=1}$ forms a system of basis in
$L^2(\Sigma,d^3x)$.} ) first-class constraints $C_r(r=1...R)$ such
that $\{C_r , C_s\}=\Sigma_{t=1}^R f_{rs}^{\ \ t}C_t$, where
$f_{rs}^{\ \ t}$ is generally a function on phase space, namely,
structure function of Poisson algebra. The algebra of classical
elementary observables $\textbf{\textsf{P}}$ is
defined as:\\ \\
\textbf{Definition 4.1.1}: \textit{The algebra of classical
elementary observables $\textbf{\textsf{P}}$ is a collection of
functions $f(m),m\in\mathcal{M}$ on the
phase space such that \\
$(1)$ $f(m)\in\textbf{\textsf{P}}$ should separate the point of
$\mathcal{M}$, i.e., for any $m\neq m'$, there exists
$f(m)\in\textbf{\textsf{P}}$, such that $f(m)\neq f(m')$;
$($analogy to the $p$ and $q$ in $\mathcal{M}=\mathrm{T}^*\mathbf{R}$.$)$\\
$(2)$ $f(m), f'(m)\in\textbf{\textsf{P}}\ \Rightarrow\ \{f(m),
f'(m)\}\in\textbf{\textsf{P}}$
$($closed under Poisson bracket$)$;\\
$(3)$ $f(m)\in\textbf{\textsf{P}}\ \Rightarrow\
\bar{f}(m)\in\textbf{\textsf{P}}$
$($closed under complex conjugate$)$. }\\ \\
So $\textbf{\textsf{P}}$ forms a sub $*$-Poisson algebra of
$C^\infty(\mathcal{M})$. In the case of
$\mathcal{M}=\mathrm{T}^*\mathbf{R}$, $\textbf{\textsf{P}}$ is
generated by the conjugate pair $(q, p)$ with $\{q,p\}=1$.

\item {\it Quantum Algebra of Elementary Observables}

Given the algebra of classical elementary observables
$\textbf{\textsf{P}}$, the quantum algebra of elementary
observables can be constructed as follows. Consider the formal
finite sequences of classical observable $(f_1...\\f_n)$ with
$f_k\in\textbf{\textsf{P}}$. Then the operations of multiplication
and involution are defined as
\begin{eqnarray}
(f_1,...,f_n)\cdot(f'_1,...,f_m')&:=&(f_1,...,f_n,f_1',...,f_m'),\nonumber\\
(f_1,..,f_n)^*&:=&(\bar{f}_n,...,\bar{f}_1).\nonumber
\end{eqnarray}
One can define the direct sum of different sequences with different
number of elements. Then the general element of the newly
constructed free $*$-algebra $F(\textbf{\textsf{P}})$ of
$\textbf{\textsf{P}}$, is formally expressed as
$\oplus^N_{k=1}(f^{(k)}_1,...f^{(k)}_{n_k})$, where
$f^{(i)}_{n_i}\in\textbf{\textsf{P}}$. Consider the elements of the
form (sequences consisting of only one element)
\begin{eqnarray}
(f+f')-(f)-(f'),\ \ (zf)-z(f),\ \
[(f),(f')]-i\hbar(\{f,f'\}),\nonumber
\end{eqnarray}
where $z\in \mathbf{C}$ and the canonical commutation bracket is
defined as
\begin{eqnarray}
[(f),(f')]:=(f)\cdot(f')-(f')\cdot(f).\nonumber
\end{eqnarray}
A 2-side ideal $\mathcal{Z}$ of $F(\textbf{\textsf{P}})$ can be
constructed from these element, and is preserved by the action of
involution $*$. One thus obtains\\ \\
\textbf{Definition 4.1.2}: \textit{The quantum algebra
$\textbf{\textsf{A}}$ of elementary observables is defined to be
the
quotient $*$-algebra $F(\textbf{\textsf{P}})/\mathcal{Z}$.}\\ \\
Note that the motivation to construct a quantum algebra of
elementary observables is to avoid the problem of operators ordering
in quantum theory so that the quantum algebra $\textbf{\textsf{A}}$
can be represented on a Hilbert space without ordering ambiguity.

\item {\it Representation of Quantum Algebra}

In order to obtain a quantum theory, we need to quantize the
classical observable in the dynamical system. The, so called,
quantization is nothing but a $*$-representation map $\pi$ from
the quantum algebra of elementary observable $\textbf{\textsf{A}}$
to the collection of linear operators $\mathcal{L}(\mathcal{H})$
on a Hilbert Space $\mathcal{H}$. Recall that a map $\pi$:
$\textbf{\textsf{A}}\rightarrow\mathcal{L}(\mathcal{H})$ is a
*-representation if and only if (1) there exists a dense subspace
$\mathcal{D}$ of $\mathcal{H}$ contained in
$\cap_{a\in\textbf{\textsf{A}}}[D(\pi(a))\cap D(\pi(a^*))]$ where
$D(\pi(a))$ is the domain of the operator $\pi(a)$ and (2) for
every $a,b\in\textbf{\textsf{A}}$ and $\lambda\in\mathbf{C}$ the
following conditions are satisfied in $\mathcal{D}$,
\begin{eqnarray}
\pi(a+b)=\pi(a)+\pi(b),&&\ \ \pi(\lambda a)=\lambda\pi(a),\nonumber\\
\pi(a\cdot b)=\pi(a)\pi(b),&&\ \ \pi(a^*)=\pi(a)^\dagger.\nonumber
\end{eqnarray}
Note that $\mathcal{L}(\mathcal{H})$ fails to be an algebra
because the domains of unbounded operators cannot be the whole
Hilbert space. However, the collection of bounded operators on
some Hilbert space is really a $*$-algebra. At the level of
quantum mechanics, the well-known Stone-Von Neumann Theorem
concludes that in quantum mechanics, there is only one strongly
continuous, irreducible, unitary representation of the Weyl
algebra, up to unitary equivalence (see, for example,
Ref.\cite{simon}). However, the conclusion of Stone-Von Neumann
cannot be generalized to the quantum field theory because the
latter has infinite many degrees of freedom (for detail, see, for
example \cite{wald1}). In quantum field theory, a representation
can be constructed by $GNS$(Gel'fand-Naimark-Segal) construction
for a quantum algebra of elementary observable
$\textbf{\textsf{A}}$, which is a unital $*$-algebra actually. The
$GNS$ construction for the representation of quantum algebra
$\mathcal{A}$ is briefly
summarized as follows.\\ \\
\textbf{Definition 4.1.3}: \textit{Given a positive linear
functional $($a state $)$ $\omega$ on $\textbf{\textsf{A}}$, the
null space $\mathcal{N}_{\omega}\in\textbf{\textsf{A}}$ with
respect to $\omega$ is defined as
$\mathcal{N}_{\omega}:=\{a\in\textbf{\textsf{A}}|\omega(a^*\cdot\
a)=0\}$, which is a left ideal in $\textbf{\textsf{A}}$. Then a
quotient map can be defined as $[.]$:
$\textbf{\textsf{A}}\rightarrow\textbf{\textsf{A}}/\mathcal{N}_{\omega}$;
$a\mapsto[a]:=\{a+b|b\in\mathcal{N}_{\omega}\}$. The
GNS-representation for $\textbf{\textsf{A}}$ with respect to
$\omega$ is a $*$-representation map: $\pi_{\omega}$:
$\textbf{\textsf{A}}\rightarrow\mathcal{L}(\mathcal{H}_{\omega})$,
where
$\mathcal{H}_{\omega}:=\langle\textbf{\textsf{A}}/\mathcal{N}_{\omega}\rangle$
and $\langle.\rangle$ denotes the completion with respect to the
naturally equipped well-defined inner product
\[<[a]|[b]>_{\mathcal{H}_{\omega}}:=\omega(a^*\cdot\ b)\] on
$\mathcal{H}_{\omega}$. This representation map is defined by
\[\pi_\omega(a)[b]:=[a\cdot b],\ \forall\ a\in\textbf{\textsf{A}}\
\mathrm{and}\ [b]\in\mathcal{H}_{\omega},\] where $\pi_\omega(a)$ is
an unbounded operator in general. Moreover, $GNS$ representation is
a cyclic representation, i.e., $\exists\
\Omega\in\mathcal{H}_{\omega}$, such that
$\langle\{\pi(a)\Omega|a\in\textbf{\textsf{A}}\}\rangle=\mathcal{H}_{\omega}$
and $\Omega$ is called a cyclic vector in the representation space.
In fact $\Omega_\omega:=[1]$ is a cyclic vector in
$\mathcal{H}_{\omega}$ and
$\langle\{\pi_\omega(a)\Omega_\omega|a\in\textbf{\textsf{A}}\}\rangle=\mathcal{H}_{\omega}$.
As a result, the positive linear functional with which we begin can
be expressed as \[\omega(a)=<\Omega_\omega|
\pi_\omega(a)\Omega_\omega>_{\mathcal{H}_{\omega}}.\] Thus a
positive linear functional on $\textbf{\textsf{A}}$ is equivalent to
a cyclic representation of $\textbf{\textsf{A}}$, which is a triple
$(\mathcal{H}_{\omega},\pi_\omega, \Omega_\omega)$. Moreover, every
non-degenerate representation is an orthogonal direct sum of
cyclic representations $($ for proof, see $\cite{conway}$ $)$ .}\\

So the kinematical Hilbert space $\mathcal{H}_{kin}$ for the system
with constrains can be obtained by $GNS$-construction. In the case
that there are gauge symmetries in our dynamical system, supposing
that there is a group $G$ acting on $\textbf{\textsf{A}}$ by
automorphisms $\alpha_g:
\textbf{\textsf{A}}\rightarrow\textbf{\textsf{A}},\ \forall\ g\in
G$, it is preferred to construct a gauge invariant representation of
$\textbf{\textsf{A}}$. So we require the positive linear functional
$\omega$ on $\textbf{\textsf{A}}$ to be gauge invariant, i.e.,
$\omega\circ\alpha_g=\omega$. Then the group $G$ is represented on
the Hilbert space $\mathcal{H}_\omega$ as:
\begin{eqnarray}
U(g)\pi_\omega(a)\Omega_\omega\:=\pi_\omega(\alpha_g(a))\Omega_\omega,\nonumber
\end{eqnarray}
and such a representation is a unitary representation of $G$. In
loop quantum gravity, it is crucial to construct a gauge invariant
and diffeomorphism invariant representation for the quantum algebra
of elementary observables.

\item {\it Implementation of the Constraints}

In the Dirac quantization programme for a system with constraints,
the constraints should be quantized as some operators in a
kinematical Hilbert space $\mathcal{H}_{kin}$. One then solves them
at quantum level to get a physical Hilbert space
$\mathcal{H}_{phys}$, that is, to find a quantum analogy $\hat{C}_r$
of the classical constraint formula $C_r$ and to solve the general
solution of the equation $\hat{C}_r\Psi=0$. However, there are
several problems in the construction of the constraint operator
$\hat{C}_r$.
\begin{itemize}
\item[(i)] $C_r$ is in general not in $\textbf{\textsf{P}}$, so there is a
factor ordering ambiguity in quantizing $C_r$ to be an operator
$\hat{C}_r$. \item[(ii)] In quantum field theory, there are
ultraviolet(UV) divergence problems in constructing operators.
However, the UV divergence can be avoided in the background
independent approach. \item[(iii)] Sometimes, quantum anomaly
appears when there are structure functions in the Poisson algebra.
Although classically we have $\{C_r , C_s\}=\Sigma_{t=1}^R f_{rs}^{\
\ t}C_t,\ \ r,s,t=1,...,R$, where $f_{rs}^{\ \ t}$ is a function on
phase space, quantum mechanically it is possible that $[\hat{C}_r ,
\hat{C}_s]\neq i\hbar\Sigma_{t=1}^R \hat{f}_{rs}^{\ \ t}\hat{C}_t$
due to the ordering ambiguity between $\hat{f}_{rs}^{\ \ t}$ and
$\hat{C}_t$. If one sets $[\hat{C}_r ,
\hat{C}_s]=\frac{i\hbar}{2}\Sigma_{t=1}^R (\hat{f}_{rs}^{\ \
t}\hat{C}_t+\hat{C}_t\hat{f}_{rs}^{\ \ t})$, for $\Psi$ satisfying
$\hat{C}_r\Psi=0$, we have
\begin{eqnarray}
[\hat{C}_r , \hat{C}_s]\Psi=\frac{i\hbar}{2}\sum_{t=1}^R
\hat{C}_t\hat{f}_{rs}^{\ \
t}\Psi=\frac{i\hbar}{2}\sum_{t=1}^R[\hat{C}_t, \hat{f}_{rs}^{\ \
t}]\Psi.\label{anomaly}
\end{eqnarray}
However, $[\hat{C}_t, \hat{f}_{rs}^{\ \ t}]\Psi$ are not necessary
to equal to zero for all $r,s,t=1...R$. If not, the problem of
quantum anomaly comes out and the new quantum constraints
$[\hat{C}_t, \hat{f}_{rs}^{\ \ t}]\Psi=0$ have to be imposed on
physical quantum states, since the classical Poisson brackets $\{C_r
, C_s\}$ are weakly equal to zero on the constraint surface
$\overline{\mathcal{M}}\subset\mathcal{M}$. Thus too many
constraints are imposed so that the physical Hilbert space
$\mathcal{H}_{phys}$ would be too small. Hence the quantum anomaly
should be avoided anyway.
\end{itemize}

\item {\it Solving the Constraints and Physical Hilbert Space}

In general the original Dirac quantization approach can not be
carried out directly, since there is usually no nontrivial
$\Psi\in\mathcal{H}_{kin}$ such that $\hat{C}_r\Psi=0$. This
happens when the constraint operator $\hat{C}_r$ has "generalized
eigenfunctions" rather than eigenfunctions. One then develops the
so-called Refined Algebraic Quantization Programme, where the
solutions of the quantum constraint can be found in the algebraic
dual space of a dense subset in $\mathcal{H}_{kin}$ (see e.g.
\cite{ALM}\cite{GM}). The quantum diffeomorphism constraint in
loop quantum gravity is solved in this approach (see section 5.2).
But the situation for the scalar constraint in general relativity
is so subtle that it is difficult to carry out the quantization
programme straightforwardly. Recently, Thiemann proposed the
method of \textbf{Master Constraint Approach} to solve the quantum
constraints \cite{thiemann3}, which seems especially suitable to
deal with the particular feature of the constraint algebra of
general relativity. A master constraint is defined as
$\textbf{M}:=\frac{1}{2}\Sigma_{r,s=1}^R\textbf{K}_{rs}C_s\bar{C}_r$
for some real positive matrix $\textbf{K}_{rs}$. Classically one
has $\textbf{M}=0$ if and only if $C_r=0$ for all $r=1...R$. So
quantum mechanically one may consider solving the \textbf{Master
Equation}: $\hat{\textbf{M}}\Psi=0$ to obtain the physical Hilbert
space $\mathcal{H}_{phys}$ instead of solving $\hat{C}_r\Psi=0,\
\forall\ r=1...R$. Because the master constraint $\textbf{M}$ is
classically positive, one has opportunities to implement it as a
self-adjoint operator on $\mathcal{H}_{kin}$. If it is indeed the
case and $\mathcal{H}_{kin}$ is separable, one can use the direct
integral representation of $\mathcal{H}_{kin}$ associated with the
self-adjoint operator $\hat{\textbf{M}}$ to obtain
$\mathcal{H}_{phys}$:
\begin{eqnarray}
\mathcal{H}_{kin}&\sim&\int_\mathbf{R}^\oplus
d\mu(\lambda)\mathcal{H}^\oplus_{\lambda}\nonumber,\\
<\Phi|\Psi>_{kin}&=&\int_\mathbf{R}d\mu(\lambda)<\Phi|\Psi>_{\mathcal{H}^\oplus_{\lambda}},\label{did}
\end{eqnarray}
where $\mu$ is a so-called spectral measure and
$\mathcal{H}^\oplus_{\lambda}$ is the (generalized) eigenspace of
$\hat{\textbf{M}}$ with the eigenvalue $\lambda$. The physical
Hilbert space is then formally obtained as
$\mathcal{H}_{phys}=\mathcal{H}^\oplus_{\lambda=0}$ with the induced
physical inner product $<\ |\
>_{\mathcal{H}^\oplus_{\lambda=0}}\ $. Now the issue of quantum anomaly is represented in terms of the size of
$\mathcal{H}_{phys}$ and the existence of sufficient numbers of
semi-classical states.

\item {\it Physical Observables}

We denote $\mathcal{M}$ as the original unconstrained phase space,
$\overline{\mathcal{M}}$ as the constraint surface, i.e.,
$\overline{\mathcal{M}}:=\{m\in\mathcal{M}|C_r(m)=0,\ \forall\
r=1...R\}$, and $\hat{\mathcal{M}}$ as the reduced phase space, i.e.
the space of orbits for gauge transformations generated by all
$C_r$. The concept of Dirac observable is defined as the
follows.\\ \\
\textbf{Definition 4.1.4}: \\
\textit{$(1)$ A function $\mathcal{O}$ on $\mathcal{M}$ is called a
weak Dirac observable if and only if the function depends only on
points of $\hat{\mathcal{M}}$, i.e.,
$\{\mathcal{O},C_r\}|_{\overline{\mathcal{M}}}=0$ for all $r=1...R$.
For the quantum version, a self-adjoint operator $\hat{\mathcal{O}}$
is a weak Dirac observable
if and only if the operator can be well defined on the physical Hilbert space.\\
$(2)$ A function $\mathcal{O}$ on $\mathcal{M}$ is called a strong
Dirac observable if and only if
$\{\mathcal{O},C_r\}|_{\mathcal{M}}=0$ for all $r=1...R$. For the
quantum version, a self-adjoint operator $\hat{\mathcal{O}}$ is a
strong Dirac observable if and only if the operator can be defined
on the kinematic Hilbert space $\mathcal{H}_{kin}$ and
$[\hat{\mathcal{O}},\hat{C}_r]=0$ in $\mathcal{H}_{kin}$ for all
$r=1...R$.}\\ \\
A physical observable is at least a weak Dirac observable. While
Dirac observables have been found in symmetry reduced models, some
even with an infinite number of degrees of freedom, it seems
extremely difficult to find them in full general relativity.
Moreover the Hamiltonian is a linear combination of first-class
constraints. So there is no dynamics in the reduced phase space, and
the meaning of time evolution of the Dirac observables becomes
subtle. However, using the concepts of partial and complete
observables \cite{rovelli1}\cite{rovelli2}\cite{rovelli}, a
systematic method to get Dirac observables is being developed, and
the problem of time in such system with a Hamiltonian
$H=\Sigma_{r=1}^R\beta_rC_r$ is being addressed.

Classically, let $f(m)$ and $\{T_r(m)\}_{r=1}^R$ be gauge
non-invariant functions (partial observables) on phase space
$\mathcal{M}$, such that $A_{sr}\equiv\{C_s,T_r\}$ is a
non-degenerate matrix on $\mathcal{M}$. A system of classical weak
Dirac observables (complete observables) $F_{f,T}^\tau$ labelled by
a collection of real parameters $\tau\equiv\{\tau_r\}_{r=1}^R$ can
be constructed as
\begin{eqnarray}
F_{f,T}^\tau:=\sum^\infty_{\{n_1\cdot\cdot\cdot
n_R\}}\frac{(\tau_1-T_1)^{n_1}\cdot\cdot\cdot(\tau_R-T_R)^{n_R}}{n_1!\cdot\cdot\cdot
n_R!}
\widetilde{X}_1^{n_1}\circ\cdot\cdot\cdot\circ\widetilde{X}_R^{n_R}(f)\nonumber,
\end{eqnarray}

where $\ \widetilde{X}_r(f):=\{\Sigma_{s=1}^R A^{-1}_{rs}C_s,\
f\}\equiv\{\widetilde{C}_r,\ f\}\ $. It can be verified that $\
[\widetilde{X}_r,\ \widetilde{X}_s]|_{\overline{\mathcal{M}}}=0$ and
$\{F_{f,T}^\tau,C_r\}|_{\overline{\mathcal{M}}}=0$, for all
$r=1...R$ (for details see \cite{dittrich} and \cite{dittrich2}).

The partial observables $\{T_r(m)\}_{r=1}^R$ may be regarded as
clock variables, and $\tau_r$ is the time parameter for $T_r$. The
gauge is fixed by giving a system of functions $\{T_r(m)\}_{r=1}^R$
and corresponding parameters $\{\tau_r\}_{r=1}^R$, namely, a section
in $\overline{\mathcal{M}}$ is selected by $T_r(m)=\tau_r$ for each
$r$, and $F_{f,T}^\tau$ is the value of $f$ on the section. To solve
the problem of dynamics, one assumes another set of canonical
coordinates $(P_1,\cdot\cdot\cdot,
P_{N-R},\Pi_1,\cdot\cdot\cdot,\Pi_{R};Q_1,\cdot\cdot\cdot,Q_{N-R},
T_1,\cdot\cdot\cdot,T_R)$ by canonical transformations in the phase
space $(\mathcal{M},\{\ ,\ \})$, where $P_s$ and $\Pi_r$ are
conjugate to $Q_s$ and $T_r$ respectively. After solving the
complete system of constraints
$\{C_r(P_i,Q_j,\Pi_s,T_t)=0\}_{r=1}^R$, the Hamiltonian $H_r$ with
respect to the time $T_r$ is obtained as $H_r:=\Pi_r(P_i,Q_j,T_t)$.
Given a system of constants $\{(\tau_0)_r\}_{r=1}^R$, for an
observable $f(P_i,Q_j)$ depending only on $P_i$ and $Q_j$, the
physical dynamics is given by \cite{dittrich}\cite{thiemann14}:
\begin{eqnarray}
(\frac{\partial}{\partial\tau_r})_{\tau=\tau_0}F_{f,T}^\tau|_{\overline{\mathcal{M}}}=F_{\{H_r,f\},T}^{\tau_0}
|_{\overline{\mathcal{M}}}
=\{F_{H_r,T}^{\tau_0},F_{f,T}^{\tau_0}\}|_{\overline{\mathcal{M}}},
\nonumber
\end{eqnarray}
where $F_{H_r,T}^{\tau_0}$ is the physical Hamiltonian function
generating the evolution with respect to $\tau_r$. Thus one has
addressed the problem of time and dynamics as a result.

\item {\it Semi-classical Analysis}

An important issue in the quantization is to check whether the
quantum constraint operators have correct classical limits. This has
to be done by using the kinematical semiclassical states in
$\mathcal{H}_{kin}$. Moreover, the physical Hilbert space
$\mathcal{H}_{phys}$ must contain enough semi-classical states to
guarantee that the quantum theory one obtains can return to the
classical theory when $\hbar\rightarrow0$. The semi-classical states
in a Hilbert space $\mathcal{H}$ should have the following
properties.\\ \\
\textbf{Definition 4.1.5}:  \textit{Given a class of observables
$\mathcal{S}$ which comprises a subalgebra in the space
$\mathcal{L}(\mathcal{H})$ of linear operators on the Hilbert
space, a family of (pure) states $\{\omega_m\}_{m\in\mathcal{M}}$
are said to be
semi-classical with respect to $\mathcal{S}$ if and only if:\\
$(1)$ The observables in $\mathcal{S}$ should have correct
semi-classical limit on semi-classical states and the fluctuations
should be small, i.e.,
\begin{eqnarray}
\lim_{\hbar\rightarrow0}|\frac{\omega_m(\hat{a})-a(m)}{a(m)}|=0,\nonumber\\
\lim_{\hbar\rightarrow0}|\frac{\omega_m(\hat{a}^2)-\omega_m(\hat{a})^2}{\omega_m(\hat{a})^2}|=0,\nonumber
\end{eqnarray}
for all $\hat{a}\in\mathcal{S}$.\\
$(2)$ The set of cyclic vectors $\Omega_m$ related to $\omega_m$ via
the $GNS$-representation
$(\pi_\omega,\mathcal{H}_\omega,\Omega_\omega)$ is dense in
$\mathcal{H}$.}\\

Seeking for semiclassical states are one of open issues of current
research in loop quantum gravity. Recent original works focus on the
construction of coherent states of loop quantum gravity in analogy
with the coherent states for harmonic oscillator system
\cite{thiemann10}\cite{thiemann11}\cite{thiemann12}\cite{thiemann13}\cite{AL2}\cite{shadow}.

\end{itemize}
The above is the general programme to quantize a system with
constraints. In the following subsection, we will apply the
programme to the theory of general relativity and restrict our
view to the representation with the properties of background
independence and spatial diffeomorphism invariance.

\subsection{Algebraic Construction of Loop Quantum Kinematics}

All prior constructions in section 3 bear analogy with
constructive quantum field theory. In this subsection we perform
the background-independent construction of algebraic quantum field
theory for loop quantum gravity. First we construct the algebra of
classical observables. Taking account of the future quantum
analogs, we define the algebra of classical observables
$\textbf{\textsf{P}}$ as the Poission $*$-subalgebra generated by
the functions of holonomies (cylindrical functions) and the fluxes
of triad fields smeared on some 2-surface. Namely, one can define
the classical algebra in analogy with geometric quantization in
finite dimensional phase space case by the so-called classical
Ashtekar-Corichi-Zapata holonomy-flux
$*$-algebra as the following \cite{unique}.\\ \\
\textbf{Definition 4.2.1}\\
\textit{The classical Ashtekar-Corichi-Zapata holonomy-flux
$*$-algebra is defined to be a vector space
$\textbf{\textsf{P}}_{\mathbf{ACZ}}:=Cyl(\overline{\mathcal{A}})\times\mathcal{V}^{\mathbf{C}}(\overline{\mathcal{A}})$,
where $\mathcal{V}^{\mathbf{C}}(\overline{\mathcal{A}})$ is the
vector space of cylindrically consistent vector fields spanned by
the vector fields $\psi Y_f(S)$ and their commutators, here the
smeared flux vector field $Y_f(S)$ is defined by
$Y_f(S):=\{\int_S\underline{\eta}_{abc}\widetilde{P}^c_if^i, \
\cdot\ \}$ for any $su(2)$-valued functions $f^i$ with compact
supports on $S$ and $\psi$ are cylindrical functions on
$\overline{\mathcal{A}}$. We equip
$\textbf{\textsf{P}}_{\mathbf{ACZ}}$
with the structure of an $*$-Lie algebra by:\\
(1) Lie bracket $\{\ ,\ \}:\
\textbf{\textsf{P}}_{\mathbf{ACZ}}\times\textbf{\textsf{P}}_{\mathbf{ACZ}}\rightarrow\
\textbf{\textsf{P}}_{\mathbf{ACZ}}$ is defined by
\begin{eqnarray}
\{(\psi,Y),\ (\psi',Y')\}:=(Y\circ\psi'-Y'\circ\psi,[Y,
Y']),\nonumber
\end{eqnarray}
for all $(\psi, Y),\ (\psi',
Y')\in\textbf{\textsf{P}}_{\mathbf{ACZ}}$ with $\psi, \psi'\in
Cyl(\overline{\mathcal{A}})$ and $Y,
Y'\in\mathcal{V}^{\mathbf{C}}(\overline{\mathcal{A}})$.\\ \\
(2) Involution: $a\mapsto\bar{a}\ \forall\
a\in\textbf{\textsf{P}}_{\mathbf{ACZ}}$ is defined by complex
conjugate of cylindrical functions and vector fields, i.e.,
$\bar{a}:=(\overline{\psi}, \overline{Y})\ \forall\ a=(\psi,
Y)\in\textbf{\textsf{P}}_{\mathbf{ACZ}}$, where
$\overline{Y}\circ\psi:=\overline{Y\circ\overline{\psi}}$.\\ \\
(3) $\textbf{\textsf{P}}_{\mathbf{ACZ}}$ admits a natural action of
$Cyl(\overline{\mathcal{A}})$ by
\begin{eqnarray}
\psi'\circ(\psi,Y):=(\psi'\psi,\psi'Y),\nonumber
\end{eqnarray}
which gives $\textbf{\textsf{P}}_{\mathbf{ACZ}}$ a module
structure.}\\ \\
The classical Ashtekar-Corichi-Zapata holonomy-flux $*$-algebra
serves as an elementary algebra in our dynamic system of gauge
field. Then one can construct the quantum algebra of elementary
observables from $\textbf{\textsf{P}}_{\mathbf{ACZ}}$ in analogy
with Definition 4.1.2.\\ \\
\textbf{Definition 4.2.2} \\
\textit{The abstract free algebra
$F(\textbf{\textsf{P}}_{\mathbf{ACZ}})$ of the classical $*$-algebra
is defined by the formal direct sum of finite sequences of classical
observables $(a_1,...,a_n)$ with
$a_k\in\textbf{\textsf{P}}_{\mathbf{ACZ}}$, where the operations of
multiplication and involution are defined as
\begin{eqnarray}
(a_1,...,a_n)\cdot(a'_1,...,a'_m)&:=&(a_1,...,a_n,a_1',...,a_m'),\nonumber\\
(a_1,..,a_n)^*&:=&(\bar{a}_n,...,\bar{a}_1).\nonumber
\end{eqnarray}
A $2$-sided ideal $\textbf{\textsf{Z}}$ can be generated by the
following elements,
\begin{eqnarray}
(a+a')-(a)-(a'),\ \ (za)-z(a),\nonumber\\
\ [(a),(a')]-i\hbar(\{a,a'\}),\ \ \ \ \ \ \ \nonumber\\
((\psi,0), a)-(\psi\circ a),\ \ \ \ \ \ \ \ \ \nonumber
\end{eqnarray}
where the canonical commutation bracket is defined by
\begin{eqnarray}
[(a),(a')]:=(a)\cdot(a')-(a')\cdot(a).\nonumber
\end{eqnarray}
Note that the ideal $\textbf{\textsf{Z}}$ is preserved by the involution $*$. \\
The quantum holonomy-flux $*$-algebra is defined by the quotient
$*$-algebra
$\textbf{\textsf{A}}=F(\textbf{\textsf{P}}_{\mathbf{ACZ}})/\textbf{\textsf{Z}}$,
which contains the unital element $1:=((1,0))$. Note that a sup-norm
has been defined by Eq.$(\ref{norm})$ for the Abelian
sub-$*$-algebra $Cyl(\overline{\mathcal{A}})$ in
$\textbf{\textsf{A}}$.}\\ \\
For simplicity, we denote the one element sequences $((\psi,0))$ and
$((0,Y))$ $\forall\ \psi\in Cyl(\overline{\mathcal{A}}),\ Y\in
\mathcal{V}^{\mathbf{C}}(\overline{\mathcal{A}})$ in
$\textbf{\textsf{A}}$ by $(\psi)$ and $(Y)$ respectively. In
particular, for all cylindrical functions $(\psi)$ and flux vector
fields $(Y_f(S))$,
\begin{eqnarray}
(\psi)^*=(\overline{\psi})\ \ \ \mathrm{and}\ \ \
(Y_f(S))^*=(Y_f(S)).\nonumber
\end{eqnarray}
Note that every element of the algebra $\textbf{\textsf{A}}$ is a
finite linear combination of elements of the form
\begin{eqnarray}
&&(\psi),\nonumber\\
&&(\psi_1)\cdot(Y_{f_{11}}(S_{11})),\nonumber\\
&&(\psi_2)\cdot(Y_{f_{21}}(S_{21}))\cdot(Y_{f_{22}}(S_{22})),\nonumber\\
&&...\nonumber\\
&&(\psi_k)\cdot(Y_{f_{k1}}(S_{k1}))\cdot(Y_{f_{k2}}(S_{k2}))\cdot...\cdot(Y_{f_{kk}}(S_{kk})),\nonumber\\
&&...\nonumber
\end{eqnarray}
Moreover, given a cylindrical function $\psi$ and a flux vector
field $Y_f(S)$, one has the relation from the commutation relation:
\begin{eqnarray}
(Y_f(S))\cdot(\psi)=i\hbar
(Y_f(S)\circ\psi)+(\psi)\cdot(Y_f(S)).\label{Y}
\end{eqnarray}
Then the kinematical Hilbert space $\mathcal{H}_{kin}$ can be
obtained properly via the $GNS$-construction for unital $*$-algebra
$\textbf{\textsf{A}}$ in the same way as in Definition 4.1.3. By
$GNS$-construction, a positive linear functional, i.e. a state
$\omega$, on $\textbf{\textsf{A}}$ defines a cyclic representation
$(\mathcal{H}_\omega,\pi_\omega,\Omega_\omega)$ for
$\textbf{\textsf{A}}$. In our case of quantum holonomy-flux
$*$-algebra, the state with both internal gauge invariance and
diffeomorphism invariance is defined for any $\psi=[\psi_\alpha]\in
Cyl(\overline{\mathcal{A}})$ and non-vanishing flux vector field
$Y_f(S)\in \mathcal{V}^{\mathbf{C}}(\overline{\mathcal{A}})$ as
\cite{unique}:
\begin{eqnarray}
&&\omega\big((\psi)\big):=\int_{SU(2)^N}
d\mu_H(A(e_1))...d\mu_H(A(e_N))\psi_\alpha(A(e_1),...,A(e_N)),\nonumber\\
&&\omega\big(\textbf{\textsf{a}}\cdot(Y_f(S))\big):=0,\ \ \
\forall\textbf{\textsf{a}}\in\textbf{\textsf{A}},\nonumber
\end{eqnarray}
where we assume that $\alpha$ contains $N$ edges. This $\omega$ is
called Ashtekar-Isham-Lewandowski state. The null space
$\textbf{\textsf{N}}_{\omega}\in\textbf{\textsf{A}}$ with respect to
$\omega$ is defined as
$\textbf{\textsf{N}}_{\omega}:=\{\textbf{\textsf{a}}\in\textbf{\textsf{A}}|\omega(\textbf{\textsf{a}}^*\cdot
\textbf{\textsf{a}})=0\}$, which is a left ideal. Then a quotient
map can be defined as:
\begin{eqnarray}
[.]:\ \textbf{\textsf{A}}&\rightarrow&\textbf{\textsf{A}}/{\textbf{\textsf{N}}_{\omega}};\nonumber\\
\textbf{\textsf{a}}&\mapsto&[\textbf{\textsf{a}}]:=\{\textbf{\textsf{a}}+\textbf{\textsf{b}}|\textbf{\textsf{b}}
\in\textbf{\textsf{N}}_{\omega}\}.\nonumber
\end{eqnarray}
The $GNS$-representation for $\textbf{\textsf{A}}$ with respect to
$\omega$ is a representation map: $\pi_{\omega}$:
$\textbf{\textsf{A}}\rightarrow\mathcal{L}(\mathcal{H}_{\omega})$
such that
$\pi_{\omega}(\textbf{\textsf{a}}\cdot\textbf{\textsf{b}})=\pi_{\omega}(\textbf{\textsf{a}})\pi_{\omega}
(\textbf{\textsf{b}})$, where
$\mathcal{H}_{\omega}:=\langle\textbf{\textsf{A}}/{\textbf{\textsf{N}}_{\omega}}\rangle=\langle
Cyl(\overline{\mathcal{A}})\rangle=\mathcal{H}_{kin}$ by
straightforward verification and the $\langle\cdot\rangle$ denotes
the completion with respect to the natural equipped inner product on
$\mathcal{H}_{\omega}$,
\begin{eqnarray}
<[\textbf{\textsf{a}}]|[\textbf{\textsf{b}}]>_{\mathcal{H}_{\omega}}:=\omega(\textbf{\textsf{a}}^*\cdot
\textbf{\textsf{b}}),\nonumber
\end{eqnarray}
which is equivalent to Eq. (\ref{inner}). The representation map
$\pi_{\omega}$ is defined by
\begin{eqnarray}
\pi_\omega(\textbf{\textsf{a}})[\textbf{\textsf{b}}]:=[\textbf{\textsf{a}}\cdot\textbf{\textsf{b}}],\
\  \forall\ \textbf{\textsf{a}}\in \textbf{\textsf{A}}, \
\mathrm{and}\ [\textbf{\textsf{b}}]\in\mathcal{H}_{\omega}.\nonumber
\end{eqnarray}
Note that $\pi_\omega(\textbf{\textsf{a}})$ is an unbounded operator
in general. It is easy to verify that
\begin{eqnarray}
\pi_\omega((Y_f(S)))[(\psi)]=i\hbar[(Y_f(S)\circ\psi)]\nonumber
\end{eqnarray}
via Eq.(\ref{Y}). Hence $\pi_\omega((Y_f(S)))[(\psi)]$ is
identical with $\hat{P}_f(S)=i\hbar Y_f(S)$ on
$\mathcal{H}_{kin}$, which can be obtained in analogy with the
method we employ at the beginning of section 3.5. Moreover, since
$\Omega_\omega:=[(1)]$ is a cyclic vector in
$\mathcal{H}_{\omega}$, the positive linear functional with which
we begin can be expressed as
\begin{eqnarray}
\omega(\textbf{\textsf{a}})=<\Omega_\omega|
\pi_\omega(\textbf{\textsf{a}})\Omega_\omega>_{\mathcal{H}_{\omega}}.\nonumber
\end{eqnarray}
Thus the Ashtekar-Isham-Lewandowski state $\omega$ on
$\textbf{\textsf{A}}$ is equivalent to a cyclic representation
$(\mathcal{H}_{\omega},\pi_\omega, \Omega_\omega)$ for
$\textbf{\textsf{A}}$, which is the Ashtekar-Isham-Lewandowski
representation for quantum holonomy-flux $*$-algebra of background
independent gauge field theory. One thus obtains the kinematical
representation of loop quantum gravity via the construction of
algebraic quantum field theory. It is important to note that the
Ashtekar-Isham-Lewandowski state is the unique state on quantum
holonomy-flux $*$-algebra $\textbf{\textsf{A}}$ invariant under
internal gauge transformations and spatial
diffeomorphisms\footnote{The proof of this conclusion depends on the
compact support property of the smear functions $f^i$ (see
\cite{unique} for detail).}, which are both automorphisms $\alpha_g$
and $\alpha_\varphi$ on $\textbf{\textsf{A}}$ such that
$\omega\circ\alpha_g=\omega$ and $\omega\circ\alpha_\varphi=\omega$.
So these gauge transformations are represented as unitary
transformations on $\mathcal{H}_{kin}$, while the cyclic vector
$\Omega_\omega$ is the unique state in $\mathcal{H}_{kin}$ invariant
under internal gauge transformations and spatial diffeomorphisms.
This is a very crucial uniqueness theorem for canonical quantization
of gauge field theory
\cite{unique}.\\ \\
\textbf{Theorem 4.2.1}: \textit{There exists exactly one internal
gauge invariant and spatial diffeomorphism invariant state
$($positive linear functional$)$ on the quantum holonomy-flux
$*$-algebra. In other words, there exists a unique internal gauge
invariant and spatial diffeomorphism invariant cyclic representation
for the quantum holonomy-flux $*$-algebra, which is called
Ashtekar-Isham-Lewandowski representation. Moreover, this
representation is irreducible with respect to an exponential version
of the quantum holonomy-flux algebra (defined in
\cite{thiemann4}), which is analogous to the Weyl algebra.} \\
\\
Hence we have finished the construction of kinematical Hilbert
space for background independent gauge field theory and
represented the quantum holonomy-flux algebra on it. Then
following the general programme presented in the last subsection,
we should impose the constraints as operators on the kinematical
Hilbert space since we are dealing with a gauge system.

\section{Quantum Gaussian and Diffeomorphism Constraints}

After constructing the kinematical Hilbert space $\mathcal{H}_{kin}$
of loop quantum gravity, one should implement the constraints on it
to obtain physical Hilbert space. Recalling the constraints
(\ref{constraint}) in generalized Palatini Hamiltonian for general
relativity and the Poission algebra (\ref{constraint algebra}) among
them, the subalgebra generated by the Gaussian constraints
$\mathcal{G}(\Lambda)$ forms a Lie algebra and a 2-sided ideal in
the constraints algebra. So in this section, we first solve the
Gaussian constraints independently of the other two kinds of
constraints and find the solution space $\mathcal{H}^G$, which is
constituted by internal gauge invariant quantum states. Then,
although the subalgebra generated by the diffeomorphism constraints
is not an ideal in the constraints algebra, we still would like to
solve them independently of the scalar constraints for the technical
convenience. At the end of this section, we will obtain the Hilbert
space $\mathcal{H}^G_{Diff}$ free of Gaussian constraints and
diffeomorphism constraints.

\subsection{Implementation of Quantum Gaussian Constraint}

Recall that the classical expression of Gaussian constraints reads
\begin{eqnarray}
\mathcal{G}(\Lambda)=\int_{\Sigma}d^3x\Lambda^iD_a\widetilde{P}^a_i
=-\int_{\Sigma}d^3x\widetilde{P}^a_iD_a\Lambda^i\equiv-P(D\Lambda),\nonumber
\end{eqnarray}
where $D_a\Lambda^i=\partial_a\Lambda^i+\epsilon^i_{\
jk}A^j_a\Lambda^k$. As the situation of triad flux, the Gaussian
constraints can be defined as cylindrically consistent vector fields
$Y_{D\Lambda}$ on $\overline{\mathcal{A}}$, which act on any
cylindrical function $f_\gamma\in Cyl_\gamma$ by
\begin{eqnarray}
Y_{D\Lambda}\circ f_\gamma(\{A(e)\}_{e\in
E(\gamma)}):=\{-P(D\Lambda),f_\gamma(\{A(e)\}_{e\in
E(\gamma)})\}.\nonumber
\end{eqnarray}
Then the Gaussian constraint operator can be defined in analogy with
the momentum operator, which acts on $f_\gamma$ as:
\begin{eqnarray}
\hat{\mathcal{G}}(\Lambda)f_\gamma(\{A(e)\}_{e\in
E(\gamma)})&:=&i\hbar Y_{D\Lambda}\circ f_\gamma(\{A(e)\}_{e\in
E(\gamma)})\nonumber\\&=&\hbar\sum_{v\in
V(\gamma)}[\Lambda^i(v)\hat{J}^{v}_i]f(\{A(e)\}_{e\in
E(\gamma)}),\nonumber
\end{eqnarray}
which is the generator of internal gauge transformations on
$Cyl_\gamma$. The kernel of the operator is easily obtained in terms
of spin-network decomposition, which is the internal gauge invariant
Hilbert space:
\begin{eqnarray}
\mathcal{H}^G=\oplus_{\alpha,\mathbf{j}}\mathcal{H}'_{\alpha,\mathbf{j},\mathbf{l}=0}\oplus\mathbf{C}.\nonumber
\end{eqnarray}
One then naturally gets the gauge invariant spin-network basis
$T_s,\ s=(\gamma(s),\mathbf{j}_s,\mathbf{i}_s)$ in $\mathcal{H}^G$
as \cite{rovelli9}\cite{ALM}\cite{baez1}:
\begin{eqnarray}
T_{s=(\gamma,\mathbf{j},\mathbf{i})}=\bigotimes_{v\in
V(\gamma)}i_v\bigotimes_{e\in E(\gamma)}\pi^{j_e}(A(e)),\ \
(j_e\neq0)\nonumber
\end{eqnarray}
by assigning a non-trivial spin representation $j$ on each edge and
a invariant tensor $i$ (intertwiner) on each vertex. In the
following context $T_s$ will also be called as spin-network states.
We denote $Cyl(\overline{\mathcal{A/G}})$ the space of finite linear
combinations of gauge invariant spin-network states, which is dense
in $\mathcal{H}^G$, and $Cyl_\gamma(\overline{\mathcal{A/G}})\subset
Cyl(\overline{\mathcal{A/G}})$ the linear span of the gauge
invariant spin network functions $T_s$ for $\gamma(s)=\gamma$. All
Yang-Mills gauge invariant operators are well defined on
$\mathcal{H}^G$. However, the condition of acting on gauge invariant
states often changes the structure of the spectrum of quantum
geometric operators. For the area operator, the spectrum depends on
certain global properties of the surface $S$ (see \cite{AL} and
\cite{area} for details). For the volume operators, non-zero
spectrum arises from at least 4-valent vertices.

\subsection{Implementation of Quantum Diffeomorphism Constraint}

Unlike the strategy in solving Gaussian constraint, one cannot
define an operator for quantum diffeomorphism constraint as the
infinitesimal generator of finite diffeomorphism transformations
(unitary operators since the measure is diffeomorphism invariant)
represented on $\mathcal{H}_{kin}$. The representation of finite
diffeomorphisms is a family of unitary operators $\hat{U}_\varphi$
acting on cylindrical functions $\psi_\gamma$ by
\begin{eqnarray}
\hat{U}_\varphi \psi_\gamma:=\psi_{\varphi\circ\gamma},
\end{eqnarray}
for any spatial diffeomorphism $\varphi$ on $\Sigma$. An 1-parameter
subgroup $\varphi_t$ in the group of spatial diffeomorphisms is then
represented as an 1-parameter unitary group $\hat{U}_{\varphi_t}$ on
$\mathcal{H}_{kin}$. However, $\hat{U}_{\varphi_t}$ is not weakly
continuous, since the subspaces $\mathcal{H}'_{\alpha(\gamma)}$ and
$\mathcal{H}'_{\alpha(\varphi_t\circ\gamma)}$ are orthogonal to each
other no matter how small the parameter $t$ is. So one always has
\begin{eqnarray}
|<\psi_\gamma|\hat{U}_{\varphi_t}|\psi_\gamma>_{kin}-<\psi_\gamma|\psi_\gamma>_{kin}|=<\psi_\gamma|\psi_\gamma>_{kin}
\neq 0, \label{weak continuous}
\end{eqnarray}
even in the limit when $t$ goes to zero. Therefore, the
infinitesimal generator of $\hat{U}_{\varphi_t}$ does not exist. In
the strategy to solve the diffeomorphism constraint, due to the Lie
algebra structure of diffeomorphism constraints subalgebra, the
so-called group averaging technique is employed. We now outline the
procedure. First, given a colored graph (a graph $\gamma$ and a
cylindrical function living on it), one can define the group of
graph symmetries $GS_\gamma$ by
\begin{eqnarray}
GS_\gamma:=Diff_\gamma/TDiff_\gamma,\nonumber
\end{eqnarray}
where $Diff_\gamma$ is the group of all diffeomorphisms preserving
the colored $\gamma$, and $TDiff_\gamma$ is the group of
diffeomorphisms which trivially acts on $\gamma$. We define a
projection map by averaging with respect to $GS_\gamma$ to obtain
the subspace in $Cyl_\gamma$ which is invariant under the
transformation of $GS_\gamma$:
\begin{eqnarray}
\hat{P}_{Diff,\gamma}\psi_\gamma:=\frac{1}{n_\gamma}\sum_{\varphi\in
GS_\gamma}\hat{U}_{\varphi}\psi_\gamma,\nonumber
\end{eqnarray}
for all cylindrical functions
$\psi_\gamma\in\mathcal{H}'_{\alpha(\gamma)}$, where $n_\gamma$ is
the number of the finite elements of $GS_\gamma$. Second, we average
with respect to all remaining diffeomorphisms which move the graph
$\gamma$. For each cylindrical function $\psi_\gamma\in
Cyl_\gamma(\overline{\mathcal{A/G}})$, there is an element
$\eta(\psi_\gamma)$ associated to it in the algebraic dual space
$Cyl^\star$ of $Cyl(\overline{\mathcal{A/G}})$, which acts on any
cylindrical function $\phi_{\gamma'}\in
Cyl_\gamma(\overline{\mathcal{A/G}})$ as:
\begin{eqnarray}
\eta(\psi_\gamma)[\phi_{\gamma'}]:=\sum_{\varphi\in
Diff(\Sigma)/Diff_\gamma}<\hat{U}_{\varphi}\hat{P}_{Diff,\gamma}\psi_\gamma|\phi_{\gamma'}>_{kin}.\nonumber
\end{eqnarray}
It is well defined since, for any given graph $\gamma'$, only finite
terms are non-zero in the summation. It is easy to verify that
$\eta(\psi_\gamma)$ is invariant under the group action of
$Diff(\Sigma)$, since
\begin{eqnarray}
\eta(\psi_\gamma)[\hat{U}_{\varphi}\phi_{\gamma'}]=\eta(\psi_\gamma)[\phi_{\gamma'}].\nonumber
\end{eqnarray}
Thus we have defined a rigging map $\eta:\
Cyl(\overline{\mathcal{A/G}})\rightarrow Cyl^\star_{Diff}$, which
maps every cylindrical function to a diffeomorphism invariant one,
where $Cyl^\star_{Diff}$ is spanned by rigged spin-network functions
$T_{[s]}\equiv\{\eta(T_{s})\},\
{[s]=([\gamma],\mathbf{j},\mathbf{i})}$ associated with
diffeomorphism classes $[\gamma]$ of graphs $\gamma$. Moreover a
Hermitian inner product can be defined on $Cyl^\star_{Diff}$ by the
natural action of the algebraic functional:
\begin{eqnarray}
<\eta(\psi_\gamma)|\eta(\phi_{\gamma'})>_{Diff}:=\eta(\psi_\gamma)[\phi_{\gamma'}].\nonumber
\end{eqnarray}
The diffeomorphism invariant Hilbert space $\mathcal{H}_{Diff}$ is
defined by the completion of $Cyl^\star_{Diff}$ with respect to
the above inner product $<\ |\
>_{Diff}$. The vacuum state and diffeomorphism invariant spin-network functions
$T_{[s]}$ form an orthonormal basis in $\mathcal{H}_{Diff}$.
Finally, we have obtained the general solutions invariant under both
internal gauge transformations and spatial diffeomorphisms.

In general relativity, the problem of observables is a subtle
issue due to the diffeomorphism invariance
\cite{rovelli6}\cite{rovelli3}\cite{rovelli4}. Now we discuss the
operators as diffeomorphism invariant observables on
$\mathcal{H}_{Diff}$. We call an operator
$\hat{\mathcal{O}}\in\mathcal{L}(\mathcal{H}_{kin})$ a strong
observable if and only if
$\hat{U}^{-1}_{\varphi}\hat{\mathcal{O}}\hat{U}_{\varphi}=\hat{\mathcal{O}},
\ \forall \ \varphi\in Diff(\Sigma)$. We call it a weak observable
if and only if $\hat{\mathcal{O}}$ leaves $\mathcal{H}_{Diff}$
invariant. Then it is easy to see that a strong observable
$\hat{\mathcal{O}}$  must be a weak one. One notices that a strong
observable $\hat{\mathcal{O}}$ can first be defined on
$\mathcal{H}_{Diff}$ by its dual operator
$\hat{\mathcal{O}}^{\star}$ as
\begin{eqnarray}
(\hat{\mathcal{O}}^{\star}\Phi_{Diff})[\psi]:=\Phi_{Diff}[\hat{\mathcal{O}}\psi],\nonumber
\end{eqnarray}
then one gets
\begin{eqnarray}
(\hat{\mathcal{O}}^{\star}\Phi_{Diff})[\hat{U}_{\varphi}\psi]=\Phi_{Diff}[\hat{\mathcal{O}}\hat{U}_{\varphi}\psi]
=\Phi_{Diff}[\hat{U}^{-1}_{\varphi}\hat{\mathcal{O}}\hat{U}_{\varphi}\psi]=(\hat{\mathcal{O}}^{\star}\Phi_{Diff})
[\psi],\nonumber
\end{eqnarray}
for any $\Phi_{Diff}\in\mathcal{H}_{Diff}$ and
$\psi\in\mathcal{H}_{kin}$. Hence
$\hat{\mathcal{O}}^{\star}\Phi_{Diff}$ is also diffeomorphism
invariant. In addition, a strong observable also has the property of
$\hat{\mathcal{O}}^{\star}\eta(\psi_\gamma)=\eta(\hat{\mathcal{O}}^\dagger\psi_\gamma)$
since, $\forall \ \phi_{\gamma'}, \psi_\gamma\in\mathcal{H}_{kin}$,
\begin{eqnarray}
&&<\hat{\mathcal{O}}^{\star}\eta(\psi_\gamma)|\eta(\phi_{\gamma'})>_{Diff}
=(\hat{\mathcal{O}}^{\star}\eta(\psi_\gamma))[\phi_{\gamma'}]=\eta(\psi_\gamma)[\hat{\mathcal{O}}\phi_{\gamma'}]\nonumber\\
&=&\sum_{\varphi\in
Diff(\Sigma)/Diff_\gamma}<\hat{U}_{\varphi}\hat{P}_{Diff,\gamma}\psi_\gamma|\hat{\mathcal{O}}
\phi_{\gamma'}>_{kin}\nonumber\\
&=&\frac{1}{n_\gamma}\sum_{\varphi\in
Diff(\Sigma)/Diff_\gamma}\sum_{\varphi'\in
GS_\gamma}<\hat{U}_{\varphi}
\hat{U}_{\varphi'}\psi_\gamma|\hat{\mathcal{O}}\phi_{\gamma'}>_{kin}\nonumber\\
&=&\frac{1}{n_\gamma}\sum_{\varphi\in
Diff(\Sigma)/Diff_\gamma}\sum_{\varphi'\in
GS_\gamma}<\hat{U}_{\varphi}
\hat{U}_{\varphi'}\hat{\mathcal{O}}^\dagger\psi_\gamma|\phi_{\gamma'}>_{kin}\nonumber\\
&=&<\eta(\hat{\mathcal{O}}^{\dagger}\psi_\gamma)|\eta(\phi_{\gamma'})>_{Diff}.\nonumber
\end{eqnarray}
Note that the Hilbert space $\mathcal{H}_{Diff}$ is still
non-separable if one considers the $C^n$ diffeomorphisms with $n>0$.
However, if one extends the diffeomorphisms to be semi-analytic
diffeomotphisms, i.e. homomorphisms that are analytic
diffeomorphisms up to finite isolate points (which can be viewed as
an extension of the classical concept to the quantum case), the
Hilbert space $\mathcal{H}_{Diff}$ would be separable
\cite{FR}\cite{AL}.

\section{Quantum Dynamics}

In this section, we consider the quantum dynamics of loop quantum
gravity. One may first consider to construct a Hamiltonian
constraint (scalar constraint) operator in $\mathcal{H}_{kin}$ or
$\mathcal{H}_{Diff}$, then attempt to find the physical Hilbert
space $\mathcal{H}_{phys}$ by solving the quantum Hamiltonian
constraint. However, difficulties arise here due to the special role
played by the scalar constraints in the constraint algebra
(\ref{constraint algebra}).  Firstly, the scalar constraints do not
form a Lie subalgebra. Hence the strategy of group average cannot be
used directly on $\mathcal{H}_{kin}$ for them. Secondly, modulo the
Gaussian constraint, there is still a structure function in the
Poisson bracket between two scalar constraints:
\begin{eqnarray}
\{\mathcal{S}(N),\mathcal{S}(M)\}=-\mathcal{V}((N\partial_bM-M\partial_bN)q^{ab}),\label{s-algebra}
\end{eqnarray}
which raises the danger of quantum anomaly in quantization.
Moreover, the diffeomorphism constraints do not form an ideal in
the quotient constraint algebra modulo the Gaussian constraints.
This fact results in that the scalar constraint operator cannot be
well defined on $\mathcal{H}_{Diff}$, as it does not commute with
the diffeomorphism transformations $\hat{U}_\varphi$. Thus the
previous construction of $\mathcal{H}_{Diff}$ seems not much
meaningful for the final construction of $\mathcal{H}_{phys}$, for
which we are seeking. However, one may still first try to
construct a Hamiltonian constraint operator in
$\mathcal{H}_{kin}$.

\subsection{Hamiltonian Constraint Operator}

The aim in this subsection is to define a quantum operator
corresponding to the Hamiltonian constraint. Its classical
expression reads:
\begin{eqnarray}
\mathcal{S}(N)&:=&\frac{\kappa\beta^2}{2}\int_\Sigma
d^3xN\frac{\widetilde{P}^a_i\widetilde{P}^b_j}{\sqrt{|\det
q|}}[\epsilon^{ij}_{\ \ k}F^k_{ab}-2(1+\beta^2)K^i_{[a}K^j_{b]}]\nonumber\\
&=&\mathcal{S}_{E}(N)-2(1+\beta^2)\mathcal{T}(N).\label{classicalH}
\end{eqnarray}
The main idea of the construction is to first express
$\mathcal{S}(N)$ in terms of the combination of Poisson brackets
between the variables which have been represented as operators on
$\mathcal{H}_{kin}$, then replace the Poisson brackets by canonical
commutators between the operators. We will use the volume functional
for a region $R\subset\Sigma$ and the extrinsic curvature functional
defined by:
\begin{eqnarray}
K&:=&\kappa\beta\int_{\Sigma}d^3x\widetilde{P}^a_iK^i_a.\nonumber
\end{eqnarray}
A key trick here is to consider the following classical identity
of the co-triad $e^i_a(x)$ \cite{thiemann1}:
\begin{eqnarray}
e^i_a(x)=\frac{(\kappa\beta)^2}{2}\underline{\eta}_{abc}\epsilon^{ijk}\frac{\widetilde{P}^b_j\widetilde{P}^c_k}
{\sqrt{\det q}}(x)=\frac{2}{\kappa\beta}\{A^i_a(x),V_R\},\nonumber
\end{eqnarray}
where $x\in R$, and the expression of the extrinsic curvature
1-form $K^i_a(x)$:
\begin{eqnarray}
K^i_a(x)=\frac{1}{\kappa\beta}\{A^i_a(x),K\}.\nonumber
\end{eqnarray}
Note that $K$ can be expressed by a Poisson bracket as
\begin{eqnarray}
K=\beta^{-2}\{\mathcal{S}_E(1),V_\Sigma\}.
\end{eqnarray}
Thus one can obtain the equivalent classical expressions of
$\mathcal{S}_{E}(N)$ and $\mathcal{T}(N)$ as:
\begin{eqnarray}
\mathcal{S}_{E}(N)&=&\frac{\kappa\beta^2}{2}\int_\Sigma
d^3xN\frac{\widetilde{P}^a_i\widetilde{P}^b_j}{\sqrt{|\det
q|}}\epsilon^{ij}_{\ \ k}F^k_{ab} \nonumber\\
&=&-\frac{2}{\kappa^2\beta}\int_\Sigma
d^3xN(x)\widetilde{\eta}^{abc}\mathrm{Tr}\big(\mathbf{F}_{ab}(x)\{\mathbf{A}_c(x),V_{R_x}\}\big),\nonumber\\
\mathcal{T}(N)&=&\frac{\kappa\beta^2}{2}\int_\Sigma
d^3xN\frac{\widetilde{P}^a_i\widetilde{P}^b_j}{\sqrt{|\det
q|}}K^i_{[a}K^j_{b]}\nonumber\\
&=&-\frac{2}{\kappa^4\beta^3}\int_\Sigma
d^3xN(x)\widetilde{\eta}^{abc}\mathrm{Tr}\big(\{\mathbf{A}_a(x),K\}\{\mathbf{A}_b(x),K\}
\{\mathbf{A}_c(x),V_{R_x}\}\big),\nonumber
\end{eqnarray}
where $\mathbf{A}_a=A_a^i\tau_i$, $\mathbf{F}_{ab}=F_{ab}^i\tau_i$,
$\mathrm{Tr}$ represents the trace of the Lie algebra matrix, and
$R_x\subset\Sigma$ denotes an arbitrary neighborhood of
$x\in\Sigma$. In order to quantize the Hamiltonian constraint as a
well-defined operator on $\mathcal{H}_{kin}$, one has to express the
classical formula of $\mathcal{S}(N)$ in terms of holonomies $A(e)$
and other variables with clear quantum analogs. There are genuine
ambiguities in this regularization procedure, which will be
summarized at the end of this subset. However, the nontrivial fact
is that there do exist well-defined strategies. We now introduce the
original strategy proposed first by Thiemann\cite{thiemann1}. Given
a triangulation $T(\epsilon)$ of $\Sigma$, where the parameter
$\epsilon$ describes how fine the triangulation is, and the
triangulation will fill out the spatial manifold $\Sigma$ when
$\epsilon\rightarrow0$. For each tetrahedron $\Delta\in
T(\epsilon)$, we use $\{s_i(\Delta)\}_{i=1,2,3}$ to denote the three
outgoing oriented segments in $\Delta$ with a common beginning point
$v(\Delta)=s(s_i(\Delta))$, and use $a_{ij}(\Delta)$ to denote the
arc connecting the end points of $s_i(\Delta)$ and $s_j(\Delta)$.
Then several loops $\alpha_{ij}(\Delta)$ are formed by
$\alpha_{ij}(\Delta):=s_i(\Delta)\circ a_{ij}(\Delta)\circ
s_j(\Delta)^{-1}$. Thus one has the identities:
\begin{eqnarray}
\{\int_{s_i(\Delta)}\mathbf{A}_a\dot{s}^a_i(\Delta),V_{R_{v(\Delta)}}\}&=&-A(s_i(\Delta))^{-1}
\{A(s_i(\Delta)),V_{R_{v(\Delta)}}\}+o(\epsilon),\label{connection1}\\
\{\int_{s_i(\Delta)}\mathbf{A}_a\dot{s}^a_i(\Delta),K\}&=&-A(s_i(\Delta))^{-1}\{A(s_i(\Delta)),K\}
+o(\epsilon),\label{connection2}\\
\int_{P_{ij}}\mathbf{F}_{ab}(x)&=&\frac{1}{2}A(\alpha_{ij}(\Delta))^{-1}
-\frac{1}{2}A(\alpha_{ij}(\Delta))+o(\epsilon^2),\label{curvature}
\end{eqnarray}
where $P_{ij}$ is the plane with boundary $\alpha_{ij}$. Note that
the above identities are constructed by taking account of internal
gauge invariance of the final formula of Hamiltonian constraint
operator. So we have the regularized expression of $\mathcal{S}(N)$
by the Riemannian sum \cite{thiemann1}:
\begin{eqnarray}
\mathcal{S}^\epsilon_{E}(N)&=&\frac{2}{3\kappa^2\beta}\sum_{\Delta\in
T(\epsilon)}N(v(\Delta))\epsilon^{ijk}
\times\nonumber\\
&&\mathrm{Tr}\big(A(\alpha_{ij}(\Delta))^{-1}A(s_k(\Delta))^{-1}\{A(s_k(\Delta)),V_{R_{v(\Delta)}}\}\big),\nonumber\\
\mathcal{T}^\epsilon(N)&=&\frac{\sqrt{2}}{6\kappa^4\beta^3}\sum_{\Delta\in
T(\epsilon)}N(v(\Delta))\epsilon^{ijk}\times\nonumber\\
&&\mathrm{Tr}\big(A(s_i(\Delta))^{-1}\{A(s_i(\Delta)),K\}A(s_j(\Delta))^{-1}\{A(s_j(\Delta)),K\}\times\nonumber\\
&&A(s_k(\Delta))^{-1}\{A(s_k(\Delta)),V_{R_{v(\Delta)}}\}\big),\nonumber\\
\mathcal{S}^\epsilon(N)&=&\mathcal{S}^\epsilon_{E}(N)-2(1+\beta^2)\mathcal{T}^\epsilon(N),\label{hamilton}
\end{eqnarray}
such that
$\lim_{\epsilon\rightarrow0}\mathcal{S}^\epsilon(N)=\mathcal{S}(N)$.
It is clear that the above regulated formula of $\mathcal{S}(N)$ is
invariant under internal gauge transformations. Since all
constituents in the expression have clear quantum analogs, one can
quantize the regulated Hamiltonian constraint as an operator on
$\mathcal{H}_{kin}$ (or $\mathcal{H}^G$) by replacing them by the
corresponding operators and Poisson brackets by canonical
commutators, i.e.,
\begin{eqnarray}
&&A(e)\mapsto\hat{A}(e),\ \ \ V_R\mapsto\hat{V}_R,\ \ \ \{\ ,\
\}\mapsto\frac{[\ ,\ ]}{i\hbar},\nonumber\\
&&\mathrm{and}\ \ \ \ \ \
K\mapsto\hat{K}^\epsilon=\frac{\beta^{-2}}{i\hbar}[\hat{\mathcal{S}}^\epsilon_E(1),\hat{V}_\Sigma].\nonumber
\end{eqnarray}
Removing the regulator by $\epsilon\rightarrow0$, it turns out
that one can obtain a well-defined limit operator on
$\mathcal{H}_{kin}$ (or $\mathcal{H}^G$) with respect to a natural
operator topology.

Now we begin to construct the Hamiltonian constraint operator in
analogy with the classical expression (\ref{classicalH}). All we
should do is to define the corresponding regulated operators on
different $\mathcal{H}'_\alpha$ separately, then remove the
regulator $\epsilon$ so that the limit operator is defined on
$\mathcal{H}_{kin}$ (or $\mathcal{H}^G$) cylindrically consistently.
In the following, given a vertex and three edges intersecting at the
vertex in a graph $\gamma$ of $\psi_\gamma\in
Cyl_\gamma(\overline{\mathcal{A/G}})$, we construct one
triangulation of the neighborhood of the vertex adapted to the three
edges. Then we average with respect to the triples of edges meeting
at the given vertex. Precisely speaking, one can make the
triangulations $T(\epsilon)$ with the following properties.
\begin{itemize}
\item The chosen triple of edges at each vertex in the graph $\gamma$ is embedded in a $T(\epsilon)$ for all
$\epsilon$, so that the vertex $v$ of $\gamma$ where the three edges
meet coincides with a vertex $v(\Delta)$ in $T(\epsilon)$.

\item For every triple of segments ($e_1,\ e_2,\ e_3$) of $\gamma$
such that $v=s(e_1)=s(e_2)=s(e_3)$, there is a tetrahedron
$\Delta\in T(\epsilon)$ such that $v=v(\Delta)=s(s_i(\Delta))$, and
$s_i(\Delta)\subset e_i,\ \forall\ i=1,2,3$. We denote such a
tetrahedron as $\Delta^0_{e_1,e_2,e_3}$.

\item For each tetrahedra $\Delta^0_{e_1,e_2,e_3}$ one can
construct seven additional tetrahedra $\Delta^\wp_{e_1,e_2,e_3},\
\wp=1,...,7$, by backward analytic extensions of $s_i(\Delta)$ so
that $U_{e_1,e_2,e_3}:=\cup_{\wp=0}^7\Delta^\wp_{e_1,e_2,e_3}$ is a
neighborhood of $v$.

\item The triangulation must be fine enough so that the
neighborhoods $U(v):=\cup_{e_1,e_2,e_3}U_{e_1,e_2,e_3}(v)$ are
disjoint for different vertices $v$ and $v'$ of $\gamma$. Thus for
any open neighborhood $U_\gamma$ of the graph $\gamma$, there exists
a triangulation $T(\epsilon)$ such that $\cup_{v\in
V(\gamma)}U(v)\subseteq U_\gamma$.

\item The distance between a vertex $v(\Delta)$ and the
corresponding arcs $a_{ij}(\Delta)$ is described by the parameter
$\epsilon$. For any two different $\epsilon$ and $\epsilon'$, the
arcs $a_{ij}(\Delta^\epsilon)$ and $a_{ij}(\Delta^{\epsilon'})$ with
respect to one vertex $v(\Delta)$ are semi-analytically
diffeomorphic with each other.

\item Taking account of all possible triangulations $T(\epsilon)$ given by different choices of the triples of edges
at each vertex in
$\gamma$, the integral over $\Sigma$ is replaced by the Riemanian
sum:
\begin{eqnarray}
\int_\Sigma\ \ \ &=&\int_{U_\gamma}\ \ \ \ +\int_{\Sigma-U_\gamma}\ ,\ \ \ \nonumber\\
\int_{U_\gamma}\ \ \ &=&\sum_{v\in V(\gamma)}\int_{U(v)}\ \ \ \ +\int_{U_\gamma-\cup_{v}U(v)}\ ,\ \ \ \nonumber\\
\int_{U(v)}\ \ \
&=&\frac{1}{E(v)}\sum_{e_1,e_2,e_3}[\int_{U_{e_1,e_2,e_3}(v)}\ \ \ \
+\int_{U(v)-U_{e_1,e_2,e_3},(v)}\ ],\nonumber
\end{eqnarray}
where $n(v)$ is the valence of the vertex $v=s(e_1)=s(e_2)=s(e_3)$,
and $E(v)\equiv \left(
\begin{array}{ll}
n(v)\\
3
\end{array} \right)$ denotes the binomial coefficient which comes from
the averaging with respect to the triples of edges meeting at given
vertex $v$. One then observes that
\begin{eqnarray}
\int_{U_{e_1,e_2,e_3}(v)}\ \ \ \
=8\int_{\Delta^0_{e_1,e_2,e_3}(v)}\ \ \,\nonumber
\end{eqnarray}
in the limit $\epsilon\rightarrow0$.

\item The triangulations for the regions
\begin{eqnarray}
&&U(v)-U_{e_1,e_2,e_3}(v),\nonumber\\
&&U_\gamma-\cup_{v\in V(\gamma)}U(v),\nonumber\\
&&\Sigma-U_\gamma, \label{*}
\end{eqnarray}
are arbitrary. These regions do not contribute to the construction
of the operator, since the commutator term
$[A(s_i(\Delta)),V_{R_{v(\Delta)}}]\psi_\gamma$ vanishes for all
tetrahedron $\Delta$ in the regions (\ref{*}).
\end{itemize}
Thus we find the regulated expression of Hamiltonian constraint
operator with respect to the triangulations $T(\epsilon)$ as
\cite{thiemann1}
\begin{eqnarray}
\hat{\mathcal{S}}^\epsilon_{E,\gamma}(N)&=&\frac{16}{3i\hbar\kappa^2\beta}\sum_{v\in
V(\gamma)}
\frac{N(v)}{{E}(v)}\sum_{v(\Delta)=v}\epsilon^{ijk}\times\nonumber\\
&&\mathrm{Tr}\big(\hat{A}(\alpha_{ij}(\Delta))^{-1}\hat{A}(s_k(\Delta))^{-1}[\hat{A}(s_k(\Delta)),
\hat{V}_{U^\epsilon_{v}}]\big),\nonumber\\
\hat{\mathcal{T}}^\epsilon_\gamma(N)&=&-\frac{4\sqrt{2}}{3i\hbar^3\kappa^4\beta^3}\sum_{v\in
V(\gamma)}\frac{N(v)}{E(v)}
\sum_{v(\Delta)=v}\epsilon^{ijk}\times\nonumber\\
&&\mathrm{Tr}\big(\hat{A}(s_i(\Delta))^{-1}[\hat{A}(s_i(\Delta)),\hat{K}^\epsilon]\hat{A}(s_j(\Delta))^{-1}
[\hat{A}(s_j(\Delta)),\hat{K}^\epsilon]\times\nonumber\\
&&\hat{A}(s_k(\Delta))^{-1}[\hat{A}(s_k(\Delta)),\hat{V}_{U^\epsilon_{v}}]\big),
\nonumber\\
\hat{\mathcal{S}}^\epsilon(N)\psi_\gamma&=&[\hat{\mathcal{S}}^\epsilon_{E,\gamma}(N)-2(1+\beta^2)
\hat{\mathcal{T}}_{\gamma}^\epsilon(N)]\psi_\gamma =\sum_{v\in
V(\gamma)}N(v)\hat{\mathcal{S}}^\epsilon_v\psi_\gamma,\nonumber
\end{eqnarray}
for any cylindrical function $\psi_\gamma\in
Cyl_\gamma(\overline{\mathcal{A/G}})$. Note that, by construction,
the operation of $\hat{\mathcal{S}}^\epsilon(N)$ on $\psi_\gamma\in
Cyl_\gamma(\overline{\mathcal{A/G}})$ is reduced to a finite
combination of that of $\hat{\mathcal{S}}^\epsilon_v$ with respect
to different vertices of $\gamma$. Hence, for each $\epsilon>0$,
$\hat{\mathcal{S}}^\epsilon(N)$ is a well-defined Yang-Mills gauge
invariant and diffeomorphism covariant operator on
$Cyl(\overline{\mathcal{A/G}})$. The family of regulated Hamiltonian
constraint operators with respect to the ordered family of graphs
are cylindrically consistent up to diffeomorphisms \cite{thiemann1}.

The last step is to remove the regulator by taking the limit
$\epsilon\rightarrow0$. However, the action of a regulated
Hamiltonian constraint operator on $\psi_\gamma$ adds arcs
$a_{ij}(\Delta)$ with a $\frac{1}{2}$-representation with respect
to each $v(\Delta)$ of $\gamma$\footnote{The Hamiltonian
constraint operator depends indeed on the choice of the
representation $j$ on the arcs $a_{ij}(\Delta)$, which is known as
one of the regularization ambiguities in the construction of
quantum dynamics. For the simplicity of the theory, one often
chooses the lowest label of representation $j=\frac{1}{2}$.},
i.e., the action of the operators family
$\hat{\mathcal{S}}^\epsilon(N)$ on cylindrical functions is
graph-changing. Thus $\hat{\mathcal{S}}^\epsilon(N)$ does not
converge with respect to the weak operator topology in
$\mathcal{H}_{kin}$ when $\epsilon\rightarrow0$, since different
$\mathcal{H}'_{\alpha(\gamma)}$ with different graphs $\gamma$ are
mutually orthogonal. Thus one has to define a weaker operator
topology to make the operator limit meaningful. By physical
motivation and the naturally available Hilbert space
$\mathcal{H}_{Diff}$, the convergence of
$\hat{\mathcal{S}}^\epsilon(N)$ holds with respect to the
so-called Uniform Rovelli-Smolin Topology \cite{urst}, where one
defines $\hat{\mathcal{S}}^\epsilon(N)$ to converge if and only if
$\Psi_{Diff}[\hat{\mathcal{S}}^\epsilon(N)\phi]$ converge for all
$\Psi_{Diff}\in Cyl^\star_{Diff}$ and $\phi\in
Cyl(\overline{\mathcal{A/G}})$. Since the value of
$\Psi_{Diff}[\hat{\mathcal{S}}^\epsilon(N)\phi]$ is actually
independent of $\epsilon$ by the fifth property of the
triangulations, the sequence converges to a nontrivial result
$\Psi_{Diff}[\hat{\mathcal{S}}^{\epsilon_0}(N)\phi]$ with
arbitrary fixed $\epsilon_0>0$. Thus one has defined a
diffeomorphism covariant, closed but non-symmetric operator,
$\hat{\mathcal{S}}(N)=\lim_{\epsilon\rightarrow0}\hat{\mathcal{S}}^\epsilon(N)=\hat{\mathcal{S}}^{\epsilon_0}(N)$,
on $\mathcal{H}_{kin}$ (or $\mathcal{H}^G$) representing the
Hamiltonian constraint. Moreover, a dual Hamiltonian constraint
operator $\hat{\mathcal{S}}'^\epsilon(N)$ is also defined on
$Cyl^\star$ depending on a specific value of $\epsilon$
\begin{eqnarray}
(\hat{\mathcal{S}}'^\epsilon(N)\Psi)[\phi]:=\Psi[\hat{\mathcal{S}}^\epsilon(N)\phi],\nonumber
\end{eqnarray}
for all $\Psi\in Cyl^\star$ and $\phi\in
Cyl(\overline{\mathcal{A/G}})$. For $\Psi_{Diff}\in
Cyl^\star_{Diff}\subset Cyl^\star$, one gets
\begin{eqnarray}
(\hat{\mathcal{S}}'(N)\Psi_{Diff})[\phi]=\Psi_{Diff}[\hat{\mathcal{S}}^\epsilon(N)\phi].\nonumber
\end{eqnarray}
which is independent of the value of $\epsilon$. Several remarks on
the Hamiltonian constraint operator are listed as follows.
\begin{itemize}
\item \textit{Finiteness of $\hat{\mathcal{S}}(N)$ on
$\mathcal{H}_{kin}$}

In ordinary quantum field theory, the continuous quantum field is
only recovered when one lets lattices spacing to approach zero,
i.e., takes the continuous cut-off parameter to its continuous
limit. However, this will produce the well-known infinity in quantum
field theory and make the Hamiltonian operator ill-defined on the
Fock space. So it seems surprising that our operator
$\hat{\mathcal{S}}(N)$ is still well defined, when one takes the
limit $\epsilon\rightarrow0$ with respect to the Uniform
Rovelli-Smolin Topology so that the triangulation goes to the
continuum. The reason behind it is that the cut-off parameter is
essentially noneffective due to the diffeomorphism invariance of our
quantum field theory. This is why there is no UV divergence in the
background independent quantum gauge field theory with
diffeomorphism invariance. On the other hand, from a convenient
viewpoint, one may think the Hamiltonian constraint operator as an
operator dually defined on a dense domain in $\mathcal{H}_{Diff}$.
However, we will see that the dual Hamiltonian constraint operator
cannot leave $\mathcal{H}_{Diff}$ invariant.

\item \textit{Implementation of Dual Quantum Constraint Algebra}

One important task is to check whether the commutator algebra
(quantum constraint algebra) among the corresponding quantum
operators of constraints both physically and mathematically
coincides with the classical constraint algebra by substituting
quantum constraint operators to classical constraint functionals and
commutators to Poisson brackets. Here the quantum anomaly has to be
avoided in the construction of constraint operators (see the
discussion for Eq.(\ref{anomaly})). First, the subalgebra of the
quantum diffeomorphism constraint algebra is free of anomaly by
construction:
\begin{eqnarray}
\hat{U}_\varphi \hat{U}_{\varphi'}\hat{U}^{-1}_\varphi
\hat{U}^{-1}_{\varphi'}=\hat{U}_{\varphi\circ\varphi'\circ\varphi^{-1}\circ\varphi'^{-1}},\nonumber
\end{eqnarray}
which coincides with the exponentiated version of the Poisson
bracket between two diffeomorphism constraints generating the
transformations $\varphi,\varphi'\in Diff(\Sigma)$. Secondly, the
quantum constraint algebra between the dual Hamiltonian constraint
operator $\mathcal{S}'(N)$ and the finite diffeomorphism
transformation $\hat{U}_\varphi$ on diffeomorphism-invariant
states coincides with the classical Poisson algebra between
$\mathcal{V}(\vec{N})$ and $\mathcal{S}(M)$. Given a cylindrical
function $\phi_\gamma$ associated with a graph $\gamma$ and the
triangulations $T(\epsilon)$ adapted to the graph $\alpha$, the
triangulations $T(\varphi\circ\epsilon)\equiv \varphi\circ
T(\epsilon)$ are compatible with the graph $\varphi\circ\gamma$.
Then we have by definition:
\begin{eqnarray}
&&\big(-([\hat{\mathcal{S}}(N),\hat{U}_\varphi])'\Psi_{Diff}\big)[\phi_\gamma]\nonumber\\
&=&([\hat{\mathcal{S}}'(N),\hat{U}'_\varphi]\Psi_{Diff})[\phi_\gamma]\nonumber\\
&=&\Psi_{Diff}[\hat{\mathcal{S}}^\epsilon(N)\phi_\gamma-\hat{\mathcal{S}}^\epsilon(N)
\phi_{\varphi\circ\gamma}]\nonumber \\
&=&\sum_{v\in
V(\gamma)}\{N(v)\Psi_{Diff}[\hat{\mathcal{S}}^\epsilon_v\phi_\gamma]-N(\varphi\circ
v)
\Psi_{Diff}[\hat{\mathcal{S}}^{\varphi\circ\epsilon}_{\varphi\circ v}\phi_{\varphi\circ\gamma}]\}\nonumber\\
&=&\sum_{v\in V(\gamma)}[N(v)-N(\varphi\circ v)]\Psi_{Diff}[\hat{\mathcal{S}}^\epsilon_v\phi_\gamma]\nonumber\\
&=&\big(\hat{\mathcal{S}}'(N-\varphi^*N)\Psi_{Diff}\big)[\phi_\gamma].\label{anomalyfree}
\end{eqnarray}
Thus there is no anomaly. However, Eq.(\ref{anomalyfree}) also
explains why the Hamiltonian constraint operator
$\hat{\mathcal{S}}(N)$ cannot leave $\mathcal{H}_{Diff}$ invariant.

Third, we compute the commutator between two Hamiltonian constraint
operators. Notice that
\begin{eqnarray}
&&[\hat{\mathcal{S}}(N),\hat{\mathcal{S}}(M)]\phi_\gamma\nonumber\\
&=&\sum_{v\in
V(\gamma)}[M(v)\hat{\mathcal{S}}(N)-N(v)\hat{\mathcal{S}}(M)]\hat{\mathcal{S}}^{\epsilon}_{v}\phi_\gamma\nonumber\\
&=&\sum_{v\in V(\gamma)}\sum_{v'\in
V(\gamma')}[M(v)N(v')-N(v)M(v')]\hat{\mathcal{S}}^{\epsilon'}_{v'}\hat{\mathcal{S}}^{\epsilon}_{v}\phi_\gamma,\nonumber
\end{eqnarray}
where $\gamma'$ is the graph changed from $\gamma$ by the action of
$\hat{\mathcal{S}}(N)$ or $\hat{\mathcal{S}}(M)$, which adds the
arcs $a_{ij}(\Delta)$ on $\gamma$, $T(\epsilon)$ is the
triangulation adapted to $\gamma$ and $T(\epsilon')$ adapted to
$\gamma'$. Since the newly added vertices by
$\hat{\mathcal{S}}^{\epsilon}_{v}$ is planar, they will never
contributes the final result. So one has
\begin{eqnarray}
&&[\hat{\mathcal{S}}(N),\hat{\mathcal{S}}(M)]\phi_\gamma\nonumber\\
&=&\sum_{v, v'\in V(\gamma), v\neq
v'}[M(v)N(v')-N(v)M(v')]\hat{\mathcal{S}}^{\epsilon'}_{v'}
\hat{\mathcal{S}}^{\epsilon}_{v}\phi_\gamma\nonumber\\
&=&\frac{1}{2}\sum_{v, v'\in V(\gamma), v\neq
v'}[M(v)N(v')-N(v)M(v')][\hat{\mathcal{S}}^{\epsilon'}_{v'}\hat{\mathcal{S}}^{\epsilon}_{v}-
\hat{\mathcal{S}}^{\epsilon'}_{v}\hat{\mathcal{S}}^{\epsilon}_{v'}]\phi_\gamma\nonumber\\
&=&\frac{1}{2}\sum_{v, v'\in V(\gamma), v\neq
v'}[M(v)N(v')-N(v)M(v')][(\hat{U}_{\varphi_{v',v}}-\hat{U}_{\varphi_{v,v'}})
\hat{\mathcal{S}}^{\epsilon}_{v'}\hat{\mathcal{S}}^{\epsilon}_{v}]\phi_\gamma,\nonumber\\
\label{hamiltonian constraint}
\end{eqnarray}
where we have used the facts that
$[\hat{\mathcal{S}}^{\epsilon}_{v},\hat{\mathcal{S}}^{\epsilon'}_{v'}]=0$
for $v\neq v'$and there exists a diffeomorphism $\varphi_{v,v'}$
such that
$\hat{\mathcal{S}}^{\epsilon'}_{v'}\hat{\mathcal{S}}^{\epsilon}_{v}=\hat{U}_{\varphi_{v',v}}
\hat{\mathcal{S}}^{\epsilon}_{v'}\hat{\mathcal{S}}^{\epsilon}_{v}$.
Obviously, we have in the Uniform Rovelli-Smolin Topology
\begin{eqnarray}
([\hat{\mathcal{S}}(N),\hat{\mathcal{S}}(M)])'\Psi_{Diff}=0\nonumber
\end{eqnarray}
for all $\Psi_{Diff}\in Cyl^\star_{Diff}$. As we have seen in
classical expression Eq.(\ref{s-algebra}), the Poisson bracket of
any two Hamiltonian constraints is given by a generator of the
diffeomrophism transformations. Therefore it is mathematically
consistent with the classical expression that two Hamiltonian
constraint operators commute on diffeomorphism states. On the other
hand, it has been shown in Refs. \cite{GL} and \cite{LM} that the
domain of dual Hamiltonian constraint operator can be extended to a
slightly larger space (habitat) in $Cyl^\star$, whose elements are
not necessary diffeomorphism invariant. While, it turns out that the
commutator between two Hamiltonian constraint operators continues to
vanish on the habitat, which seems to be problematic. Fortunately,
the quantum operator corresponding to the right hand side of
classical Poisson bracket (\ref{s-algebra}) also annihilates every
state in the habitat \cite{GL}, so the quantum constraint algebra is
consistent at this level. But it is not clear whether the quantum
constraint algebra, especially the commutator between two
Hamiltonian constraint is consistent with the classical one on some
larger space in $Cyl^\star$ containing more diffeomorphism variant
states. So further work on the semi-classical analysis is needed to
test the classical limit of Eq.(\ref{hamiltonian constraint}). The
way to do it is to look for some suitable semi-classical states for
calculating the classical limit of the operators. However, due to
the graph-changing property of the Hamiltonian constraint operator,
the semi-classical analysis of the Hamiltonian constraint operator
and the quantum constraint algebra is still an open issue so far.

\item \textit{General Regularization Scheme of Hamiltonian Constraint}

In Ref.\cite{AL}, a general scheme of regulation is introduced for
the quantization of Hamiltonian constraint, which includes
Thiemann's regularization as a specific choice. Such a general
regularization can be summarized as follows:

First, we assigns a partition of $\Sigma$ into cells $\Box$ of
arbitrary shape. In every cell of the partition we define edges
$s_J$, $J=1,...,n_s$ and loops $\beta_i$, $i=1,...,n_\beta$, where
$n_s$, $n_\beta$ may be different for different cells. One uses
$\epsilon$ to represent the scale of the cell $\Box$. Then one fixes
an arbitrary chosen representation $\rho$ of $SU(2)$ for the
calculation of the holonomies in Eqs.
(\ref{connection1})-(\ref{curvature}). This structure is called a
\textit{permissible classical regulator} if the regulated
Hamiltonian constraint expression with respect to this partition has
correct limit when $\epsilon\rightarrow0$.

Secondly, one assigns the diffeomorphism covariant property to the
regulator and lets the partition adapted to the choice of the graph.
Given a cylindrical function $\psi_\gamma\in
Cyl^3_\gamma(\overline{\mathcal{A/G}})$, the partition is
sufficiently refined so that every vertex $v\in V(\gamma)$ is
contained in exact one cell of the partition. If $(\gamma,\ v)$ is
diffeomorphic to $(\gamma',\ v')$ then, for every $\epsilon$ and
$\epsilon'$, the quintuple $(\gamma,\ v,\ \Box,\ (s_J),\ (\beta_i))$
is diffeomorphic to the quintuple $(\gamma',\ v',\ \Box',\ (s'_J),\
(\beta'_i))$, where $\Box$ and $\Box'$ are the cells in the
partitions with respect to $\gamma$ and $\gamma'$ respectively,
containing $v$ and $v'$ respectively.

As a result, the Hamiltonian constraint operator in this general
regularization scheme is expressed as:
\begin{eqnarray}
\hat{\mathcal{S}}^\epsilon_{E,\gamma}(N)&=&\sum_{v\in
V(\gamma)}\frac{N(v)}{i\hbar\kappa^2\beta}
\sum_{i,J}C^{iJ}\mathrm{Tr}\big((\rho[A(\beta_i)]-\rho[A(\beta^{-1}_i)])\rho[A(s_J^{-1})][\rho[A(s_J)],\
\hat{V}]\big),\nonumber\\
\hat{\mathcal{T}}^\epsilon_\gamma(N)&=&\sum_{v\in V(\gamma)}\frac{i
N(v)}{\hbar^3\kappa^4\beta^3}\sum_{I,J,K}T^{IJK}\mathrm{Tr}\big(\rho[A(s_I^{-1})][\rho[A(s_I)],\
\hat{K}]\rho[A(s_J^{-1})][\rho[A(s_J)],\
\hat{K}]\nonumber\\ &\times&\rho[A(s_K^{-1})][\rho[A(s_K)],\ \hat{V}]\big),\nonumber\\
\hat{\mathcal{S}}^\epsilon(N)\psi_\gamma&=&[\hat{\mathcal{S}}^\epsilon_{E,\gamma}(N)-2(1+\beta^2)
\hat{\mathcal{T}}_{\gamma}^\epsilon(N)]\psi_\gamma,\nonumber
\end{eqnarray}
where $C^{iJ}$ and $T^{IJK}$ are fixed constants independent of the
value of $\epsilon$. After removing the regulator $\epsilon$ via
diffeomorphism invariance the same as we did above, one obtains a
well-defined diffeomorphism covariant operator on
$\mathcal{H}_{kin}$ (or $\mathcal{H}^G$) in the sense of Uniform
Rovelli-Smolin Topology, or dually defines the operator on some
suitable domain in $Cyl^\star$. Note that such a general scheme of
construction exhibits that there are a great deal of freedom in
choosing the regulators, so that there are considerable ambiguities
in our quantization for seeking a proper quantum dynamics for
gravity, which is still an open issue today.

\end{itemize}

\subsection{Inclusion of Matter Field}

The quantization technique for the Hamiltonian constraint can be
generalized to quantize the Hamiltonian of matter fields coupled
to gravity \cite{thiemann7}. As an example, in this subsection we
consider the situation of background independent quantum dynamics
of a real massless scale field coupled to gravity. The coupled
generalized Palatini action reads \cite{han}
\begin{eqnarray}
S[e_{K}^{\beta},\omega_{\alpha}^{\
IJ},\phi]=S_{p}[e_{K}^{\beta},\omega_{\alpha}^{\
IJ}]+S_{KG}[e_{K}^{\beta},\phi],\nonumber
\end{eqnarray}
where
\begin{eqnarray}
S_{p}[e_{K}^{\beta},\omega_{\alpha}^{\ IJ}]
=\frac{1}{2\kappa}\int_{M}d^4x(e)
e_{I}^{\alpha}e_{J}^{\beta}(\Omega_{\alpha\beta}^{\ \
IJ}+\frac{1}{2\beta}\epsilon^{IJ}_{\ \ KL}\Omega_{\alpha\beta}^{\
\ KL}),\nonumber\\
S_{KG}[e_{K}^{\beta},\phi]=-\frac{\alpha_M}{2}\int_{M}d^4x(e)
\eta^{IJ}e^{\alpha}_{I}e^{\beta}_{J}(\partial_\alpha\phi)\partial_\beta\phi,\nonumber
\end{eqnarray}
here the real number $\alpha_M$ is the coupling constant. After
3+1 decomposition and Legendre transformation, one obtains the
total Hamiltonian of the coupling system on the 3-manifold
$\Sigma$ as:
\begin{eqnarray}
\mathcal{H}_{tot}=\Lambda^iG_i+N^aC_a+NC,\nonumber
\end{eqnarray}
where $\Lambda^i$, $N^a$ and $N$ are Lagrange multipliers, and the
three constraints in the Hamiltonian are expressed as
\cite{AR}\cite{han}:
\begin{eqnarray}
G_i&=&D_a\widetilde{P}^a_i\ :=\
\partial_a\widetilde{P}^a_i+\epsilon_{ij}^{\
\ k}A_a^i\widetilde{P}^a_k,\\
C_a&=&\widetilde{P}^b_iF_{ab}^i-A^i_aG_i+\widetilde{\pi}\partial_a\phi,\\
C&=&\frac{\kappa\beta^2}{2\sqrt{|\det
q|}}\widetilde{P}^a_i\widetilde{P}^b_j[\epsilon^{ij}_{\ \
k}F^k_{ab}-2(1+\beta^2)K^i_{[a}K^j_{b]}]\nonumber\\
&+&\frac{1}{\sqrt{|\det
q|}}[\frac{\kappa^2\beta^2\alpha_M}{2}\delta^{ij}\widetilde{P}^{a}_{i}\widetilde{P}^{b}_{j}(\partial_a
\phi)\partial_b\phi
+\frac{1}{2\alpha_M}\widetilde{\pi}^2],\label{kgconstraint}
\end{eqnarray}
here $\widetilde{\pi}$ denotes the momentum conjugate to $\phi$:
\begin{eqnarray}
\widetilde{\pi}:=\frac{\partial\cal
L}{\partial\dot{\phi}}=\frac{\alpha_M}{N}\sqrt{|\det
q|}(\dot{\phi}-N^a\partial_a\phi).\nonumber
\end{eqnarray}
Thus one has the elementary Poisson brackets
\begin{eqnarray}
\{A^i_a(x),\widetilde{P}^b_j(y)\}&=&\delta^a_b\delta^i_j\delta(x-y),\nonumber\\
\{\phi(x),\widetilde{\pi}(y)\}&=&\delta(x-y).\nonumber
\end{eqnarray}
Note that the second term of the Hamiltonian constraint
(\ref{kgconstraint}) is just the Hamiltonian of the real scalar
field.

Then we look for the background independent representation for the
real scalar field coupled to gravity, following the polymer
representation of the scalar field \cite{ALS}. The classical
configuration space, $\mathcal{U}$, consists of all real-valued
smooth functions $\phi$ on $\Sigma$. Given a set of finite number
of points $X=\{x_1,...,x_N\}$ in $\Sigma$, a equivalent relation
can be defined by: given two scalar field $\phi_1,
\phi_2\in\mathcal{U}$, $\phi_1\sim\phi_2$ if and only if $\exp[i
\lambda_i\phi_1(x_i)]=\exp[i \lambda_j\phi_2(x_j)]$ for all
$x_i\in X$ and all real number $\lambda_j$. Since one can define a
projective family with respect to the sets of points (graphs for
scalar field), a projective limit $\overline{\mathcal{U}}$, which
is a compact topological space, is obtained as the quantum
configuration space of scalar field. Next, we denote by
$Cyl_X(\overline{\mathcal{U}})$ the vector space generated by
finite linear combinations of the following functions of $\phi$:
\begin{eqnarray}
T_{X,\bf{\lambda}}(\phi):=\prod_{x_j\in X}\exp[i
\lambda_j\phi(x_j)],\nonumber
\end{eqnarray}
where $\bf{\lambda}\equiv (\lambda_1, \lambda_2, \cdot\cdot\cdot,
\lambda_N)$ are arbitrary non-zero real numbers assigned at each
point. It is obvious that $Cyl_X(\overline{\mathcal{U}})$ has the
structure of a $*$-algebra. The vector space
$Cyl(\overline{\mathcal{U}})$ of all cylindrical functions on
$\mathcal{U}$ is defined by the linear span of $T_0=1$ and
$T_{X,\bf{\lambda}}$. Completing $Cyl(\overline{\mathcal{U}})$ with
respect to the sup norm, one obtains a unital Abelian $C$*-algebra
$\overline{Cyl(\overline{\mathcal{U}})}$. Thus one can use the GNS
structure to construct its cyclic representations. A preferred
positive linear functional $\omega_0$ on
$\overline{Cyl(\overline{\mathcal{U}})}$ is defined by
\begin{eqnarray}
\omega_0(T_{X,\bf{\lambda}})=\left\{%
\begin{array}{ll}
    1 & \hbox{if $\lambda_j=0$ $\forall j$} \\
    0 & \hbox{otherwise}, \\
\end{array}%
\right.\ \nonumber
\end{eqnarray}
which defines a diffeomorphism-invariant faithful Borel measure
$\mu$ on $\overline{\mathcal{U}}$ as
\begin{eqnarray}\label{smeasure}
\int_\mathcal{U}d\mu(T_{X,\bf{\lambda}})=\left\{%
\begin{array}{ll}
    1 & \hbox{if $\lambda_j=0$ $\forall j$} \\
    0 & \hbox{otherwise}. \\
\end{array}%
 \right.\
\end{eqnarray}
Thus one obtains the Hilbert space,
$\mathcal{H}^{KG}_{kin}:=L^2(\overline{\mathcal{U}}, d\mu)$, of
square integrable functions on $\overline{\mathcal{U}}$ with respect
to $\mu$. The inner product can be expressed explicitly as:
\begin{equation}
<T_c|T_{c'}>^{KG}_{kin}=\delta_{cc'},\nonumber
\end{equation}
where the label $c:=(X, \bf{\lambda})$ is called scalar-network. As
one might expect, the quantum configuration space
$\overline{\mathcal{U}}$ is just the Gel'fand spectrum of
$\overline{Cyl(\overline{\mathcal{U}})}$. More concretely, for a
single point set $X_0\equiv \{x_0\}$,
$Cyl_{X_0}(\overline{\mathcal{U}})$ is the space of all almost
periodic functions on a real line $\mathbf{R}$. The Gel'fand
spectrum of the corresponding $C$*-algebra
$\overline{Cyl_{X_0}(\overline{\mathcal{U}})}$ is the Bohr
compactificaiton $\overline{\mathbf{R}}_{x_0}$ of $\mathbf{R}$
\cite{ALS}, which is a compact topological space such that
$\overline{Cyl_{X_0}(\overline{\mathcal{U}})}$ is the $C$*-algebra
of all continuous functions on $\overline{\mathbf{R}}_{x_0}$. Since
$\mathbf{R}$ is densely embedded in $\overline{\mathbf{R}}_{x_0}$,
$\overline{\mathbf{R}}_{x_0}$ can be regarded as a completion of
$\mathbf{R}$.

It is clear from Eq.(\ref{smeasure}) that an orthonomal basis in
$\mathcal{H}^{KG}_{kin}$ is given by the scalar vacuum $T_0=1$ and
the so-called scalar-network functions $T_c(\phi)$. So the total
kinematical Hilbert space $\mathcal{H}_{kin}$ is the direct product
of the kinematical Hilbert space $\mathcal{H}_{kin}^{GR}$ for
gravity and the kinematical Hilbert space for real scalar field,
i.e.,
$\mathcal{H}_{kin}:=\mathcal{H}_{kin}^{GR}\otimes\mathcal{H}_{kin}^{KG}$.
Then the spin-scalar-network state $T_{s,c}\equiv T_s(A)\otimes
T_c(\phi)\in Cyl_{\gamma(s)}(\overline{\mathcal{A/G}})\otimes
Cyl_{X(c)}(\overline{\mathcal{U}})\equiv Cyl_{\gamma(s,c)}$ is a
gravity-scalar cylindrical function on graph
$\gamma(s,c)\equiv\gamma(s)\cup X(c)$. Note that generally $X(c)$
may not coincide with the vertices of the graph $\gamma(s)$. It is
straightforward to see that all of these functions constitutes a
orthonormal basis in $\mathcal{H}_{kin}$ as
\begin{eqnarray}
<T_{s'}(A)\otimes T_{c'}(\phi)|T_s(A)\otimes
T_c(\phi)>_{kin}=\delta_{s's}\delta_{c'c}\ .\nonumber
\end{eqnarray}
Note that none of $\mathcal{H}_{kin}$, $\mathcal{H}^{GR}_{kin}$ and
$\mathcal{H}^{KG}_{kin}$ is a separable Hilbert space.

Given a pair $(x_0, \lambda_0)$, there is an elementary
configuration for the scalar field, the so-called point holonomy,
\begin{eqnarray}
U(x_0,\lambda_0):=\exp[i\lambda_0\phi(x_0)].\nonumber
\end{eqnarray}
It corresponds to a configuration operator $\hat{U}(x_0,\lambda_0)$,
which acts on any cylindrical function $\psi(\phi)\in
Cyl_{X(c)}(\overline{\mathcal{U}})$ by
\begin{equation}
\hat{U}(x_0,\lambda_0)\psi(\phi)=U(x_0,\lambda_0)\psi(\phi).\nonumber
\end{equation}
All these operators are unitary. But since the family of operators
$\hat{U}(x_0,\lambda)$ fails to be weakly continuous in $\lambda$,
there is no field operator $\hat{\phi}(x)$ on
$\mathcal{H}^{KG}_{kin}$. The momentum functional smeared on a
3-dimensional region $R\subset\Sigma$ is expressed by
\begin{eqnarray}
\pi(R):=\int_R d^3x \widetilde{\pi}(x).\nonumber
\end{eqnarray}
The Poisson bracket between the momentum functional and a point
holonomy can be easily calculated to be
\begin{eqnarray}
\{\pi(R), U(x,\lambda)\}=-i\lambda\chi_R(x)U(x, \lambda),\nonumber
\end{eqnarray}
where $\chi_R(x)$ is the characteristic function for the region $R$.
So the momentum operator is defined by the action on scalar-network
functions $T_{c=(X,\bf{\lambda})}$ as
\begin{eqnarray}
\hat{\pi}(R)T_{c}(\phi):=i\hbar\{\pi(R),
T_c(\phi)\}=\hbar[\sum_{x_j\in
X}\lambda_j\chi(x_j)]T_c(\phi).\nonumber
\end{eqnarray}
Now we can impose the quantum constraints on $\mathcal{H}_{kin}$ and
consider the quantum dynamics. First, the Gaussian constraint can be
solved independently of $\mathcal{H}_{kin}^{KG}$, since it only
involves gravitational field. It is also expected that the
diffeomorphism constraint can be implemented by the group averaging
strategy in the similar way as that in the case of pure gravity.
Given a spatial diffeomorphism transformation $\varphi$, an unitary
transformation $\hat{U}_\varphi$ was induced by $\varphi$ in the
Hilbert space $\mathcal{H}_{kin}$, which is expressed as
\begin{eqnarray}
\hat{U}_\varphi
T_{s=(\gamma(s),\mathbf{j},\mathbf{i}),c=(X(c),\mathbf{\lambda})}
=T_{\varphi\circ
s=(\varphi(\gamma(s)),\mathbf{j},\mathbf{i}),\varphi\circ
c=(\varphi(X(c)),\mathbf{\lambda})}.\nonumber
\end{eqnarray}
Then the differomorphism invariant spin-scalar-network functions are
defined by group averaging as
\begin{eqnarray}
T_{[s,c]}:=\frac{1}{n_{\gamma(s,c)}}\sum_{\varphi\in
Diff(\Sigma)/Diff_{\gamma(s,c)}}\sum_{\varphi'\in
GS_{\gamma(s,c)}}\hat{U}_{\varphi}\hat{U}_{\varphi'}T_{s,c},
\end{eqnarray}
where $Diff_{\gamma}$ is the set of diffeomorphisms leaving the
colored graph $\gamma$ invariant, $GS_{\gamma}$ denotes the graph
symmetry quotient group $Diff_{\gamma}/TDiff_{\gamma}$ where
$TDiff_{\gamma}$ is the set of the diffeomorphisms which is trivial
on the graph $\gamma$, and $n_\gamma$ is the number of elements in
$GS_\gamma$. Following the standard strategy in quantization of pure
gravity, an inner product can be defined on the vector space spanned
by the diffeomorphism invariant spin-scalar-network functions(and
the vacuum states for gravity, scalar and both respectively) such
that they form an orthonormal basis as:
\begin{eqnarray}
<T_{[s,c]}|T_{[s',c']}>_{Diff}:=T_{[s,c]}[T_{s',c'\in[s',c']}]=\delta_{[s,c],[s',c']}.
\end{eqnarray}
After the completion procedure, we obtain the expected Hilbert space
of diffeomorphism invariant states for the scalar field coupled to
gravity, which is denoted by $\mathcal{H}_{Diff}$. Then the only
nontrivial task is the implementation of the Hamiltonian constraint
$\mathcal{S}(N)$. One thus needs to define a corresponding
Hamiltonian constraint operator on $\mathcal{H}_{kin}$. While the
gravitational part of
\begin{eqnarray}
\mathcal{S}(N):=\int_\Sigma d^3x NC\nonumber
\end{eqnarray}
is a well-defined operator $\hat{\mathcal{S}}_{GR}(N)$ by the
Uniform Rovelli-Smolin Topology, the crucial point in this
subsection is to define an operator corresponding to the Hamiltonian
functional $\mathcal{S}_{KG}(N)$ of the scalar field, which can be
decomposed into two parts
\begin{eqnarray}
\mathcal{S}_{KG}(N)&=&\mathcal{S}_{KG,\phi}(N)+\mathcal{S}_{KG,Kin}(N),\nonumber
\end{eqnarray}
where
\begin{eqnarray}
\mathcal{S}_{KG,\phi}(N)&=&\frac{\kappa^2\beta^2\alpha_M}{2}\int_\Sigma
d^3xN\frac{1}{\sqrt{|\det
q|}}\delta^{ij}\widetilde{P}^{a}_{i}\widetilde{P}^{b}_{j}(\partial_a\phi)\partial_b\phi,\nonumber\\
\mathcal{S}_{KG,Kin}(N)&=&\frac{1}{2\alpha_M}\int_\Sigma
d^3xN\frac{1}{\sqrt{|\det q|}}\widetilde{\pi}^2\nonumber.
\end{eqnarray}
We use the identities, for $x\in R$
\begin{eqnarray}
\widetilde{P}^a_i=\frac{1}{2\kappa\beta}\widetilde{\eta}^{abc}\epsilon_{ijk}e^j_be^k_c\
\ \ \ \mathrm{and} \ \ \ \
e^i_a(x)=\frac{2}{\kappa\beta}\{A^i_a(x),V_{U_x}\}.\nonumber
\end{eqnarray}
Hence
\begin{eqnarray}
\widetilde{P}^a_i(x)=\frac{2}{\kappa^3\beta^3}\widetilde{\eta}^{abc}\epsilon_{ijk}
\{A^j_b(x),V_{U_x}\}\{A^k_c(x),V_{U_x}\}.\nonumber
\end{eqnarray}
Then the expressions of $\mathcal{S}_{KG,\phi}(N)$ and
$\mathcal{S}_{KG,Kin}(N)$ can be regulated via a point-splitting
strategy
\begin{eqnarray}
\mathcal{S}^\epsilon_{KG,\phi}(N)&=&\frac{\kappa^2\beta^2\alpha_M}{2}\int_\Sigma
d^3y\int_\Sigma
d^3xN^{1/2}(x)N^{1/2}(y)\chi_\epsilon(x-y)\delta^{ij}\times\nonumber\\
&&\frac{1}{\sqrt{V_{U^\epsilon_x}}}\widetilde{P}^{a}_{i}(x)(\partial_a\phi(x))\frac{1}{\sqrt{V_{U^\epsilon_y}}}
\widetilde{P}^{b}_{j}(y)\partial_b\phi(y)\nonumber\\
&=&\frac{32\alpha_M}{\kappa^4\beta^4}\int_\Sigma d^3y\int_\Sigma
d^3xN^{1/2}(x)N^{1/2}(y)\chi_\epsilon(x-y)\delta^{ij}\times\nonumber\\
&&\widetilde{\eta}^{aec}(\partial_a\phi(x))\mathrm{Tr}\big(\tau_i\{\mathbf{A}_e(x),V_{U^\epsilon_x}^{3/4}\}
\{\mathbf{A}_c(x),V_{U^\epsilon_x}^{3/4}\}\big)\times\nonumber\\
&&\widetilde{\eta}^{bfd}(\partial_b\phi(y))\mathrm{Tr}\big(\tau_j\{\mathbf{A}_f(y),V_{U^\epsilon_y}^{3/4}\}
\{\mathbf{A}_d(y),V_{U^\epsilon_y}^{3/4}\}\big),\nonumber\\
\mathcal{S}_{KG,Kin} &=&\frac{1}{2\alpha_M}\int_\Sigma
d^3x\int_\Sigma
d^3yN^{1/2}(x)N^{1/2}(y)\widetilde{\pi}(x)\widetilde{\pi}(y)\times\nonumber\\
&&\int_\Sigma
d^3u\frac{\det(e_a^i(u))}{(V_{U^\epsilon_u})^{3/2}}\int_\Sigma
d^3w\frac{\det(e_a^i(w))}{(V_{U^\epsilon_w})^{3/2}}\chi_\epsilon(x-y)\chi_\epsilon(u-x)\chi_\epsilon(w-y)\nonumber\\
&=&\frac{1}{2\alpha_M}\frac{2^8}{9(\kappa\beta)^6}\int_\Sigma
d^3x\int_\Sigma
d^3yN^{1/2}(x)N^{1/2}(y)\widetilde{\pi}(x)\widetilde{\pi}(y)\times\nonumber\\
&&\int_\Sigma d^3u\
\widetilde{\eta}^{abc}\mathrm{Tr}\big(\{\mathbf{A}_a(u),\sqrt{V_{U^\epsilon_u}}\}
\{\mathbf{A}_b(u),\sqrt{V_{U^\epsilon_u}}\}\{\mathbf{A}_c(u),\sqrt{V_{U^\epsilon_u}}\}\big)\times\nonumber\\
&&\int_\Sigma d^3w\
\widetilde{\eta}^{def}\mathrm{Tr}\big(\{\mathbf{A}_d(w),\sqrt{V_{U^\epsilon_w}}\}
\{\mathbf{A}_e(w),\sqrt{V_{U^\epsilon_w}}\}\{\mathbf{A}_f(w),\sqrt{V_{U^\epsilon_w}}\}\big)\times\nonumber\\
&&\chi_\epsilon(x-y)\chi_\epsilon(u-x)\chi_\epsilon(w-y),\nonumber
\end{eqnarray}
where the matrices $\mathbf{A}_a\equiv A_a^i\tau_i$,
$\chi_\epsilon(x-y)$ is the characteristic function of a box
containing $x$ with scale $\epsilon$ such that
$\lim_{\epsilon\rightarrow0}\chi_\epsilon(x-y)/{\epsilon^3}=\delta(x-y)$,
and $V_{U^\epsilon_x}$ is the volume of the box. Introducing a
triangulation $T(\epsilon)$ of $\Sigma$ by tetrahedrons $\Delta$, we
notice the following useful identities
\begin{eqnarray}
\{\int_{s(\Delta)}dt\
\mathbf{A}_a\dot{s}^a(t),V_{U^\epsilon_{s(s(\Delta))}}^{3/4}\}&=&-A(s(\Delta))^{-1}\{A(s(\Delta)),
V_{U^\epsilon_{s(s(\Delta))}}^{3/4}\}
+o(\epsilon),\nonumber
\end{eqnarray}
and
\begin{eqnarray}
\int_{s(\Delta)} dt\ \partial_a\phi\dot{s}^a(t)&=&\frac{1}{i\lambda}
U(s(s(\Delta)),\lambda)^{-1}[U(t(s(\Delta)),\lambda)-U(s(s(\Delta)),\lambda)]+o(\epsilon)
\nonumber
\end{eqnarray}
for nonzero $\lambda$, where $s(s(\Delta))$ and $t(s(\Delta))$
denote respectively the beginning and end points of segment
$s(\Delta)$ with scale $\epsilon$ associated with a tetrahedron
$\Delta$. Regulated on the triangulation, the scalar field part of
the classical Hamiltonian constraint reads
\begin{eqnarray}
\mathcal{S}^\epsilon_{KG,\phi}(N)
&=&-\frac{4\alpha_M}{9\kappa^4\beta^4}\sum_{\Delta'\in
T(\epsilon)}\sum_{\Delta\in
T(\epsilon)}N^{1/2}(v(\Delta))N^{1/2}(v(\Delta'))\chi_\epsilon(v(\Delta)-v(\Delta'))
\delta^{ij}\times\nonumber\\
&&\epsilon^{lmn}\frac{1}{\lambda}U(v(\Delta),\lambda)^{-1}[U(t(s_l(\Delta)),\lambda)-U(v(\Delta),\lambda)]
\times\nonumber\\
&&\mathrm{Tr}\big(\tau_iA(s_m(\Delta))^{-1}\{A(s_m(\Delta)),V_{U^\epsilon_{v(\Delta)}}^{3/4}\}A(s_n(\Delta))^{-1}
\{A(s_n(\Delta)),V_{U^\epsilon_{v(\Delta)}}^{3/4}\}\big)\times\nonumber\\
&&\epsilon^{kpq}\frac{1}{\lambda}U(v(\Delta'),\lambda)^{-1}[U(t(s_k(\Delta')),\lambda)-U(v(\Delta'),\lambda)]
\times\nonumber\\
&&\mathrm{Tr}\big(\tau_jA(s_p(\Delta'))^{-1}\{A(s_p(\Delta')),V_{U^\epsilon_{v(\Delta')}}^{3/4}\}A(s_q(\Delta'))^{-1}
\{A(s_q(\Delta')),V_{U^\epsilon_{v(\Delta')}}^{3/4}\}\big),\nonumber\\
\mathcal{S}^\epsilon_{KG,Kin}(N)
&=&\frac{16}{81\alpha_M(\kappa\beta)^6}\sum_{\Delta\in
T(\epsilon)}\sum_{\Delta'\in
T(\epsilon)}N^{1/2}(v(\Delta))N^{1/2}(v(\Delta')){\pi}(\Delta){\pi}(\Delta')\times\nonumber\\
&&\sum_{\Delta''\in
T(\epsilon)}\epsilon^{imn}\mathrm{Tr}\big(A(s_i(\Delta''))^{-1}
\{A(s_i(\Delta'')),\sqrt{V_{U^\epsilon_{v(\Delta'')}}}\}\times\nonumber\\
&&A(s_m(\Delta''))^{-1}\{A(s_m(\Delta'')),\sqrt{V_{U^\epsilon_{v(\Delta'')}}}\}\times\nonumber\\
&&A(s_n(\Delta''))^{-1}\{A(s_n(\Delta'')),\sqrt{V_{U^\epsilon_{v(\Delta'')}}}\}\big)\times\nonumber\\
&&\sum_{\Delta'''\in
T(\epsilon)}\epsilon^{jkl}\mathrm{Tr}\big(A(s_j(\Delta'''))^{-1}
\{A(s_j(\Delta''')),\sqrt{V_{U^\epsilon_{v(\Delta''')}}}\}\times\nonumber\\
&&A(s_k(\Delta'''))^{-1}\{A(s_k(\Delta''')),\sqrt{V_{U^\epsilon_{v(\Delta''')}}}\}\times\nonumber\\
&&A(s_l(\Delta'''))^{-1}\{A(s_l(\Delta''')),\sqrt{V_{U^\epsilon_{v(\Delta''')}}}\}\big)\times\nonumber\\
&&\chi_\epsilon(v(\Delta)-v(\Delta'))\chi_\epsilon(v(\Delta'')-v(\Delta))\chi_\epsilon(v(\Delta''')-v(\Delta')).
\label{classicalH}
\end{eqnarray}
Note that the above regularization is explicitly dependent on the
parameter $\lambda$, which will leads to a kind of quantization
ambiguity of the real scalar field dynamics in polymer-like
representation. Introducing a partition $\mathcal{P}$ of the
3-manifold $\Sigma$ into cells $C$, we can smear the essential
"square roots" of $\mathcal{S}^\epsilon_{KG,\phi}$ and
$\mathcal{S}^\epsilon_{KG,Kin}$ in one cell $C$ respectively and
promote them as regulated operators in $\mathcal{H}_{kin}$ with
respect to triangulations $T(\epsilon)$ depending on
spin-scalar-network state $T_{s,c}$ as
\begin{eqnarray}
\hat{W}^{\epsilon,C}_{\gamma(s,c),\phi,i}&=&\sum_{v\in
V(\gamma(s,c))}\frac{\chi_C(v)}{E(v)}\sum_{v(\Delta)=v}\hat{h}^{\epsilon,\Delta}_{\phi,v,i},\nonumber\\
\hat{W}^{\epsilon,C}_{\gamma(s,c),Kin}&=&\sum_{v\in
V(\gamma(s,c))}\frac{\chi_C(v)}{E(v)}\sum_{v(\Delta)=v}\hat{h}^{\epsilon,\Delta}_{Kin,v},
\label{sqarerootH}
\end{eqnarray}
where $\chi_C(v)$ is the characteristic function of the cell $C$,
and
\begin{eqnarray}
\hat{h}^{\epsilon,\Delta}_{\phi,v,i}&:=&\frac{i}{\hbar^2}\epsilon^{lmn}\frac{1}{\lambda(v)}\hat{U}(v,\lambda(v))^{-1}
[\hat{U}(t(s_l(\Delta)),\lambda(v))
-\hat{U}(v,\lambda(v))]\times\nonumber\\
&&\mathrm{Tr}\big(\tau_i\hat{A}(s_m(\Delta))^{-1}[\hat{A}(s_m(\Delta)),\hat{V}_{U^\epsilon_{v}}^{3/4}]
\hat{A}(s_n(\Delta))^{-1}[\hat{A}(s_n(\Delta)),\hat{V}_{U^\epsilon_{v}}^{3/4}]\big),\nonumber\\
\hat{h}^{\epsilon,\Delta}_{Kin,v}&:=&\frac{1}{(i\hbar)^3}\hat{\pi}(v)\epsilon^{lmn}\mathrm{Tr}
\big(\hat{A}(s_l(\Delta))^{-1}[\hat{A}(s_l(\Delta)),
\sqrt{\hat{V}_{U^\epsilon_{v}}}]\times\nonumber\\
&&\hat{A}(s_m(\Delta))^{-1}[\hat{A}(s_m(\Delta)),\sqrt{\hat{V}_{U^\epsilon_{v}}}]\times\nonumber\\
&&\hat{A}(s_n(\Delta))^{-1}[\hat{A}(s_n(\Delta)),\sqrt{\hat{V}_{U^\epsilon_{v}}}]\big).\label{ambiguity}
\end{eqnarray}
Both operators in (\ref{sqarerootH}) and their adjoint operators are
cylindrically consistent up to diffeomorphisms. Thus there are two
densely defined operators $\hat{W}^{C}_{\phi,i}$ and
$\hat{W}^{C}_{Kin}$ in $\mathcal{H}_{kin}$ associated with the two
consistent families of (\ref{sqarerootH}). We now give several
remarks on their properties.
\begin{itemize}
\item {\textit{Removal of regulator $\epsilon$}}

It is not difficult to see that the action of the operator
$\hat{W}^{\epsilon,C}_{\gamma(s,c),\phi,i}$ on a spin-scalar-network
function $T_{s,c}$ is graph-changing. It adds finite number of
vertices with representation $\lambda(v)$ at $t(s_i(\Delta))$ with
distance $\epsilon$ from the vertex $v$. Recall that the action of
the gravitational Hamiltonian constraint operator on a spin network
function is also graph-changing. As a result, the family of
operators $\hat{W}^{\epsilon,C}_{\gamma(s,c),\phi,i}$ also fails to
be weakly converged when $\epsilon\rightarrow0$. However, due to the
diffeomorphism covariant properties of the triangulation, the limit
operator can be well-defined via the uniform Rovelli-Smolin
topology, or equivalently, the operator can be dually defined on
diffeomorphism invariant states. But the dual operator cannot leave
$\mathcal{H}_{Diff}$ invariant.

\item {\textit{Quantization ambiguity}}

As a main difference of the dynamics in polymer-like
representation from that in U(1) group representation
\cite{thiemann7}, a continuous label $\lambda$ appears explicitly
in the expression of (\ref{sqarerootH}). Hence there is an
one-parameter quantization ambiguity due to the real scalar field.
Recall that the construction of gravitational Hamiltonian
constraint operator also has a similar ambiguity due to the choice
of the representations $j$ of the edges added by its action. A
related quantization ambiguity also appears in the dynamics of
loop quantum cosmology \cite{boj6}.
\end{itemize}

By taking the limit $\mathcal{P}\rightarrow\Sigma$ so that
$C\rightarrow v$, the quantum Hamiltonian constraint
$\hat{\mathcal{S}}^\epsilon_{KG}(N)$ of scalar field is expressed
as:
\begin{eqnarray}
\hat{\mathcal{S}}^{\epsilon}_{KG,\gamma(s,c)}(N)&:=&\sum_{v\in
V(\gamma(s,c))}N(v) \big[64\times
\frac{4\alpha_M}{9\kappa^4\beta^4}\delta^{ij}(\hat{W}^{\epsilon,v}_{\phi,i})^\dagger\hat{W}^{\epsilon,v}_{\phi,j}\nonumber\\
&+&8^4\times
\frac{16}{81\alpha_M(\kappa\beta)^6}(\hat{W}^{\epsilon,v}_{Kin})^\dagger\hat{W}^{\epsilon,v}_{Kin}\big],\label{scalarconstraint}
\end{eqnarray}
where the operators $\hat{W}^{\epsilon,v}_{\phi,i}$ and
$\hat{W}^{\epsilon,v}_{Kin}$ are the inductive limit of the
consistent family $\{\hat{W}^{\epsilon,v}_{\gamma,\phi,i}\}_\gamma$
and $\{\hat{W}^{\epsilon,v}_{\gamma,Kin}\}_\gamma$, and
$(\hat{W}^{\epsilon,v}_{\phi,i})^\dagger$ and
$(\hat{W}^{\epsilon,v}_{Kin})^\dagger$ are their adjoint
respectively. Hence the family of Hamiltonian constraint operators
(\ref{scalarconstraint}) is also cylindrically consistent up to a
diffeomorphism, and the regulator $\epsilon$ can be removed via the
uniform Rovelli-Smollin topology, or equivalently the limit operator
dually acts on diffeomorphism invariant states as
\begin{eqnarray}
(\hat{\mathcal{S}}'_{KG}(N)\Psi_{Diff})[f]=
\lim_{\epsilon\rightarrow0}\Psi_{Diff}[\hat{\mathcal{S}}^{\epsilon}_{KG}(N)f],
\end{eqnarray}
for any $f\in Cyl(\overline{\mathcal{A}})\otimes
Cyl(\overline{\mathcal{U}})$. Similar to the dual of
$\hat{\mathcal{S}}_{GR}(N)$, the operator
$\hat{\mathcal{S}}'_{KG}(N)$ fails to commute with the dual of
finite diffeomorphism transformation operators, unless the smearing
function $N(x)$ is a constant function over $\Sigma$. In fact, one
can define a self-adjoint Hamlitonian operator from
$\mathcal{S}'_{KG}(1)$ for the polymer scalar field in the
diffeomorphism invariant Hilbert space $\mathcal{H}_{Diff}$
\cite{HM}. From Eq.(\ref{scalarconstraint}), it is not difficult to
prove that for positive $N(x)$ the Hamiltonian constraint operator
$\hat{\mathcal{S}}_{KG}(N)$ of scalar field is positive and
symmetric in $\mathcal{H}_{kin}$ and hence has a unique self-adjoint
extension \cite{HM}. Note that there is an 1-parameter ambiguity in
our construction of $\hat{\mathcal{S}}_{KG}(N)$ due to the real
scalar field, which is manifested as the continuous parameter
$\lambda$ in the expression of
$\hat{h}^{\epsilon,\Delta}_{\phi,v,i}$ in (\ref{ambiguity}). Thus
the total Hamiltonian constraint operator of scalar field coupled to
gravity has been obtained as
\begin{eqnarray}
\hat{\mathcal{S}}(N)=\hat{\mathcal{S}}_{GR}(N)+\hat{\mathcal{S}}_{KG}(N).\label{Hconstraint}
\end{eqnarray}
Again, there is no UV divergence in this quantum Hamiltonian
constraint. Recall that, in standard quantum field theory the UV
divergence can only be cured by renormalization procedure, in which
one has to multiply the Hamiltonian by a suitable power of the
regulating parameter $\epsilon$ artificially. While, now $\epsilon$
has naturally disappeared from the expression of
(\ref{Hconstraint}). So renormalization is not needed for the
polymer-like scalar field coupled to gravity, since quantum gravity
has been presented as a natural regulator. This result heightens our
confidence that the issue of divergence in quantum fields theory can
be cured in the framework of loop quantum gravity.

Now we have obtained the desired matter-coupled quantum Hamiltonian
constraint equation
\begin{eqnarray}
-\big(\hat{\mathcal{S}}'_{KG}(N)\Psi_{Diff}\big)[f]=\big(\hat{\mathcal{S}}_{GR}'(N)
\Psi_{Diff}\big)[f].\label{evo constr}
\end{eqnarray}
Comparing it with the well-known Sch\"{o}rdinger equation for a
particle,
\begin{eqnarray}
{i\hbar\frac{\partial}{\partial
t}}\psi(x,t)=H(\hat{x},\widehat{-i\hbar\frac{\partial}{\partial
x}})\psi(x,t),\nonumber
\end{eqnarray}
where $\psi(x,t)\in L^2(\mathbf{R},dx)$ and $t$ is a parameter
labeling time evolution, one may take the viewpoint that the
matter field constraint operator $\hat{\mathcal{S}}_{KG}'(N)$
plays the role of $i\hbar\frac{\partial}{\partial t}$. Then $\phi$
appears as the parameter labeling the evolution of the
gravitational field state. In the reverse viewpoint, gravitational
field would become the parameter labeling the evolution of the
quantum matter field. Note that such an idea has been successfully
applied in loop quantum cosmology model to understand the quantum
nature of big bang in the deep Planck regime
\cite{APS}\cite{APS1}.

\subsection{Master Constraint Programme}

Although the Hamiltonian constraint operator introduced in Section
6.1 is densely defined on $\mathcal{H}_{kin}$ and diffeomorphism
covariant, there are still several problems unsettled which are
listed below.
\begin{itemize}

\item It is unclear whether the commutator between two Hamiltonian
constraint operators resembles the classical Poisson bracket
between two Hamiltonian constraints. Hence it is doubtful whether
the quantum Hamiltonian constraint produces the correct quantum
dynamics with correct classical limit \cite{GL}\cite{LM}.

\item The dual Hamiltonian constraint operator does not leave the
Hilbert space $\mathcal{H}_{Diff}$ invariant. Thus the inner
product structure of $\mathcal{H}_{Diff}$ cannot be employed in
the construction of physical inner product.

\item Classically the collection of Hamiltonian constraints do not
form a Lie algebra. So one cannot employ group average strategy in
solving the Hamiltonian constraint quantum mechanically, since the
strategy depends on group structure crucially.
\end{itemize}
However, if one could construct an alternative classical
constraint algebra, giving the same constraint phase space, which
is a Lie algebra (no structure function), where the subalgebra of
diffeomorphism constraints forms an ideal, then the programme of
solving constraints would be much improved at a basic level. Such
a constraint Lie algebra was first introduced by Thiemann in
\cite{thiemann3}. The central idea is to introduce the master
constraint:
\begin{eqnarray}
\textbf{M}:=\frac{1}{2}\int_\Sigma
d^3x\frac{|\widetilde{C}(x)|^2}{\sqrt{|\det
q(x)|}},\label{mconstraint}
\end{eqnarray}
where $\widetilde{C}(x)$ is the scalar constraint in
Eq.(\ref{scalar}). One then gets the master constraint algebra:
\begin{eqnarray}
\{\mathcal{V}(\vec{N}),\ \mathcal{V}(\vec{N}')\}&=&\mathcal{V}([\vec{N},\vec{N}']),\nonumber\\
\{\mathcal{V}(\vec{N}),\ \textbf{M}\}&=&0,\nonumber\\
\{\textbf{M},\ \textbf{M}\}&=&0.\nonumber
\end{eqnarray}
The master constraint programme has been well tested in various
examples \cite{thiemann9}\cite{DT2}\cite{DT3}\cite{DT4}\cite{DT5}.
In the following, we extend the diffeomorphism transformations such
that the Hilbert space $\mathcal{H}_{Diff}$ is separable. This
separability of $\mathcal{H}_{Diff}$ and the positivity and the
diffeomorphism invariance of $\textbf{M}$ will be working together
properly and provide us with powerful functional analytic tools in
the programme to solve the constraint algebra quantum mechanically.
The regularized version of the master constraint can be expressed as
\begin{eqnarray}
\textbf{M}^{\epsilon}:=\frac{1}{2}\int_\Sigma d^3y \int_\Sigma
d^3x\chi_\epsilon(x-y)\frac{\widetilde{C}(y)}{\sqrt{V_{U_y^\epsilon}}}
\frac{\widetilde{C}(x)}{\sqrt{V_{U^\epsilon_{x}}}}.\nonumber
\end{eqnarray}
Introducing a partition $\mathcal{P}$ of the 3-manifold $\Sigma$
into cells $C$, we have an operator $\hat{H}^\epsilon_{C,\gamma}$
acting on any cylindrical function $f_\gamma\in
Cyl_\gamma(\overline{\mathcal{A/G}})$ in $\mathcal{H}^G$ as
\begin{equation}
\hat{H}^\epsilon_{C,\gamma}\ f_\gamma=\sum_{v\in
V(\gamma)}\frac{\chi_C(v)}{E(v)}\sum_{v(\Delta)=v}\hat{h}^{\epsilon,\Delta}_{v}
f_\gamma,\label{so1}
\end{equation}
via a family of state-dependent triangulations $T(\epsilon)$ on
$\Sigma$, where $\chi_C(v)$ is the characteristic function of the
cell $C(v)$ containing a vertex $v$ of the graph $\gamma$. Note that
both $\hat{H}^\epsilon_{C,\gamma}$ and its adjoint are cylindrically
consistent up to diffeomorphisms, and the expression of
$\hat{h}^{\epsilon,\Delta}_{v}$ reads
\begin{eqnarray}
\hat{h}^{\epsilon,\Delta}_{v}&=&\frac{16}{3i\hbar\kappa^2\beta}\epsilon^{ijk}\mathrm{Tr}
\big(\hat{A}(\alpha_{ij}(\Delta))^{-1}
\hat{A}(s_k(\Delta))^{-1}[\hat{A}(s_k(\Delta)),
\sqrt{\hat{V}_{U^\epsilon_{v}}}]\big)\nonumber\\
&&+2(1+\beta^2)\frac{4\sqrt{2}}{3i\hbar^3\kappa^4\beta^3}\epsilon^{ijk}\mathrm{Tr}\big(\hat{A}(s_i(\Delta))^{-1}
[\hat{A}(s_i(\Delta)),\hat{K}^\epsilon]\nonumber\\
&&\hat{A}(s_j(\Delta))^{-1}
[\hat{A}(s_j(\Delta)),\sqrt{\hat{V}_{U^\epsilon_{v}}}]\hat{A}(s_k(\Delta))^{-1}[\hat{A}(s_k(\Delta)),
\hat{K}^\epsilon]\big).\label{hath}
\end{eqnarray}
Note that $\hat{h}^{\epsilon,\Delta}_{v}$ is similar to that
involved in the regulated Hamiltonian constraint operator in section
6.1, while the only difference is that now the volume operator is
replaced by its quare-root in Eq.(\ref{hath}). Hence the action of
$\hat{H}^\epsilon_{C,\gamma}$ on $f_\gamma$ adds arcs
$a_{ij}(\Delta)$ with 1/2-representation with respect to each
$v(\Delta)$ of $\gamma$. Thus, for each $\epsilon>0$,
$\hat{H}^\epsilon_{C,\gamma}$ is a Yang-Mills gauge invariant and
diffeomorphism covariant operator defined on
$Cyl_\gamma(\overline{\mathcal{A/G}})$. The family of such operators
with respect to different graphs is cylindrically consistent up to
diffeomorphisms and hence can give a limit operator $\hat{H}_{C}$
densely defined on $\mathcal{H}^G$ by the uniform Rovelli-Smollin
topology. The adjoint operator
$(\hat{H}^{\epsilon}_{C,\gamma})^\dagger$ can be well defined in
${\cal H}^G$ as
\begin{equation}
(\hat{H}^\epsilon_{C,\gamma})^\dagger=\sum_{v\in
V(\gamma)}\frac{\chi_C(v)}{E(v)}\sum_{v(\Delta)=v}(\hat{h}^{\epsilon,\Delta}_{v})^\dagger,
\end{equation}
such that the limit operators $\hat{H}_{C}$ and
$(\hat{H}_{C})^\dagger$ in the uniform Rovelli-Smolin topology
satisfy
\begin{eqnarray}
<g_{\gamma'}, \hat{H}_C f_{\gamma}>_{kin}&=&<g_{\gamma'},
\hat{H}_{C,\gamma} f_{\gamma}>_{kin}=<(\hat{H}_{C,\gamma})^\dagger
g_{\gamma'},f_{\gamma}>_{kin}\nonumber\\
&=&<(\hat{H}_{C})^\dagger
g_{\gamma'},f_{\gamma}>_{kin}=<(\hat{H}_{C})^\dagger_{\gamma'}g_{\gamma'},f_{\gamma}>_{kin},
\end{eqnarray}
where $\hat{H}_C$ and $(\hat{H}_{C})^\dagger$ are respectively the
inductive limits of $\hat{H}_{C,\gamma}$ and
$(\hat{H}_{C,\gamma})^\dagger$. Then a master constraint operator,
$\hat{\mathbf{M}}$, acting on any $\Psi_{Diff}\in{\cal H}_{Diff}$
can be defined as \cite{HM2}
\begin{equation}
(\hat{\mathbf{M}}\Psi_{Diff})[f_\gamma]:=\lim_{\mathcal{P}\rightarrow
\Sigma;\epsilon,\epsilon'\rightarrow\mathrm{0}}\Psi_{Diff}[\sum_{C\in\mathcal{P}}\frac{1}{2}\hat{H}^\epsilon_{C}
(\hat{H}^{\epsilon'}_{C})^\dagger f_\gamma],\label{master}
\end{equation}
for any $f_\gamma$ is a finite linear combination of spin-network
function. Note that $\hat{H}^\epsilon_{C}
(\hat{H}^{\epsilon'}_{C})^\dagger f_\gamma$ is also a finite linear
combination of spin-network functions on an extended graph with the
same skeleton of $\gamma$, hence the value of
$(\hat{\textbf{M}}\Psi_{Diff})[f_\gamma]$ is finite for any given
$\Psi_{Diff}$. Thus $\hat{\textbf{M}}\Psi_{Diff}$ lies in the
algebraic dual of the space of cylindrical functions. Furthermore,
we can show that $\hat{\mathbf{M}}$ leaves the diffeomorphism
invariant distributions invariant. For any diffeomorphism
transformation $\varphi$ on $\Sigma$,
\begin{eqnarray}
(\hat{U}'_\varphi\hat{\textbf{M}}\Psi_{Diff})[f_\gamma]&=&\lim_{\mathcal{P}\rightarrow
\Sigma;\epsilon,\epsilon'\rightarrow\mathrm{0}}\Psi_{Diff}[\sum_{C\in\mathcal{P}}\frac{1}{2}\hat{H}^\epsilon_{C}
(\hat{H}^{\epsilon'}_{C})^\dagger \hat{U}_\varphi
f_\gamma]\nonumber\\
&=&\lim_{\mathcal{P}\rightarrow
\Sigma;\epsilon,\epsilon'\rightarrow\mathrm{0}}\Psi_{Diff}[\hat{U}_\varphi\sum_{C\in\mathcal{P}}
\frac{1}{2}\hat{H}^{\varphi^{-1}(\epsilon)}_{\varphi^{-1}(C)}
(\hat{H}^{\varphi^{-1}(\epsilon')}_{\varphi^{-1}(C)})^\dagger f_\gamma]\nonumber\\
&=&\lim_{\mathcal{P}\rightarrow
\Sigma;\epsilon,\epsilon'\rightarrow\mathrm{0}}\Psi_{Diff}[\sum_{C\in\mathcal{P}}\frac{1}{2}\hat{H}^\epsilon_{C}
(\hat{H}^{\epsilon'}_{C})^\dagger f_\gamma],
\end{eqnarray}
where in the last step, we used the fact that the diffeomorphism
transformation $\varphi$ leaves the partition invariant in the limit
$\mathcal{P}\rightarrow\sigma$ and relabel $\varphi(C)$ to be $C$.
So we have the result
\begin{eqnarray}
(\hat{U}'_\varphi\hat{\textbf{M}}\Psi_{Diff})[f_\gamma]=(\hat{\textbf{M}}\Psi_{Diff})[f_\gamma].\label{diff}
\end{eqnarray}
So given a diffeomorphism invariant spin-network state $T_{[s]}$,
the resulted state $\hat{\textbf{M}}T_{[s]}$ must be a
diffeomorphism invariant element in the algebraic dual of
$Cyl(\overline{\mathcal{A/G}})$, which can be formally expressed as
\begin{eqnarray}
\hat{\textbf{M}}T_{[s]}=\sum_{[s_1]}c_{[s_1]}T_{[s_1]}.\nonumber
\end{eqnarray}
Thus for any $T_{s_2}$, one has
\begin{eqnarray}
\lim_{\mathcal{P}\rightarrow
\Sigma;\epsilon,\epsilon'\rightarrow\mathrm{0}}T_{[s]}[\sum_{C\in\mathcal{P}}\frac{1}{2}\hat{H}^\epsilon_{C}
(\hat{H}^{\epsilon'}_{C})^\dagger
T_{s_2}]=\sum_{[s_1]}c_{[s_1]}T_{[s_1]}[T_{s_2}],\nonumber
\end{eqnarray}
where the cylindrical function
$\sum_{C\in\mathcal{P}}\frac{1}{2}\hat{H}^{\epsilon'}_{C}
(\hat{H}^{\epsilon}_{C})^\dagger T_{s_2}$ is a finite linear
combination of spin-network functions on some graphs $\gamma\ '$
with the same skeleton of $\gamma(s_2)$ up to finite number of arcs.
Hence fixing the diffeomorphism equivalent class $[s]$, only for
spin-networks $s_2$ which lie in a finite number of diffeomorphism
equivalent classes the left hand side of the last equation can be
non-zero. So there should also be only finite number of classes
$[s_1]$ in the right hand side such that the corresponding
coefficients $c_{[s_1]}$ are non-zero. As a result,
$\hat{\textbf{M}}T_{[s]}$ is a finite linear combination of
diffeomorphism invariant spin-network states and hence lies in the
Hilbert space of diffeomorphism invariant states
$\mathcal{H}_{Diff}$ for any $[s]$. Therefore $\hat{\textbf{M}}$ is
densely defined on $\mathcal{H}_{Diff}$. Moreover, given two
diffeomorphism invariant spin-network functions $T_{[s_1]}$ and
$T_{[s_2]}$, a straightforward calculation can give the matrix
elements of $\hat{\mathbf{M}}$ as \cite{HM2}\cite{HM}
\begin{eqnarray}
&&<T_{[s_1]}|\hat{\textbf{M}}|T_{[s_2]}>_{Diff}\nonumber\\
&=&\lim_{\mathcal{P}\rightarrow
\Sigma;\epsilon,\epsilon'\rightarrow\mathrm{0}}\sum_{C\in\mathcal{P}}\frac{1}{2}
\overline{T_{[s_2]}[\hat{H}^\epsilon_{C}(\hat{H}^{\epsilon'}_{C})^\dagger T_{s_1\in[s_1]}]}\nonumber\\
&=&\sum_{[s]}\sum_{v\in
V(\gamma(s\in[s]))}\frac{1}{2}\lim_{\epsilon,\epsilon'\rightarrow\mathrm{0}}
\overline{T_{[s_2]}[\hat{H}^{\epsilon}_{v}
T_{s\in[s]}]}T_{[s_1]}[\hat{H}^{\epsilon'}_{v}T_{s\in[s]}]\nonumber\\
&=&\sum_{[s]}\sum_{v\in
V(\gamma(s\in[s]))}\frac{1}{2}\overline{(\hat{H}'_v T_{[s_2]})[
T_{s\in[s]}]}(\hat{H}'_v T_{[s_1]}) [ T_{s\in[s]}].\label{master2}
\end{eqnarray}
From Eq.(\ref{master2}) and the fact that the master constraint
operator $\hat{\mathbf{M}}$ is densely defined on
$\mathcal{H}_{Diff}$, it is obvious that $\hat{\mathbf{M}}$ is a
positive and symmetric operator in ${\cal H}_{Diff}$. Therefore, the
quadratic form $Q_{\mathbf{M}}$ associated with $\hat{\mathbf{M}}$
is closable \cite{rs}. The closure of $Q_{\mathbf{M}}$ is the
quadratic form of a unique self-adjoint operator
$\hat{\overline{\mathbf{M}}}$, called the Friedrichs extension of
$\hat{\mathbf{M}}$. We relabel $\hat{\overline{\mathbf{M}}}$ to be
$\hat{\mathbf{M}}$ for simplicity. From the construction of
$\hat{\mathbf{M}}$, the qualitative description of the kernel of the
Hamiltonian constraint operator in Ref.\cite{thiemann5} can be
transcribed to describe the solutions to the equation:
$\hat{\mathbf{M}}\Psi_{Diff}=0$. In particular, the diffeomorphism
invariant cylindrical functions based on at most 2-valent graphs are
obviously normalizable solutions. In conclusion, there exists a
positive and self-adjoint operator $\hat{\mathbf{M}}$ on
$\mathcal{H}_{Diff}$ corresponding to the master constraint
(\ref{mconstraint}), and zero is in the point spectrum of
$\hat{\mathbf{M}}$.

Note that the quantum constraint algebra can be easily checked to be
anomaly free. i.e.,
\begin{eqnarray}
[\hat{\mathbf{M}},\hat{U}'_\varphi]=0,\ \ \
[\hat{\mathbf{M}},\hat{\mathbf{M}}]=0.\nonumber
\end{eqnarray}
which is consistent with the classical master constraint algebra in
this sense. As a result, the difficulty of the original Hamiltonian
constraint algebra can be avoided by introducing the master
constraint algebra, due to the Lie algebra structure of the latter.
It can be seen that zero is in the spectrum of $\hat{\mathbf{M}}$
\cite{thiemann15}, so the further task is to obtain the physical
Hilbert space $\mathcal{H}_{phys}$ which is the kernel of the master
constraint operator with some suitable physical inner product, and
the issue of quantum anomaly is represented in terms of the size of
$\mathcal{H}_{phys}$ and the existence of semi-classical states.
Note that the master constraint programme can be straightforwardly
generalized to include matter fields \cite{HM}. We list some open
problems in master constraint programme for further research.
\begin{itemize}

\item Kernel of Master Constraint Operator

Since the master constraint operator $\hat{\textbf{M}}$ is
self-adjoint, it is a practical problem to find the DID of
$\mathcal{H}_{Diff}$:
\begin{eqnarray}
\mathcal{H}_{Diff}&\sim&\int^\oplus
d\mu(\lambda)\mathcal{H}^\oplus_\lambda,\nonumber\\
<\Phi|\Psi>_{Diff}&=&\int_\mathbf{R}d\mu(\lambda)<\Phi|\Psi>_{\mathcal{H}^\oplus_{\lambda}},\nonumber
\end{eqnarray}
where $\mu(\lambda)$ is the spectral measure with respect to the
master constraint operator $\hat{\mathbf{M}}$. It is expected that
one can identify $\mathcal{H}^\oplus_{\lambda=0}$ with the physical
Hilbert space. However, as discussed in Ref.\cite{thiemann9}, such a
prescription would be ambiguous in the case where zero is only in
the continuous spectrum. Also certain physical information would be
lost in the case where zero is an embedded eigenvalue. The
prescription is unambiguous only if zero is an isolated eigenvalue,
in which case however the whole machinery of the DID is not need at
all because
$\mathcal{H}^\oplus_{\lambda=0}\subset\mathcal{H}_{Diff}$ and the
physical inner product coincides with the kinematical
(differomorphism invariant) one. To cure the problem, some improved
prescriptions are proposed also in Ref.\cite{thiemann9}, where one
decomposes the measure with respect to the spectrum types before
direct integral decomposition. Then certain ambiguities can be
canceled by some physical criterions, such as, a complete subalgebra
of bounded Dirac observables should be represented irreducibly as
self-adjoint operators on the physical Hilbert space, and the
resulting physical Hilbert space should admit a sufficient number of
semiclassical states. Nonetheless, due to the complicated structure
of master constraint operator, it is certainly difficult to manage
the spectrum analysis and direct integral decomposition. On the
other hand, in the light of the self-adjointness of the master
constraint operator and the Lie-algebra structure of the constraint
algebra, a formal group average strategy was introduced in
Ref.\cite{thiemann3} as another possible way to get the physical
Hilbert space, which posses potential relation with the
path-integral formulation. However, so far such a strategy is still
formal.

\item Dirac Observables

Classically, one can prove that a function $\mathcal{O}\in
C^\infty(\mathcal{M})$ is a weak observable with respect to the
scalar constraint if and only if
\begin{eqnarray}
\{\mathcal{O},\{\mathcal{O},\textbf{M}\}\}|_{\overline{\mathcal{M}}}=0.\nonumber
\end{eqnarray}
We define $\mathcal{O}$ to be a strong observable with respect to
the scalar constraint if and only if
\begin{eqnarray}
\{\mathcal{O},\textbf{M}\}|_{\mathcal{M}}=0,\nonumber
\end{eqnarray}
and to be a ultra-strong observable if and only if
\begin{eqnarray}
\{\mathcal{O},\mathcal{S}(N)\}|_{\mathcal{M}}=0.\nonumber
\end{eqnarray}
In quantum version, an observable $\hat{\mathcal{O}}$ is a weak
Dirac observable if and only if $\hat{\mathcal{O}}$ leaves
$\mathcal{H}_{phys}$ invariant, while $\hat{\mathcal{O}}$ is now
called a strong Dirac observable if and only if
$\hat{\mathcal{O}}$ commutes with the master constraint operator
$\hat{\textbf{M}}$. Given a bounded self-adjoint operator
$\hat{\mathcal{O}}$ defined on $\mathcal{H}_{Diff}$, for instance,
a spectral projection of some observables leaving
$\mathcal{H}_{Diff}$ invariant, if the uniform limit exists, the
bounded self-adjoint operator defined by group averaging
\begin{eqnarray}
\widehat{[\mathcal{O}]}:=\lim_{T\rightarrow\infty}\frac{1}{2T}\int_{-T}^{T}dt\
\hat{U}(t)^{-1} \hat{\mathcal{O}}\hat{U}(t)\nonumber
\end{eqnarray}
commutes with $\hat{\textbf{M}}$ and hence becomes a strong Dirac
observable on the physical Hilbert space.

\item Testing the Classical Limit of Master Constraint Operator

One needs to construct spatial diffeomorphism invariant
semiclassical states to calculate the expectation value and
fluctuation of the master constraint operator. If the results
coincide with the classical values up to $\hbar$ corrections, one
can go ahead to finish our quantization programme with confidence.
It is encouraging that within the so-called Algebraic Quantum
Gravity framework \cite{GT1}, the correct classical limit of a
master constraint operator is recently obtained
\cite{GT2}\cite{GT3}.

\end{itemize}

\subsection{ADM Energy Operator of Loop Quantum Gravity}

To solve the dynamical problem in loop quantum gravity, one may
consider to find a suitable Hamiltonian operator, in order to
settle up the problem of time. A strategy for that is to seek for
an operator corresponding to the ADM energy for asymptotically
flat spacetime, which equivalently takes the form \cite{thiemann8}
\begin{eqnarray}
E_{ADM}=\lim_{S\rightarrow\partial\Sigma}-2\kappa\beta^2\int_{S}dS
\frac{n_a}{\sqrt{|\det
q|}}\widetilde{P}_i^a\partial_b\widetilde{P}_j^b\delta^{ij},\label{ADM}
\end{eqnarray}
where $n_a$ is the normal co-vector of a close 2-sphere $S$ and
$dS$ is the coordinate volume element on $S$ induced from that of
a asymptotically Cartesian coordinate system on $\Sigma$.

In Ref.\cite{thiemann8}, Thiemann quantized the ADM energy
(\ref{ADM}) to obtain a positive semi-definite and self-adjoint
operator $\hat{E}_{ADM}$ as
\begin{eqnarray}
\hat{E}_{ADM}f_\alpha:=2\hbar^2\kappa\beta^2\sum_{v\in
V(\alpha)\cap\partial\Sigma}\frac{1}{\hat{V}_v}\delta^{ij}\hat{J}^v_i\hat{J}^v_jf_\alpha,\nonumber
\end{eqnarray}
which is defined on an extension of $\mathcal{H}_{kin}$ allowing for
edges without compact support (see the infinite tensor product
extension of kinematical Hilbert space \cite{thiemann13}). Since the
volume operator $\hat{V}_v$ commutes with the "total angular
momentum" operator $[\hat{J}^v]=\delta^{ij}\hat{J}^v_i\hat{J}^v_j$,
these two operators can be simultaneous diagonalized with respect to
certain linear combinations of spin network states. The eigenvalues
of $\hat{E}_{ADM}$ are of the form
$2\hbar^2\kappa\beta^2\Sigma_{v\in
V(\alpha)\cap\partial\Sigma}j_v(j_v+1)/\lambda_v$, where $\lambda_v$
is the eigenvalue of $\hat{V}_v$. Thus we may think that the spin
quantum numbers of spin network states are playing the role of the
occupied numbers of Fock states in quantum field theory, which
provide a non-linear Fork decomposition for loop quantum gravity.
This motivates us to call the future quantum dynamics of loop
quantum gravity as Quantum Spin Dynamics (QSD)
\cite{thiemann1}\cite{thiemann16}\cite{thiemann5}\cite{thiemann6}\cite{thiemann7}\cite{thiemann8}.

Moreover, $\hat{E}_{ADM}$ trivially commutes with all constraint
operators, since the gauge transformations are trivial at
$\partial\Sigma$. Hence $\hat{E}_{ADM}$ is a true quantum Dirac
observable. Then a meaningful time parameter can be selected by
the continuous one-parameter unitary group generated by
$\hat{E}_{ADM}$, which leads to a "Schr\"{o}dinger equation" for
QSD as:
\begin{eqnarray}
i\hbar\frac{\partial}{\partial t}f=\hat{E}_{ADM}f\ \ .\nonumber
\end{eqnarray}

\section{Applications and Advances}

This section is devoted as a summary of the applications and some
recent advances which are not discussed in the main content of the
article. After providing a guidance for beginners to references in
the current research of loop quantum cosmology, quantum black holes,
and black hole entropy calculation, the basic ideas of coherent
states construction will be sketched. We refer to Refs. \cite{AL},
\cite{inside} and \cite{lecture} for more concrete exploration. Some
key open problems in the current research of loop quantum gravity
will also be raised in our discussion.

\subsection{Symmetric Models and Black Hole Entropy}

It is well known that the most difficulty in general relativity is
the singularity problem. The presence of singularities, such as the
big bang and black holes, is widely believed to be a signal that
classical general relativity has been pushed beyond the domain of
its validity. Can loop quantum gravity at the present stage resolve
the singularity problem?

As the full quantum dynamics of loop quantum gravity has not been
solved completely, one then deals with the singularity problem in
certain symmetric models by applying the ideas and techniques from
loop quantum gravity. For simplifications, one generally freezes all
but a finite number of degrees of freedom by imposing the suitable
symmetry condition \cite{BK}. The symmetry-reduced models can also
provide a mathematically simple arena to test the ideas and
constructions in the full loop quantum gravity theory. The
singularity problem was first considered in the so-called loop
quantum cosmology models by imposing spatially homogeneity and (or)
isotropy. The seminar work by Bojowald \cite{bojowald} shows that
the big bang singularity is absent in loop quantum cosmology
\cite{ABL}. The result then leads to a new understanding on the
initial condition problem in quantum cosmology
\cite{boj}\cite{boj1}. Another remarkable result is that the loop
quantum cosmological modification of Friedmann equation may cure the
fine turning problem of the inflation potential, so that the
inflation can arise naturally and exit gracefully due to the quantum
geometry effect \cite{boj2}\cite{DH}. Recently, semiclassical states
are used to understand the quantum evolution of the universe across
the deep Planck regime \cite{APS}. It turns out that the classical
big bang is replaced by a quantum big bounce \cite{APS1}\cite{APS2}.
The predictions from loop quantum cosmology are reliable in the
sense that the quantum dynamics of both the homogeneous with
isotropic and with anisotropic models are proved to have correct
classical limits \cite{ABL}\cite{TM}. Loop quantum cosmology is
currently a very active research field. One may see Refs.\cite{boj3}
and \cite{asht} for brief overviews. For readers who want to know
the fundamental structure of loop quantum cosmology, we refer to
Refs. \cite{ABL} and \cite{velh2}. Also, a comprehensive review in
this field has already appeared \cite{boj6}.

By imposing spatially spherical symmetry, one can study
nonhomogeneous models, such as the Schwarzschild black hole
\cite{singular}, where the techniques from loop quantum gravity are
also employed \cite{boj4}. The treatment of these models is thus
quite similar to that of loop quantum cosmology. It turns out that
the interiors of the black holes are also singularity-free due to
the quantum geometric properties
\cite{singular}\cite{mod}\cite{boj5}\cite{AB2}. One may further
study the "end state" of the gravitational collapse of matter fields
inside a black hole \cite{BG}\cite{collapse} and black hole
evaporation \cite{AB}. One can find the basic framework and recent
results of loop quantum black hole in Ref.\cite{boj06}. There are
still appealing issues which one may consider about the quantum
black holes. The investigation in this direction has just started.
Besides the above noticeable models, there are also some other
symmetric models, such as the Husain-Kuchar model \cite{ALM} and
static spacetimes \cite{ma3}, which have been studied from the
constructions of loop quantum gravity.

Another very puzzling issue in general relativity is the
thermodynamics of black holes
\cite{bekenstein}\cite{BCH}\cite{wald1}. The black hole entropy
formula brings together the three pillars of fundamental physics:
general relativity, quantum theory and statistical mechanics.
However, the formula itself is obtained by a rather hodge-podge
mixture of classical and semi-classical ideas. Can one use loop
quantum gravity to calculate the microscopic degrees of freedom
which account for the black hole entropy?

We now turn to the black hole entropy calculation in loop quantum
gravity. Recall that the definition of the event horizon of a black
hole in general relativity concerns the global structure of the
spacetime \cite{wald}. However, to account for black hole entropy by
statistical calculations in loop quantum gravity, one needs to
define locally the notion of a horizon, which can assume that the
black hole itself is in equilibrium while the exterior geometry is
not forced to be time independent. This is the so-called isolated
horizon classically defined by Ashtekar et al (see Ref.\cite{AFK}
for a precise definition). It turns out that the zeroth and the
first laws of black-hole mechanics can be naturally extended to type
II isolated horizons \cite{AFK}\cite{ABL2}, where the horizon
geometry is axi-symmetric. If one considers the spacetimes which
contain an isolated horizon as an internal boundary, the action
principle and the Hamiltonian description are well defined. Note
that, in contrast with the symmetry-reduced models, here the phase
space has an infinite number of degrees of freedom.

In quantum kinematical setting, it is natural to begin with a total
Hilbert space $\mathcal{H}=\mathcal{H}_V\times\mathcal{H}_S$, where
$\mathcal{H}_V$ is built from suitable functions of generalized
connections in the bulk and $\mathcal{H}_S$ from suitable functions
of generalized surface connections. The horizon boundary condition
can then be imposed as an operator equation on $\mathcal{H}$. Taking
account of the structure of the surface term in the symplectic
structure, this quantum boundary condition implies that
$\mathcal{H}_S$ is the Hilbert space of a $U(1)$ Chern-Simons theory
on a punctured 2-sphere \cite{ABK2}\cite{AEP}. To calculate entropy,
one constructs the micro-canonical ensemble by considering only the
subspace of the bulk theory with a fixed area of the horizon (a
similar idea was raised in an earlier paper by Rovelli \cite{rov}).
Employing the spectrum (\ref{spectrum}) of the area operator in
$\mathcal{H}_V$, a detail analysis can estimates the number of
Chern-Simons surface states consistent with the given area. One thus
obtains the (black hole) horizon entropy, whose leading term is
indeed proportional to the horizon area \cite{ABK2}. However, the
expression of the entropy agrees with the Hawking-Bekenstein formula
only if one chooses a particular Barbero-Immirzi parameter $\beta_0$
(see Ref.\cite{lewandowski1} for a recent discussion on the choice
of $\beta_0$). The nontrivial fact is that this theory with fixed
$\beta_0$ can yield the Hawking-Bekenstein value of entropy of all
isolated horizons, irrespective of the values of charges, angular
momentum and cosmology constant, the amount of distortion or hair
\cite{AL}. The sub-leading term has also been calculated and shown
to be proportional to the logarithm of the horizon area \cite{KM}.
Note that in the entropy calculation the quantum Gauss and
diffeomorphism constraints are crucially used, while the final
result is insensitive to the details of how the Hamiltonian
constraint is imposed. There is an excellent review on this subject
in Ref.\cite{AL}.

\subsection{Construction of Coherent States}

As shown in section 6, both the Hamiltonian constraint operator
$\hat{\mathcal{S}}(N)$ and the master constraint operator
$\hat{\textbf{M}}$ can be well defined in the framework of loop
quantum gravity. However, since the Hilbert spaces
$\mathcal{H}_{kin}$ and $\mathcal{H}_{Diff}$, the operators
$\hat{\mathcal{S}}(N)$ and $\hat{\textbf{M}}$ are constructed in
such ways that are drastically different from usual quantum field
theory, one has to check whether the constraint operators and the
corresponding algebras have correct classical limits with respect to
suitable semiclassical states. In order to find the suitable
semiclassical states and check the classical limit of the theory,
the idea of a non-normalizable coherent state defined by a
generalized Laplace operator and its heat kernel was introduced for
the first time in Ref.\cite{ALM2}. Recently, kinematical coherent
states are constructed in two different approaches. One leads to the
so-called complexifier coherent states proposed by Thiemann et al
\cite{thiemann10}\cite{thiemann11}\cite{thiemann12}\cite{thiemann13}.
The other is promoted by Varadarajan
\cite{var1}\cite{var2}\cite{var3} and developed by Ashtekar et al
\cite{AL2}\cite{shadow}.

The complexifier approach is somehow motivated by the coherent state
construction for compact Lie groups \cite{hall}. One begins with a
positive function $C$ (complexifier) on the classical phase space
and arrives at a "coherent state" $\psi_m$, which more possibly
belongs to the dual space $Cyl^\star$ rather than
$\mathcal{H}_{kin}$. However, one may consider the so-called
"cut-off state" of $\psi_m$ with respect to a finite graph as a
graph-dependent coherent state in $\mathcal{H}_{kin}$
\cite{thiemann2}. By construction, the coherent state $\psi_m$ is an
eigenstate of an annihilation operator coming also from the
complexifier $C$ and hence has desired semiclassical properties
\cite{thiemann11}\cite{thiemann12}. We now sketch the basic idea of
its construction. Given the Hilbert space $\mathcal{H}$ for a
dynamical system with constraints and a subalgerba of observables
$\mathcal{S}$ in the space $\mathcal{L}(\mathcal{H})$ of linear
operators on $\mathcal{H}$, the semiclassical states with respect to
$\mathcal{S}$ are defined as Definition 2.2.5. Kinematical coherent
states $\{\Psi_m\}_{m\in\mathcal{M}}$ are semiclassical states which
in addition satisfy the annihilation operator property
\cite{thiemann10}\cite{thiemann2}, namely there exists certain
non-self-adjoint operator $\hat{z}=\hat{a}+i\lambda\hat{b}$ with
$\hat{a},\ \hat{b}\in\mathcal{S}$ and certain squeezing parameter
$\lambda$, such that
\begin{eqnarray}
\hat{z}\, \Psi_m=z(m)\Psi_m. \label{annihilation}
\end{eqnarray}
Note that Eq.(\ref{annihilation}) implies that the minimal
uncertainty relation is saturated for the pair of elements
$(\hat{a},\ \hat{b})$, i.e.,
\begin{eqnarray}
\Psi_m([\hat{a}-\Psi_m(\hat{a})]^2)=\Psi_m([\hat{b}-\Psi_m(\hat{b})]^2)=\frac{1}{2}|\Psi_m([\hat{a},
\hat{b}])|.
\end{eqnarray}
Note also that coherent states are usually required to satisfy the
additional peakedness property, namely for any $m\in\mathcal{M}$ the
overlap function $|<\Psi_m,\Psi_{m'}>|$ is concentrated in a phase
volume $\frac{1}{2}|\Psi_m([\hat{q}, \hat{p}])|$, where $\hat{q}$ is
a configuration operator and $\hat{p}$ a momentum operator. So the
central stuff in the construction is to define a suitable
"annihilation operator" $\hat{z}$ in analogy with the simplest case
of harmonic oscillator. A powerful tool named as "complexifier" is
introduced in Ref.\cite{thiemann10} to define a meaningful $\hat{z}$
operator which can give rise to kinematical coherent states for a general quantum system.\\ \\
\textbf{Definition 7.2.1}: \textit{Given a phase space
$\mathcal{M}=\mathrm{T}^*\mathcal{C}$ for some dynamical system
with configuration coordinates $q$ and momentum coordinates $p$, a
complexifier, $C$, is a positive smooth function on $\mathcal{M}$, such that\\
$(1)$ $C/\hbar$ is dimensionless;\\
$(2)$ $\lim_{||p||\rightarrow\infty}\frac{|C(m)|}{||p||}=\infty$ for
some suitable norm on the space of the momentum;\\
$(3)$ Certain complex coordinates $(z(m), \bar{z}(m))$ of
$\mathcal{M}$ can be constructed from $C$. }
\\ \\
Given a well-defined complexifier $C$ on phase space $\mathcal{M}$,
the programme for constructing coherent states associated with $C$
can be carried out as the following.
\begin{itemize}
\item {\it Complex polarization}

The condition (3) in Definition 7.3.1 implies that the complex
coordinate $z(m)$ of $\mathcal{M}$ can be constructed via
\begin{eqnarray}
z(m):=\sum_{n=0}^\infty\frac{i^n}{n!}\{q,C\}_{(n)}(m),\label{complex}
\end{eqnarray}
where the multiple Poisson bracket is inductively defined by
$\{q,C\}_{(0)}=q,\ \{q,C\}_{(n)}=\{\{q,C\}_{(n-1)},C\}$. One will
see that $z(m)$ can be regarded as the classical version of an
annihilation operator.

\item {\it Defining annihilation operator}

After the quantization procedure, a Hilbert space
$\mathcal{H}=L^2(\mathcal{C}, d\mu)$ with a suitable measure $d\mu$
on a suitable configuration space $\mathcal{C}$ can be constructed.
It is reasonable to assume that $C$ can be defined as a positive
self-adjoint operator $\hat{C}$ on $\mathcal{H}$. Then a
corresponding operator $\hat{z}$ can be defined by transforming the
Poisson brackets in Eq.(\ref{complex}) into commutators, i.e.,
\begin{eqnarray}
\hat{z}:=\sum_{n=0}^\infty\frac{i^n}{n!}\frac{1}{(i\hbar)^n}[\hat{q},\hat{C}]_{(n)}=e^{-\hat{C}/\hbar}\hat{q}
e^{\hat{C}/\hbar},
\end{eqnarray}
which is called as an \textit{annihilation operator}.

\item {\it Constructing coherent states}

Let $\delta_{q'}(q)$ be the $\delta$-distribution on $\mathcal{C}$
with respect to the measure $d\mu$. Since $\hat{C}$ is assumed to be
positive and self-adjoint, the conditions (1) and (2) in Definition
7.3.1 imply that $e^{-\hat{C}/\hbar}$ is a well-defined "smoothening
operator". So it is quite possible that the heat kernel evolution of
the $\delta$-distribution, $e^{-\hat{C}/\hbar}\delta_{q'}(q)$, is a
square integrable function in $\mathcal{H}$, which is even analytic.
Then one may analytically extend the variable $q'$ in
$e^{-\hat{C}/\hbar}\delta_{q'}(q)$ to complex values $z(m)$ and
obtain a class of states $\psi'_m$ as
\begin{eqnarray}
\psi'_m(q):=[e^{-\hat{C}/\hbar}\delta_{q'}(q)]_{q'\rightarrow
z(m)},\label{coherent}
\end{eqnarray}
such that one has
\begin{eqnarray}
\hat{z}\,
\psi'_m(q):=[e^{-\hat{C}/\hbar}\hat{q}\delta_{q'}(q)]_{q'\rightarrow
z(m)}=[q'e^{-\hat{C}/\hbar}\delta_{q'}(q)]_{q'\rightarrow
z(m)}=z(m)\, \psi'_m(q).
\end{eqnarray}
Hence $\psi'_m$ is automatically an eigenstate of the annihilation
operator $\hat{z}$. So it is natural to define coherent states
$\psi_m(q)$ by normalizing $\psi'_m(q)$.\\
\end{itemize}
One may check that all the coherent state properties usually
required are likely to be satisfied by the above complexfier
coherent states $\psi_m(q)$ \cite{thiemann2}. As a simple example,
in the case of one-dimensional harmonic oscillator with Hamiltonian
$H=\frac{1}{2}(\frac{p^2}{2m}+\frac{1}{2}m\omega^2q^2)$, one may
choose the complexifier $C=p^2/(2m\omega)$. It is straightforward to
check that the coherent state constructed by the above procedure
coincides with the usual harmonic oscillator coherent state up to a
phase \cite{thiemann2}. So the complexifier coherent state can be
considered as a suitable generalization of the concept of usual
harmonic oscillator coherent state.

The complexifer approach can be used to construct kinematical
coherent states in loop quantum gravity. Given a suitable
complexifier $C$, for each analytic path $e\subset\Sigma$ one can
define
\begin{eqnarray}
A^{\mathbf{C}}(e):=\sum_{n=0}^\infty\frac{i^n}{n!}\{A(e),C\}_{(n)},\label{cconnection}
\end{eqnarray}
where $A(e)\in SU(2)$ is assigned to $e$. As the complexifier $C$ is
assumed to give rise to a positive self-adjoint operator $\hat{C}$
on the kinematical Hilbert space $\mathcal{H}_{kin}$, one further
supposes that $\hat{C}/\hbar T_s=\tau\lambda_sT_s$, where $\tau$ is
a so-called classicality parameter, $\{T_s(A)\}_s$ form a system of
basis in $\mathcal{H}_{kin}$ and are analytic in
$A\in\overline{\mathcal{A}}$. Moreover the $\delta$-distribution on
the quantum configuration space $\overline{\mathcal{A}}$ can be
formally expressed as $\delta_{A'}(A)=\sum_s
T_s(A')\overline{T_s(A)}$. Thus by applying Eq.(\ref{coherent}) one
obtains coherent states
\begin{eqnarray}
\psi'_{A^\mathbf{C}}(A)=(e^{-\hat{C}/\hbar})\delta_{A'}(A)|_{A'\rightarrow
A^\mathbf{C}}=\sum_se^{-\tau\lambda_s}T_s(A^{\mathbf{C}})\overline{T_s(A)}.\label{cstate}
\end{eqnarray}
However, since there are uncountable infinite number of terms in the
expression (\ref{cstate}), the norm of $\psi'_{A^\mathbf{C}}(A)$
would in general be divergent. So $\psi'_{A^\mathbf{C}}(A)$ is
generally not an element of $\mathcal{H}_{kin}$ but rather an
distribution on a dense subset of $\mathcal{H}_{kin}$. In order to
test the semiclassical limit of quantum geometric operators on
$\mathcal{H}_{kin}$, one may further consider the "cut-off state" of
$\psi'_{A^\mathbf{C}}(A)$ with respect to a finite graph $\gamma$ as
a graph-dependent coherent state in $\mathcal{H}_{kin}$
\cite{thiemann2}. So the key input in the construction is to choose
a suitable complexifer. There are vast possibilities of choice. For
example, a candidate complexifier $C$ is considered in
Ref.\cite{lecture} such that the corresponding operator acts on
cylindrical functions $f_{\gamma}$ by
\begin{eqnarray}
(\hat{C}/\hbar)f_{\gamma}=\frac{1}{2}(\sum_{e\in
E(\gamma)}l(e)\hat{J}_e^2)f_{\gamma},
\end{eqnarray}
where $\hat{J}_e^2$ is the Casimir operator defined by
Eq.(\ref{casimir}) associated to the edge $e$, the positive numbers
$l(e)$ satisfy $l(e\circ e')=l(e)+l(e')$ and $l(e^{-1})=l(e)$. Then
it can be shown from Eq.(\ref{cconnection}) that $A^{\mathbf{C}}(e)$
is an element of $SL(2, \mathbf{C})$. So the classical
interpretation of the annihilation operators is simply the
generalized complex $SU(2)$ connections. It has been shown in Refs.
\cite{thiemann11} and \cite{thiemann12} that the "cut-off state" of
the corresponding coherent state,
\begin{eqnarray}
\psi_{A^\mathbf{C},\gamma}(A)=\psi'_{A^\mathbf{C},\gamma}(A)/||\psi'_{A^\mathbf{C},\gamma}(A)||,
\end{eqnarray}
has desired semiclassical properties, where
\begin{eqnarray}
\psi'_{A^\mathbf{C},\gamma}(A):=\sum_{s,\gamma(s)=\gamma}e^{-\frac{1}{2}\sum_{e\in
E(\gamma(s))}l(e)j_e (j_e+1)}T_s(A^{\mathbf{C}})\overline{T_s(A)}.
\end{eqnarray}
But unfortunately, these cut-off coherent states cannot be directly
used to test the semiclassical limit of the Hamiltonian constraint
operator $\hat{\mathcal{S}}(N)$, since $\hat{\mathcal{S}}(N)$ is
graph-changing so that its expectation values with respect to these
cut-off states are always zero! So further work in this approach is
expected in order to overcome the difficulty. Anyway, the
complexifier approach provides a clean construction mechanism and
manageable calculation method for semiclassical analysis in loop
quantum gravity.

We now turn to the second approach. As we have seen, loop quantum
gravity is based on quantum geometry, where the fundamental
excitations are 1-dimensional polymer-like. On the other hand, low
energy physics is based on quantum field theories which are
constructed in a flat spacetime continuum. The fundamental
excitations of these fields are 3-dimensional, typically
representing wavy undulations on the background Minkowskian
geometry. The core strategy in this approach is then to relate the
polymer excitations of quantum geometry to Fock states used in low
energy physics and to locate Minkowski Fock states in the background
independent framework. Since quantum Maxwell field can be
constructed in both Fock representation and polymer-like
representation, one first gains insights from the comparison between
the two representations, then generalizes the method to quantum
geometry. A "Laplacian operator" can be defined on
$\mathcal{H}_{kin}$ \cite{ALM2}\cite{AL2}, from which one may define
a candidate coherent state $\Phi_0$, also in $Cyl^\star$,
corresponding to the Minkowski spacetime. To calculate the
expectation values of kinematical operators, one considers the
so-called "shadow state" of $\Phi_0$, which is the restriction of
$\Phi_0$ to a given finite graph. However, the construction of
shadow states is subtly different from that of cut-off states.

We will only describe the simple case of Maxwell field to illustrate
the ideas of construction \cite{var1}\cite{var2}\cite{AL}. Following
the quantum geometry strategy discussed in Sec.4, the quantum
configuration space $\overline{\mathbf{A}}$ for the polymer
representation of the $U(1)$ gauge theory can be similarly
constructed. A generalized connection
$\mathbf{A}\in\overline{\mathbf{A}}$ assigns each oriented analytic
edge in $\Sigma$ an element of $U(1)$. The space
$\overline{\mathbf{A}}$ carries a diffeomorphism and gauge invariant
measure $\mu_0$ induced by the Haar measure on $U(1)$, which give
rise to the Hilbert space,
$\mathcal{H}_0:=L^2(\overline{\mathbf{A}}, d\mu_0)$, of polymer
states. The basic operators are holonomy operators
$\hat{\mathbf{A}}(e)$ labeled by 1-dimensional edges $e$, which act
on cylindrical functions by multiplication, and smeared electric
field operators $\hat{E}(g)$ for suitable test 1-forms $g$ on
$\Sigma$, which are self-adjoint. Note that, since the gauge group
$U(1)$ is Abelian, it is more convenient to smear the electric
fields in 3 dimensions \cite{AL}. The eigenstates of $\hat{E}(g)$,
so-called flux network states $\mathcal{N}_{\alpha, \vec{n}}$,
provide an orthonormal basis in $\mathcal{H}_0$, which are defined
for any finite graph $\alpha$ with $N$ edges as:
\begin{equation}
\mathcal{N}_{\alpha,\vec{n}}(\mathbf{A}):=[{\mathbf{A}}(e_1)]^{n_1}
[\mathbf{A}(e_2)]^{n_2}\cdot\cdot\cdot[\mathbf{A}(e_N)]^{n_N},
\end{equation}
where $\vec{n}\equiv(n_1,\cdot\cdot\cdot,n_N)$ assigns an integer
$n_I$ to each edge $e_I$. The action of $\hat{E}(g)$ on the flux
network states reads
\begin{equation}
\hat{E}(g)\, \mathcal{N}_{\alpha, \vec{n}}=-\hbar(\sum_I
n_I\int_{e_I}g)\mathcal{N}_{\alpha, \vec{n}}.
\end{equation}
In this polymer-like representation, cylindrical functions are
finite linear combinations of flux network states and span a dense
subspace of $\mathcal{H}_0$. Denote $\mathbf{Cyl}$ the set of
cylindrical functions and $\mathbf{Cyl}^\star$ its algebraic dual.
One then has a triplet $\mathbf{Cyl}\subset\mathcal{H}_0\subset
\mathbf{Cyl}^\star$ in analogy with the case of loop quantum
gravity.

The Schr\"{o}dinger or Fock representation of the Maxwell field, on
the other hand, depends on the Minkowski background metric. Here the
Hilbert space is given by $\mathcal{H}_F=L^2(\mathcal{S}', d\mu_F)$,
where $\mathcal{S}'$ is the appropriate space of tempered
distributions on $\Sigma$ and $\mu_F$ is the Gaussian measure. The
basic operators are connections $\hat{\mathbf{A}}(f)$ smeared in 3
dimensions with suitable vector densities $f$ and smeared electric
fields $\hat{E}(g)$. But $\hat{\mathbf{A}}(e)$ fail to be well
defined. To resolve this tension between the two representations,
one proceeds as follows. Let $\vec{x}$ be the Cartesian coordinates
of a point in $\Sigma=\mathbf{R}^3$. Introduce a test function by
using the Euclidean background metric on $\mathbf{R}^3$,
\begin{eqnarray}
f_r(\vec{x})=\frac{1}{(2\pi)^{3/2}r^3} \exp(-|\vec{x}|^2/2r^2),
\end{eqnarray}
which approximates the Dirac delta function for small $r$. The
Gaussian smeared form factor for an edge $e$ is defined as
\begin{equation}
X^a_{(e,r)}(\vec{x}):=\int_e ds\,f_r(\vec{e}(s)-\vec{x})\dot{e}^a.
\end{equation}
Then one can define a smeared holonomy for $e$ by
\begin{equation}
\mathbf{A}_{(r)}(e):=\exp[-i\int_{\mathbf{R}^3}X^a_{(e,r)}(\vec{x})A_a(\vec{x})],
\end{equation}
where $A_a(\vec{x})$ is the $U(1)$ connection 1-form of the Maxwell
field on $\Sigma$. Similarly one can define Gaussian smeared
electric fields by
\begin{equation}
E_{(r)}(g):=\int_{\mathbf{R}^3}g_a(\vec{x})\int_{\mathbf{R}^3}f_r(\vec{y}-\vec{x})E^a(\vec{y}).
\end{equation}
In this way one obtains two Poission brackets algebras. One is
formed by smeared holonomies and electric fields with
\begin{eqnarray}
\{\mathbf{A}_{(r)}(e), \mathbf{A}_{(r)}(e')\}=0=\{E(g), E(g')\}\\
\nonumber \{\mathbf{A}_{(r)}(e),
E(g)\}=-i(\int_{\mathbf{R}^3}X^a_{(e,r)}g_a)\, \mathbf{A}_{(r)}(e).
\end{eqnarray}
The other is formed by unsmeared holonomies and Gaussian smeared
electric fields with
\begin{eqnarray}
\{\mathbf{A}(e), \mathbf{A}(e')\}=0=\{E_{(r)}(g), E_{(r)}(g')\}\\
\nonumber \{\mathbf{A}(e),
E_{(r)}(g)\}=-i(\int_{\mathbf{R}^3}X^a_{(e,r)}g_a)\, \mathbf{A}(e).
\end{eqnarray}
Obviously, there is an isomorphism between them,
\begin{eqnarray}
I_r: \, (\mathbf{A}_{(r)}(e), E(g)) \mapsto (\mathbf{A}(e),
E_{(r)}(g)).
\end{eqnarray}
Using the isomorphism $I_r$, one can pass back and forth between the
polymer and the Fock representations. Specifically, the image of the
Fock vacuum can be shown to be the following element of
$\mathbf{Cyl}^\star$ \cite{var1}\cite{var2},
\begin{equation}
(V|\, = \, \sum_{\alpha,{\vec n}}\, \exp [ -\frac{\hbar}{2}
\sum_{IJ} G_{IJ}n_I n_J ]\, (\mathcal{N}_{\alpha, \vec{n}}|
,\label{Fvacuum}
\end{equation}
where $(\mathcal{N}_{\alpha, \vec{n}}|\in\mathbf{Cyl}^\star$ maps
the flux network function $|\mathcal{N}_{\alpha, \vec{n}}\rangle$ to
one and every other flux network functions to zero. While the states
$(\mathcal{N}_{\alpha, \vec{n}}|$ do not have any knowledge of the
underlying Minkowskian geometry, this information is coded in the
matrix $G_{IJ}$ associated with the edges of the graph $\alpha$,
given by \cite{AL}
\begin{equation}
G_{IJ}\ =  \int_{e_I}dt \dot{e}^a_I(t) \int_{e_J}dt' \dot{e_J}^b
(t')\, \int d^3x\, \delta_{ab}(\vec{x})\,
[f_r(\vec{x}-\vec{e}_I(t))\,
|\Delta|^{-\frac{1}{2}}\,f(\vec{x},\vec{e}_J(t'))],
\end{equation}
where $\delta_{ab}$ is the flat Euclidean metric and $\Delta$ its
Laplacian. Therefore, one can single out the Fock vacuum state
directly in the polymer representation by invoking Poincar\'e
invariance without any reference to the Fock space. Similarly, one
can directly locate in $\mathbf{Cyl}^\star$ all coherent states as
the eigenstates of the exponentiated annihilation operators. Since
$\mathbf{Cyl}^\star$ does not have an inner product, one uses the
notion of shadow states to do semiclassical analysis in the polymer
representation. From Eq.(\ref{Fvacuum}), the action of the Fock
vacuum $(V|$ on $\mathcal{N}_{\alpha, \vec{n}}$ reads
\begin{eqnarray}
(V|\mathcal{N}_{\alpha, \vec{n}}\rangle\
=\int_{\overline{\mathbf{A}}_\alpha}d\mu^0_\alpha\,
\overline{V}_\alpha \mathcal{N}_{\alpha, \vec{n}},
\end{eqnarray}
where the state $V_\alpha$ is in the Hilbert space
$\mathcal{H}_\alpha$ for the graph $\alpha$ and given by
\begin{equation}
V_\alpha (\mathbf{A})\, = \, \sum_{\vec n}\, \exp [ -\frac{\hbar}{2}
\sum_{IJ} G_{IJ}n_I n_J ]\, \mathcal{N}_{\alpha,
\vec{n}}(\mathbf{A}) .
\end{equation}
Thus for any cylindrical functions $\varphi_\alpha$ associated with
$\alpha$,
\begin{equation}
(V|\varphi_\alpha\rangle\ =\langle V_\alpha |\varphi_\alpha\rangle,
\end{equation}
where the inner product in the right hand side is taken in
$\mathcal{H}_\alpha$. Hence $V_\alpha(\mathbf{A})$ are referred to
as "shadows" of $(V|$ on the graphs $\alpha$. The set of all shadows
captures the full information in $(V|$. By analyzing shadows on
sufficiently refined graphs, one can introduce criteria to test if a
given element of $\mathbf{Cyl}^\star$ represents a semi-classical
state \cite{AL}. It turns out that the state $(V|$ does satisfy this
criterion and hence can be regarded as semi-classical in the polymer
representation.

The mathematical and conceptual tools gained from simple models like
the Maxwell fields are currently being used to construct
semiclassical states of quantum geometry. A candidate kinematical
coherent state corresponding to the Minkowski spacetime has been
proposed by Ashtekar and Lewandowki in the light of a "Laplacian
operator" \cite{AL2}\cite{AL}. However, the detail structure of this
coherent state is yet analyzed and there is no a priori guarantee
that it is indeed a semiclassical state.

One may find comparisons of the two approaches from both sides
\cite{complexifier}\cite{AL}. It turns out that the Varadarajan's
Laplacian coherent state for polymer Maxwell field can also be
derived from Thiemann's complexifier method. However, one cannot
find a complexifier to get the coherent state proposed by Ashtekar
et al for loop quantum gravity. Both approaches have their own
virtues and need further developments. The complexifier approach
provides a clean construction mechanism and manageable calculation
method, while the Laplacian operator approach is related closely
with the well-known Fock vacuum state. We expect that a judicious
combination of the two approaches may lead to significant progress
in semiclassical analysis of loop quantum gravity.

\subsection{Semiclassical Analysis and Quantum Dynamics}

Although powerful tools have been developed to construct
semiclassical states, the analysis of the classical limits of the
Hamiltonian constraint operator and the corresponding constraints
algebra has not been carried out. Although the semiclassical
analysis of a master constraint operator is being carried out in
the framework of Algebraic Quantum Gravity proposed by Giesel and
Thiemann \cite{GT1}\cite{GT2}\cite{GT3}, one still needs
diffeomorphism invariant coherent states in $\mathcal{H}_{Diff}$
(see Refs. \cite{complexifier} and \cite{ABC} for recent progress
in this aspect) to do semiclassical analysis of the master
constraint operator in loop quantum gravity. Moreover, a crucial
question of the semiclassical analysis is whether there are enough
physical semiclassical states in certain unknown physical Hilbert
space of loop quantum gravity, which may correspond to all
classical solutions of the Einstein equation. This is the final
theoretical criterion for any candidate theory of quantum gravity
with general relativity as its classical limit. The physical
semiclassical states are also relevant, if one wishes to use the
full theory rather than symmetric models to analyze cosmology and
black holes. In the matter coupled to gravity content, one would
like to check whether the coupled quantum system approaches
quantum field theory in curved spacetime in suitable semiclassical
limit. This issue is being studied at the kinematical level
\cite{ST1}\cite{ST2}.

In the light of the canonical quantization of loop quantum
gravity, the so-called spin foams are devised as histories traced
out by "time evolution" of spin networks, which provide a
path-integral approach to quantum dynamics \cite{baez}. One
expects that the path integral can be used to compute "transition
amplitudes" and extract physical states, which may shed new light
on the quantum Hamiltonian constraint and on the physical inner
product. In the successful Barrett-Crane model and its various
modifications \cite{BC1}\cite{BC2}, one regards classical general
relativity as a topological field theory (the so-called BF
theory), supplemented with an algebraic constraint. An interesting
discovery in this approach is that a certain modified version of
the Barrett-Crane model is equivalent to a manageable group field
theory \cite{perez1}\cite{PR}\cite{CPR}. It then turns out that
the sum over geometries for a fixed discrete topology is finite.
For a detail exploration of spin foam models, we refer to the
recent review article \cite{perez} and references therein.

Although many developments in spin foam approach are very
interesting from a mathematical physics perspective, their
significance to quantum gravity is still less clear \cite{AL}. An
obvious weakness in most of these works is that the discrete
topology is fixed, whence the the issue of summing over all
topologies remains largely unexplored. However, it is expected that
a judicious comparison of methods from the canonical treatment of
the Hamiltonian constraint and spin foam models may promote the
research in both approaches. In fact, there are considerable
attempts to calculate particle scattering amplitude in
non-perturbative quantum gravity by combining the methods from the
two approaches \cite{MR1}\cite{MR2}. There are also other approaches
to deal with the quantum dynamics such as, the Vassiliev knot
invariants approach \cite{GP1} and the "consistent discretization"
approach \cite{GP2}\cite{GP3}. Here we will not introduce their
concrete ideas. One may find the detail exploration of the former in
Refs. \cite{BG1} and \cite{BG2}, and a recent summary for the latter
in Ref.\cite{GP}.

In summary, the full treatments of the semiclassical analysis and
quantum dynamics are entangled with each other and expected to be
settled together. These are the core open problems in loop quantum
gravity, which are now under investigations.

\section*{Acknowledgments}

The authors would like to thank Abhay Ashtekar and Jerzy Lewandowski
for careful reading of the original manuscript and many valuable
comments and suggestions. Especially they thank Carlo Rovelli and
Thomas Thiemann for their advanced lectures at BNU and many
enlightening discussions. Muxin Han would like to acknowledge all
the members in the relativity group at BNU for their kind support,
especially thank You Ding, Xuefei Gong, Li Qin, and Hongbao Zhang
for their kind help and smooth cooperation. Also, Muxin Han
appreciates Dr. Andrzej Oko{\l}\'{o}w for the enlightening
discussions about the quantum algebra of loop quantum gravity. This
work is a part of projects 10205002 and 10675019 supported by NSFC.
Muxin Han would also like to acknowledge support from Undergraduate
Research Foundation of BNU, fellowship and assistantship of LSU,
Hearne Foundation of LSU, and funding from Advanced Research and
Development Activity.

\end{document}